\documentclass{article}
\newcommand{\define}{\buildrel \Delta \over =}
\usepackage{amsmath}
\usepackage{amssymb}
\usepackage{graphicx}
\usepackage{pslatex}
\usepackage{epstopdf}
\usepackage{subfig}
\usepackage{epsfig}

\begin{document}
%
\title{Information Sources on a Bratteli Diagram}

\author{John~C.~Kieffer\thanks{John C.\ Kieffer is with Dept.\ of Electrical \& Computer Engineering, University of Minnesota Twin Cities, Minneapolis, Minnesota USA. e-mail: kieffer@umn.edu.}}

\maketitle

\begin{abstract}
A Bratteli diagram is a type of graph in which the vertices
are split into finite subsets occupying an infinite sequence of levels, starting with a bottom level
and moving to successively higher levels along edges connecting
consecutive levels. An information source on a Bratteli diagram
consists of a sequence of PMFs  on the vertex sets at each level
that are compatible under edge transport. 
By imposing a regularity condition on the Bratteli diagram,
we obtain various results for its information sources including
ergodic and entropy rate decomposition theorems, a Shannon-Mcmillan-Breiman
theorem, and lossless and lossy source coding theorems.
Proof methodology exploits the Vershik transformation on the
path space of a Bratteli diagram.
Some results for finite alphabet stationary sequential information sources
are seen to be a special case of the results of this paper. 
\end{abstract}


%

\section{Introduction}

Let $A$ be a finite set and let $A^{\infty}$ be the product measurable space 
of one-sided sequences from $A$. The family of stationary sequential information sources with alphabet $A$ can be viewed
as the set ${\cal P}(A^{\infty})$ of probability measures on $A^{\infty}$
that are preserved by the shift transformation $T_A$ on $A^{\infty}$. 
We also have the family of 
ergodic sources ${\cal P}_e(A^{\infty})$, which consists of those sources in ${\cal P}(A^{\infty})$
that are trivial on the sigma-field of $T_A$-invariant measurable subsets of $A^{\infty}$. 
Each source $\mu$ in ${\cal P}(A^{\infty})$ has an entropy rate, defined as the entropy of 
the dynamical system $(A^{\infty},T_A,\mu)$. In classical information source theory, there is
a well-established body of results regarding the information sources in ${\cal P}(A^{\infty})$. 
These results include the ergodic decomposition theorem \cite{graydavisson}, 
the entropy rate decomposition theorem \cite{jacobs} \cite[Thm.\ 2.4.1]{grayEI}, the 
Shannon-McMillan-Breiman theorem \cite{mcmillan} \cite{breiman}, universal
lossless source coding theorems \cite{davisson}, and a fixed-length lossy source coding theory \cite{partha}.

\par

Efforts have been expended in developing results analogous to the results just cited, for information sources going
beyond the family of sequential sources ${\cal P}(A^{\infty})$. 
For example, progress along these
lines has occurred for   asymptotically mean stationary information sources 
 \cite{graykieffer} \cite[Chap. 4]{grayEI},  for random fields on trees and on
the lattice ${\mathbb Z}^d$ for $d>1$ \cite{thouvenot}\cite{yeberger}, for certain quantum source  families  \cite{kalt} \cite{bjelakovic},
and for random graphical structures \cite{choi}. 
\par
This paper is concerned with the further development  of information source theory on graphs.
The type of graph considered is called a Bratteli diagram, formally defined below. 
Originally, Bratteli diagrams were 
put forward \cite{bratteli} as a tool to resolve structural questions concerning certain operator algebras (AF-algebras). 
Subsequently, they were studied as a means to obtain models of measurable dynamical systems \cite{vershik0}\cite{vershik} and certain 
topological dynamical systems \cite{HPS}.  In recent years, there has been extensive work on
ergodic-theoretic questions arising from a Bratteli diagram $D$; the
papers \cite{frick}\cite{melapetersen}\cite{bez}\cite{bez2}\cite{dooley} are representative of this 
ergodic-theoretic line of research. 
\par 
The vertices of a Bratteli diagram $D$ occupy a countably infinite number of levels, namely, 
a level $0$, a level $1$, a level $2$, etc. Edges of $D$ can only connect
vertices between consecutive levels. We see in Section II that
there is a path space $\Omega_D$ associated with $D$ consisting of certain infinite paths, each path starting at level $0$ and ascending
consecutively from level to level following edges connecting levels. There is also
a natural one-to-one transformation $T$
of $\Omega_D$ onto itself called the Vershik transformation. A Bratteli diagram $D$ induces
various dynamical systems called Bratteli-Vershik systems, which are the triples $(\Omega_D,T,P)$ in which
$P$ is a probability measure on $\Omega_D$ preserved by $T$.
As we shall see, each Bratteli-Vershik system gives rise to
a certain information source which we call a Bratteli-Vershik source. We shall obtain results for Bratteli-Vershik sources analogous
to classical source coding theory results obtained for the sequential source family ${\cal P}(A^{\infty})$. 
In analyzing a Bratteli-Vershik source, one exploits how the Vershik transformation $T$ acts on the path space $\Omega_D$,
whereas, in analyzing a sequential source in ${\cal P}(A^{\infty})$, one exploits how the shift transformation
$T_A$ acts on the sequence space $A^{\infty}$. Since the action of $T$ on $\Omega_D$ is unlike the action
of $T_A$ on $A^{\infty}$, proof methodologies for obtaining Bratteli-Vershik source results differ from the
methodologies for obtaining sequential source results, even though the two types of results are analogous.  
\par
{\it Notation and Terminology.} 
\begin{itemize}
\item $|S|$ denotes
the cardinality of finite set $S$, $S^*$ denotes the set of all finite-length strings
with entries from $S$, and $|s|$ denotes the length of string $s\in S^*$. If $s_i$
is a string in $S^*$ for $1\leq i \leq n$, then $s_1s_2\cdots s_n \in S^*$ denotes
the string which is the concatenation of $s_1,s_2,\cdots,s_n$. 
$S^*$ is a semigroup under the concatenation operation $(s_1,s_2)\to s_1s_2$.
\item   The terminology {\it random object} designates a measurable function mapping
a probability space into some measurable space; {\it random variable} designates
a random object which takes its values in the real line (or extended real line). 
\item A finite set $S$ is taken to be the measurable space with measurable sets all subsets of $S$,
and if $\mu$ is a PMF on $S$, we shall also use $\mu$ to denote the probability measure on $S$
induced by $\mu$. 
\item The Shannon entropy $H(\mu)$ of a PMF $\mu$ on finite set $S$ 
is defined by 
$$H(\mu) \define \sum_{s\in S,\;\mu(s)>0}-\mu(s)\log_2\mu(s).$$
Furthermore, if $X$ is an $S$-valued random object on probability space $(\Omega,\cal F,P)$ whose PMF is $\mu$, then $H(X) \define H(\mu)$, or we write $H_P(X)$ for $H(X)$ if the probability measure $P$ is to be emphasized.
All elementary properties of entropy (such as concavity and subadditivity) are assumed throughout.
\end{itemize}
\par

{\it Definition: Bratteli diagram.} Suppose we have a graph $(V,E)$ such that
\begin{itemize}
\item {\bf (a.1):} The set of vertices $V$ is countably infinite and decomposes as a disjoint union
$V = \bigcup_{n=0}^{\infty}V_n$ in which each $V_n$ is finite and nonempty. $V_n$ is referred
to as the set of vertices of $V$ at level $n$.
\item {\bf (a.2):} The set of edges $E$ is countably infinite and decomposes as a disjoint union
$E = \bigcup_{n=0}^{\infty}E_n$ in which each $E_n$ is finite and nonempty.
\item {\bf (a.3):} For each $n\geq 0$, each edge $e$ in $E_n$ connects a vertex in $V_n$ (called
the {\it source vertex} of $e$ and denoted $s(e)$) with
a vertex in $V_{n+1}$ (called the {\it range vertex} of $e$ and denoted $r(e)$). Thus,
the edges in $E_n$ connect level $n$ vertices with level $n+1$ vertices.
\item {\bf (a.4):} For each $n\geq 0$ and each vertex  $v\in V_n$, there
is at least one edge $e\in E_n$ such that $s(e)=v$.
\item {\bf (a.5):} For each $n\geq 1$ and each vertex $v\in V_n$, the set
of edges $E(v)\define \{e\in E_{n-1}:r(e)=v\}$ is nonempty. 
\end{itemize}
Then $D=(V,E)$ is called a {\it Bratteli diagram}. Two Bratteli diagrams are said to be isomorphic if they are isomorphic as graphs. 
\par
Figures 1(a) and 1(b) illustrate two examples of Bratteli diagrams discussed in the following subsection.
\par

\subsection{Bratteli diagrams via multisets} 
The multiset concept is a generalization of the concept of set and is a useful concept in dealing with Bratteli diagrams. Recall that a set is an unordered collection
of distinct elements. A multiset is an unordered collection of elements
in which repetitions of elements are allowed. 
Each distinct element $s$ of a multiset $M$ appears in $M$ with 
a certain multiplicity $m(s|M)$. The cardinality $|M|$ of $M$ is the total number of elements of $M$
including multiplicities, that is, $|M|$ is the sum of $m(s|M)$ over the distinct elements $s$ of $M$. 
For example, for the multiset $M=\{a,a,a,b,b\}$, we have
$m(a|M)=3$, $ m(b|M)=2$, and $|M|=m(a|M)+m(b|M)=5$. 
\par

Let $D=(V,E)$ be a Bratteli diagram. $V^+$ denotes the set of vertices $\cup_{n\geq 1}V_n$. 
$\{E(v):v\in V^+\}$ forms a partition of the edge set $E$. 
For each $v\in V^+$, 
we define $M_D(v)$ to be the multiset of cardinality $|E(v)|$ whose set of distinct elements
is $\{s(e):e\in E(v)\}$ and the multiplicity of each vertex $v'$ in this set is
$$m(v'|M_D(v))  = |\{e\in E(v):s(e)=v'\}|.$$
Suppose that one knows the levels $\{V_n:n\geq 0\}$ of the vertex set of a Bratteli diagram $D=(V,E)$
as well as the multisets $\{M_D(v):v\in V^+\}$. Then $D$ is uniquely determined up to isomorphism.
Consequently, we sometimes specify a Bratteli diagram $D=(V,E)$
by specifying its vertex set levels $\{V_n\}$ and the multisets associated with
the vertices in $V^+$. We do this in Examples 1.1 and 1.2 which
follow.
\par

\begin{figure}[htb]
\centering
\includegraphics[width=\textwidth]{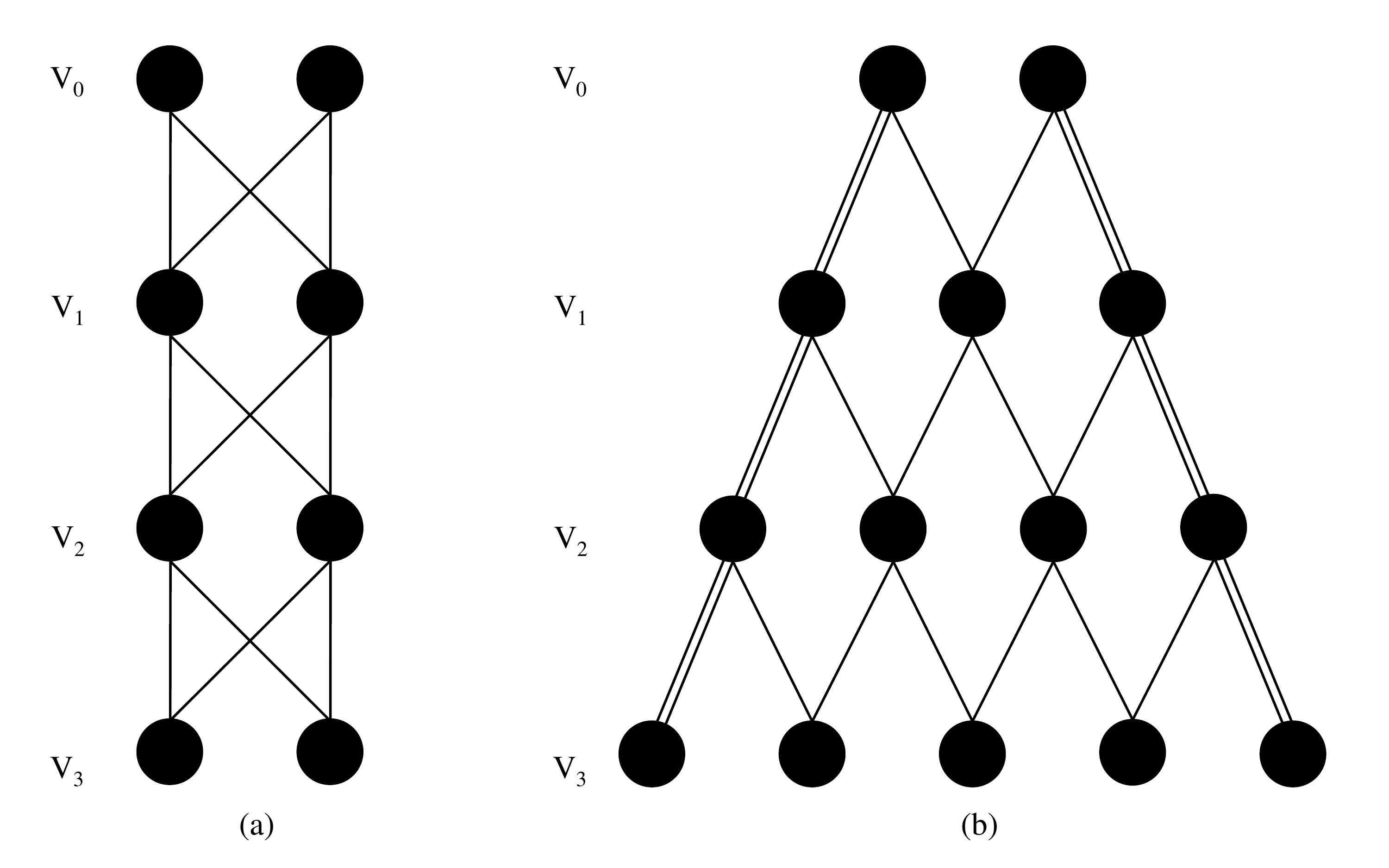}
\caption{Levels 0-3 of regular Bratteli diagrams in Ex. 1.1 (a) and Ex. 1.2 (b).}
 \end{figure}

\textbf{Example  1.1.}  Consider the Bratteli diagram $D=(V,E)$ in which
$$V_n = \{v_0(n),v_1(n)\},\;\;n\geq 0,$$
$$M_D(v) = \{v_0(n-1),v_1(n-1)\},\;\;v\in V_n,\;\;n\geq 1.$$
Figure 1(a) depicts levels 0-3 of $D$ (consisting of $V_0$ through $V_3$ and their connecting edges). 
\par

\textbf{Example  1.2.}  Let $D=(V,E)$ be the Bratteli diagram 
such that
$$V_n = \{v_0(n), v_1(n),\cdots,v_{n+1}(n)\},\;\;n\geq 0$$
and such that
\begin{eqnarray*}
M_D(v_0(n)) &=& \{ v_0(n-1), v_0(n-1)\},\\
M_D(v_i(n)) &=& \{v_{i-1}(n-1), v_i(n-1)\},\;\;0<i<n+1\\
M_D(v_{n+1}(n))&=& \{v_n(n-1), v_n(n-1)\}
\end{eqnarray*}
for $n\geq 1$.
Levels 0-3 of $D$ are depicted in Figure 1(b), from which it is seen that
$D$ is based on Pascal's triangle in a natural way. 
\par

\subsection{Bratteli-Vershik information sources} 
Let $D=(V,E)$ be a Bratteli diagram.
Suppose $n\geq 0$ and $\lambda$ is a PMF on $V_{n+1}$. We perform the following two-step random experiment
which creates a $(V_{n+1}\times V_n)$-valued random pair $(X_{n+1},X_n)$.
\begin{itemize}
\item In Step 1, random vertex $X_{n+1}\in V_{n+1}$ is selected whose PMF is $\lambda$. 
\item In Step 2, given
$X_{n+1}=v$, an edge is selected randomly from $E(v)$ (according to the equiprobable
distribution on $E(v)$) and then random $X_n\in V_n$ is the  source vertex of this edge. 
\end{itemize}
Let $[\lambda]$ denote the PMF of $X_n$ on $V_n$. It is easily worked out that
$$[\lambda](v) =  \sum_{e\in E_{n}:s(e)=v}|E(r(e))|^{-1}\lambda((r(e)),\;\;v\in V_{n}.$$
In this way, the edges in $E_n$ are used to transport PMF $\lambda$ on $V_{n+1}$
into PMF $[\lambda]$ on $V_n$.  We define ${\cal S}(D)$ to be the set of all functions $\mu:V\to[0,1]$ such that
the restriction $\mu_n$ of $\mu$ to $V_n$ is a PMF ($n\geq 0$) and
$$\mu_n = [\mu_{n+1}],\;\;n\geq 0.$$
The preceding equation constitutes a consistency relationship among the set of 
PMFs $\{\mu_n:n\geq 0\}$ comprising $\mu$, vis-a-vis the diagram $D$. 
The members of ${\cal S}(D)$ are called the {\it Bratteli-Vershik information sources  on $D$} (B-V sources on $D$).
It is straightforward to show that ${\cal S}(D)$ is a convex set, that is, if
$\mu^1,\mu^2$ are any pair of distinct elements of ${\cal S}(D)$, 
$$S(\mu^1,\mu^2) \define \{\alpha\mu^1+(1-\alpha)\mu^2: \alpha\in(0,1)\} \subset {\cal S}(D).$$
The extreme points of ${\cal S}(D)$ are the elements of ${\cal S}(D)$ not belonging  
to any of the sets $S(\mu^1,\mu^2)$.
We let ${\cal S}_e(D)$ denote the set of all extreme points of ${\cal S}(D)$.  
The members of ${\cal S}_e(D)$ are the {\it ergodic} Bratteli-Vershik  information sources on $D$ (ergodic
B-V sources on $D$). 

\par
We point out  some examples of B-V sources. For the Bratteli diagram $D=(V,E)$ of Example 1.1, we have ${\cal S}(D)={\cal S}_e(D)$,
each consisting of the unique Bratteli-Vershik source 
 $\mu:V\to[0,1]$ in which $\mu(v)=1/2$ for every $v\in V$. For the Bratteli diagram $D=(V,E)$ of Example 1.2, 
it can be shown that there are exactly two ergodic
Bratteli-Vershik sources, namely, the source $\sigma:V\to[0,1]$ for which
$$\sigma(v_0(n)) = 1,\;\;n\geq 0,$$
and the source $\tau:V\to[0,1]$ for which
$$\tau(v_{n+1}(n)) = 1,\;\;n\geq 0.$$
${\cal S}(D)$ is then the convex hull of $\{\sigma,\tau\}$. 
Examples 1.3-1.4 presented later on give Bratteli diagrams for which there are uncountably many
ergodic B-V sources.
\par

{\it Topological Spaces ${\cal S}(D)$ and  ${\cal S}_e(D)$.}  The source
families ${\cal S}(D)$ and  ${\cal S}_e(D)$ are each topological spaces with the topologies
they inherit from the Cartesian product topology on $[0,1]^V$. ${\cal S}_e(D)$ is also
a measurable space whose measurable sets are the Borel sets with
respect to the topology on ${\cal S}_e(D)$. For each $v\in V$, the mapping
$\sigma\to \sigma(v)$ from ${\cal S}_e(D)$ into the real line is continuous and therefore is
a Borel measurable mapping.
\par

{\bf Lemma 1.1.} Let $D=(V,E)$ be any Bratteli diagram. Then:
\begin{itemize}
\item ${\cal S}(D)$  is non-empty and compact.
\item ${\cal S}_e(D)$ is a countable intersection of open subsets of ${\cal S}(D)$.
\item For each $\mu\in{\cal S}(D)$, there exists a probability measure $\lambda_{\mu}$ on
${\cal S}_e(D)$ such that
$$\mu(v) = \int_{{\cal S}_e(D)}\sigma(v)d\lambda_{\mu}(\sigma),\;\;v\in V.$$
\end{itemize}
\par

{\it Proof.} Let Bratteli diagram $D=(V,E)$ be given. ${\mathbb R}^V$ is the 
vector space of all real-valued mappings on $V$. ${\mathbb R}^V$ is a locally convex
topological vector space under the Cartesian product topology. $[0,1]^V$ 
is a compact subset of ${\mathbb R}^V$. For each $N\geq 1$, let ${\cal S}_N(D)$
be the set of all $\mu\in [0,1]^V$ such that 
the restriction $\mu_n$ of $\mu$ to $V_n$ is a PMF ($n\geq 0$) and
$\mu_n = [\mu_{n+1}]_D$ holds for $0 \leq n < N$. Each ${\cal S}_N(D)$ is non-empty
and is a closed and therefore a compact subset of $[0,1]^V$. Consequently, ${\cal S}(D)$
is also non-empty and compact because it is the intersection of the ${\cal S}_N(D)$'s and
the sequence of sets $\{{\cal S}_N(D):N\geq 1\}$ is a nested sequence. The rest of Lemma 1.1 follows
by applying the metrizable form of Choquet's theorem \cite[p.\ 14]{phelps} to
the compact convex metrizable topological space ${\cal S}(D)$. 
\par

{\bf Remark.} The {\it ergodic decomposition theorem} is said to hold for Bratteli diagram $D$ if the
probability measure $\lambda_{\mu}$ on ${\cal S}_e(D)$ in Lemma 1.1 is unique for each B-V source $\mu$ on $D$.
In the following subsection, we introduce types of Bratteli diagrams for which the ergodic decomposition
theorem holds.

\subsection{Some Types of Bratteli Diagrams}

Let $\beta$ be an integer $\geq 2$. Let $S$ be a finite set and let $x$ be a string in the
semigroup $S^*$ such that 
$|x|/\beta$ is an integer. We define the $\beta$-decomposition of $x$ to be the
$\beta$-tuple $(x[0],x[1],\cdots,x[\beta-1])$ in which each entry $x[i]$ is a string in $S^*$
of length $|x|/\beta$ and the factorization
$$x=x[0]x[1]\cdots x[\beta-1]$$
holds.
\par

{\it Canonical Bratteli Diagrams.} Let $\beta$ be an integer $\geq 2$. A Bratteli diagram $D=(V,E)$ is said to be
{\it $\beta$-canonical} if
\begin{itemize}
\item $|E(v)|=\beta$ for each $v\in V^+$.
\item $V_n \subset V_0^{\beta^n}$ for each $n\geq 1$.
\item For each $x\in V^+$, $M_D(x)=\{x[0],x[1],\cdots,x[\beta-1]\}$, where
$(x[0],x[1],\cdots,x[\beta-1])$ is the $\beta$-decomposition of $x$.
\end{itemize}
A Bratteli diagram is said to be canonical if it is $\beta$-canonical for some $\beta\geq 2$.
\par

{\bf Example 1.3.} Let $A$ be a finite non-empty set and let $\beta\geq 2$. Then we have
the $\beta$-canonical Bratteli diagram $D_{\beta}(A) = (V,E)$ in which
$V_n = A^{\beta^n}$ for $n\geq 0$. For each sequential source $\alpha\in {\cal P}(A^{\infty})$, let $\alpha^{\dagger}:V\to[0,1]$ be the mapping whose restriction to $V_n$ is the marginal PMF of $\alpha$ on $A^{\beta^n}$ ($n\geq 0$).
It is easy to show that $\alpha^{\dagger}$ is a Bratteli-Vershik source on $D$. 
The mapping $\alpha\to \alpha^{\dagger}$ is thus
an embedding  (i.e., a 
one-to-one mapping) of ${\cal P}(A^{\infty})$ into ${\cal S}(D_{\beta}(A))$.
In Sec.\ II, we make use of this embedding to show that some universal source coding theorems 
for the sequential source family ${\cal P}(A^{\infty})$ are derivable
from universal source coding theorems for Bratteli-Vershik source family ${\cal S}(D_{\beta}(A))$.
In this way, we can view some parts of sequential source coding theory as a special case of B-V source coding theory.
\par

{\it Regular Bratteli Diagrams.} Let $\beta$ be an integer $\geq 2$.  A Bratteli diagram is defined to be
{\it $\beta$-regular} if it is isomorphic to a $\beta$-canonical Bratteli diagram. 
A Bratteli diagram is regular if it is a $\beta$-regular diagram for some $\beta\geq 2$. 
A $\beta$-canonical Bratteli diagram is $\beta$-regular because it is isomorphic to itself.
We develop a necessary and sufficient condition for a Bratteli diagram to be regular,
by which we can easily  determine whether or not a non-canonical diagram is regular.
If $M$ is a multiset,
we let $N(M)$ be the number of distinct  orderings of  the elements of $M$. For example $N(\{a,a,b\})=3$ since we have the three orderings $aab,aba,baa$. If $S(M)$ is the set consisting of the distinct elements
of multiset $M$, then
$$N(M) = \frac{|M|!}{\prod_{s\in S(M)}m(s|M)!}.$$
\par
{\bf Lemma 1.2.} Let $\beta$ be an integer $\geq 2$. A Bratteli diagram $D=(V,E)$ is {\it $\beta$-regular} if
and only if
\begin{itemize}
\item {\bf (b.1):} $|E(v)|=\beta$ for every $v\in V^+$.
\item {\bf (b.2):} For any multiset $M \in \{M_D(v):v\in V^+\}$, 
$$|\{v\in V^+: M_D(v)=M\}| \leq N(M).$$
\end{itemize}
\par

{\bf Remarks.} The diagram in Example 1.4 is regular because it is canonical. The diagrams in Examples 1.1-1.3 all
satisfy (b.1)-(b.2) with $\beta=2$, and thus these diagrams are also regular by Lemma 1.2.
\par

Conditions (b.1)-(b.2) are invariant under isomorphism, and it is easy to see that
these conditions hold for any $\beta$-canonical Bratteli diagram. Thus, (b.1)-(b.2) hold
for any $\beta$-regular Bratteli diagram. 
The more interesting part of proving Lemma 1.2, accomplished next, is to
show that properties (b.1)-(b.2) imposed on a Bratteli diagram $D$ imply that $D$ is isomorphic to a 
canonical diagram. 

\par
{\it Isomorphisms via Indexings.} Let $\beta$ be an integer $\geq 2$. Let $D=(V,E)$ be a
Bratteli diagram satisfying conditions (b.1)-(b.2). An {\it indexing} of $D$ is any function 
$I:E\to\{0,1,\cdots,\beta-1\}$ satisfying the
following two properties:
\begin{itemize}
\item {\bf (c.1):} $I(E(v)) = \{0,1,\cdots,\beta-1\}$ for $v\in V^+$.
\item {\bf (c.2):} If $v,v'$ are distinct vertices in $V^+$, there exist edges $e\in E(v)$ and
$e'\in E(v')$ such that $I(e)=I(e')$ and $s(e)\not = s(e')$.
\end{itemize}
Indexings of $D$ exist because of property (b.2); Figure 2 illustrates indexings of the Bratteli diagrams
in Examples 1.1-1.2. Fix any indexing $I$ of $D$. We show how to use $I$ to construct an isomorphism
between $D$ and some $\beta$-canonical Bratteli diagram $D'=(V',E')$. Such an isomorphism will
consist of a pair of mappings $(\eta,\tau)$ in which $\eta$ is a one-to-one mapping of
$V$  onto $V'$, $\tau$ is a one-to-one mapping of $E$ onto $E'$, and the property
\begin{equation}
 s(\tau(e))=\eta(s(e)),\;\;r(\tau(e)) = \eta(r(e)),\;\;e\in E
\label{3mar2016eq1}\end{equation}
holds. Let $n\geq 1$. 
For each $v\in V_n$, 
let $G_v$ be the context-free grammar with set of non-terminal symbols $V_1\cup V_2\cup\cdots\cup V_n$,
set of terminal symbols $V_0$, start symbol $v$, and production rules of the form
$$u \to (s(e_0(u)),s(e_1(u)),\cdots,s(e_{\beta-1}(u))),$$
where $u$ is a non-terminal symbol and $e_i(u)$ is the edge $e\in E(u)$ for which $I(e)=i$. The language 
$L(G_v)$ of $G_v$ is a singleton set $L(G_v)=\{x(G_v)\}$, and $x(G_v)$ 
belongs to the set $V_0^{\beta^n}$. Let $D'=(V',E')$ be the $\beta$-canonical grammar in which
$$V_0'=V_0,$$
$$V_n' = \{x(G_v):v\in V_n\},\;\;n\geq 1,$$
$$E'(x) = \{e_0'(x),e_1'(x),\cdots,e_{\beta-1}'(x)\},\;\;x\in (V')^+,$$
where
$$s(e_0'(x)) = x[i],\;\;i=0,1,\cdots,\beta-1.$$
Let $\eta:V\to V'$ be the mapping 
$$\eta(v) = v,\;\;v\in V_0,$$
$$\eta(v) = x(G_v),\;\;v\in V^+.$$
and let $\tau:E\to E'$ be the mapping
$$\tau(e_i(v)) = e_i'(\eta(v)),\;\;v\in V^+,\;\;i=0,1,\cdots,\beta-1.$$
Both of these mappings are one-to-one and onto.
Because of the property
$$\eta(v) =\eta(s(e_0(v)))\eta(s(e_1(v)))\cdots \eta(s(e_{\beta-1}(v))),\;\;v\in V^+,$$
 (\ref{3mar2016eq1}) holds 
and thus $(\eta,\tau)$ is an isomorphism between $D$ and $D'$.
We will refer to the function $\eta$ constructed above as $\eta_I$ to denote
its dependence upon indexing $I$.
\par

\begin{figure}[htb]
\centering
\includegraphics[width=\textwidth]{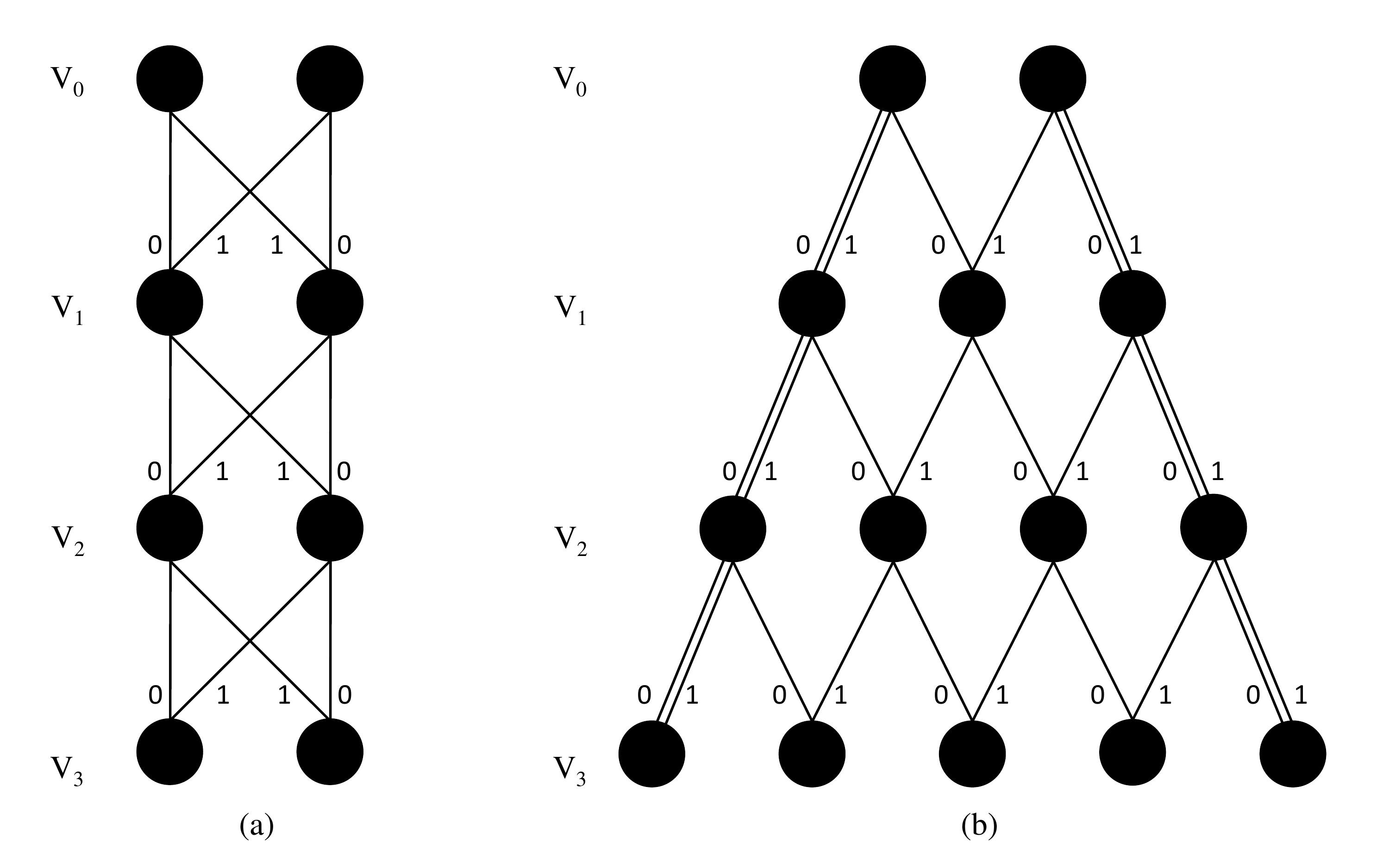}
\caption{Indexings of diagrams of Ex. 1.1 and Ex. 1.2.}
 \end{figure}
\par

For example, if $D=(V,E)$ is the $2$-regular Bratteli diagram of Ex.\ 1.2, if
we label the vertices depicted in Fig.\ 1(b) left-to-right as
$$V_0=\{a,b\},\;\;V_1=\{c,d,e\},\;\;V_2=\{f,g,h,i\},\;\;V_3=\{j,k,l,m,n\},$$
and if $I$ is the indexing of $D$ in Fig.\ 2, 
then we compute $\eta_I(l)$ via three rounds of substitutions as follows:\newpage
$$l$$
$$\downarrow$$
$$gh$$
$$\downarrow$$
$$cdde$$
$$\downarrow$$
$$\eta_I(l)=aaababbb$$
\par

{\it Background on Simplicial Grids.} A simplicial grid in a Euclidean space $\mathbb E$ is a sequence
$\{K_n:n\geq 0\}$ in which (a) $K_n$ is a finite simplicial complex of simplexes
in $\mathbb E$ for $n\geq 0$ and (b) $K_n$ is a subdivision of $K_{n-1}$ for $n\geq 1$. 
As we shall see below, certain simplicial grids in Euclidean spaces induce regular Bratteli diagrams in a natural way,
but first we present some background on simplicial grids.
Given simplicial grid $\{K_n:n\geq 0\}$ in $\mathbb E$,
its underlying space is the compact subset $\Theta$ of $\mathbb E$ such that
$$\Theta = \bigcup_{\sigma\in K_n}\sigma,\;\;n\geq 0.$$
For each $n\geq 0$, the vertex set $V(K_n)$ of $K_n$ is the set of all points $x$
such that $x$ is a vertex of
some simplex in $K_n$. 
For each $x\in\Theta$ and $n\geq 0$, there is a unique PMF $p_{x|K_n}$ on $V(K_n)$
such that 
$$x = \sum_{v\in V(K_n)}vp_{x|K_n}(v)$$
and $\{v\in V(K_n):p_{x|K_n}(v)>0\}$ is the set of vertices of a simplex in $K_n$.
$p_{x|K_n}$ is called the {\it barycentric distribution} of $x$ with respect to $K_n$. 
\par

{\it Regular Diagrams Induced by Grids.} Let $\beta$ be an integer $\geq 2$. Simplicial grid $\{K_n:n\geq 0\}$ 
in Euclidean space $\mathbb E$ is defined to be $\beta$-admissible 
if for each $n\geq 1$ and $x\in V(K_n)$, the barycentric distribution of $x$
with respect to $K_{n-1}$ takes its values in the set $\beta^{-1}\{0,1,2,\cdots,\beta\}$, which allows us 
to define the unique multiset
$$M(x) \define \{x_1,x_2,\cdots,x_{\beta}\}$$ 
whose distinct entries form the vertex set
of some simplex in $K_{n-1}$, and
$$x = \beta^{-1}(x_1+x_2+\cdots+x_{\beta}).$$
A $\beta$-admissible simplicial grid $\{K_n:n\geq 0\}$ induces the $\beta$-regular
Bratteli diagram $D=(V,E)$ in which
\begin{itemize}
\item {\bf (d.1):} For $n\geq 0$, 
$$V_n \define V(K_n) \times \{n\} = \{(x,n): x\in V(K_n)\}.$$
\item {\bf (d.2):} For $(x,n) \in V_n$ and $n\geq 1$,
\begin{eqnarray*}
M_D((x,n)) &\define& M(x)\times \{n-1\}\\
&=& \{(x_1,n-1),(x_2,n-1),\cdots,(x_{\beta},n-1)\}
\end{eqnarray*}
\end{itemize}
If $\Theta$ is the underlying space of $\{K_n\}$, then for each $\theta\in\Theta$ define $\mu^{\theta}:V\to[0,1]$ to be the mapping
$$\mu^{\theta}((x,n)) \define p_{{\theta}|K_n}(x),\;\;x\in V(K_n),\;\;n\geq 0.$$
It can be shown that
$$\{p^{\theta}: \theta\in\Theta\} = {\cal S}_e(D).$$
\par

{\it Example 1.4.} We illustrate the Bratteli diagram induced
by a well known simplicial grid called the Kuhn grid \cite{kuhn}\cite{bey}. 
Fix any integer $\beta\geq 2$. 
Let ${\mathbb E}$ be the Euclidean space ${\mathbb R}^k$, where $k\geq 1$.
 Let $S\subset {\mathbb E}$ be the hypercube
$$S = \{(x_1,x_2,\cdots,x_k): 0 \leq x_i \leq 1,\;\;i=1,2,\cdots,k\}.$$
Let $\Pi_k$ be the set of all permutations of $\{1,2,\cdots,k\}$.
For each $\pi\in \Pi_k$, let $\sigma_{\pi}$ be the simplex
$$\sigma_{\pi} \define \{(x_1,\cdots,x_k)\in S:x_{\pi(1)} \leq x_{\pi(2)}\leq\cdots\leq x_{\pi(k)}\}.$$
The set $K_0$ consisting of the $k!$ simplexes $\sigma_{\pi}$ 
and their faces is a simplicial complex. 
For each $n\geq 1$, let $K_n$ be the simplicial complex consisting
of all simplexes of form 
$$\beta^{-n}(\sigma_{\pi} + z),\;\;\pi\in\Pi_k; z\in {\mathbb Z}^k,$$
that are subsets of $S$, 
together with all their faces. $\{K_n:n\geq 0\}$ is the $(\beta,k)$ Kuhn grid, whose
underlying space is the hypercube $S$.  
For each $n\geq 0$, the vertex set $V(K_n)$ of $K_n$ 
consists of the $(\beta^n+1)^k$ points in $S$ of the form
$(i_1,i_2,\cdots,i_k)/\beta^n$ in which each $i_j\in\{0,1,\cdots,\beta^n\}$. 
Once the multisets $\{M(x):x\in V(K_1)\}$ are determined, then each multiset $M(u)$ for
 $u\in V(K_n)$ and $n\geq 2$ can be determined by translation and scaling as follows:
pick $z\in {\mathbb Z}^k$ such that
$$u \in \beta^{n-1}(S+z) \subset S;$$
then $x=\beta^{n-1}u-z$ belongs to $V(K_1)$ and
$$M(u) = \beta^{n-1}(M(x)+z).$$ 
The $\beta$-regular Bratteli diagram $D=(V,E)$ induced by the Kuhn grid
is now determined according to (d.1)-(d.2). For the $(\beta,k)=(3,2)$ Kuhn grid,
Fig.\ 3 illustrates the subgraph
$(V_0\cup V_1,E_0)$ of the induced diagram $D=(V,E)$, where, for example,  
$I\to B,B,C$ in Fig.\ 3 means that $M_D(I)=\{B,B,C\}$ and consequently the three
edges in $E(I)$ have source vertices $B,B,C$, respectively.
\par

\begin{figure}[t]
\centering
\includegraphics[width=\textwidth]{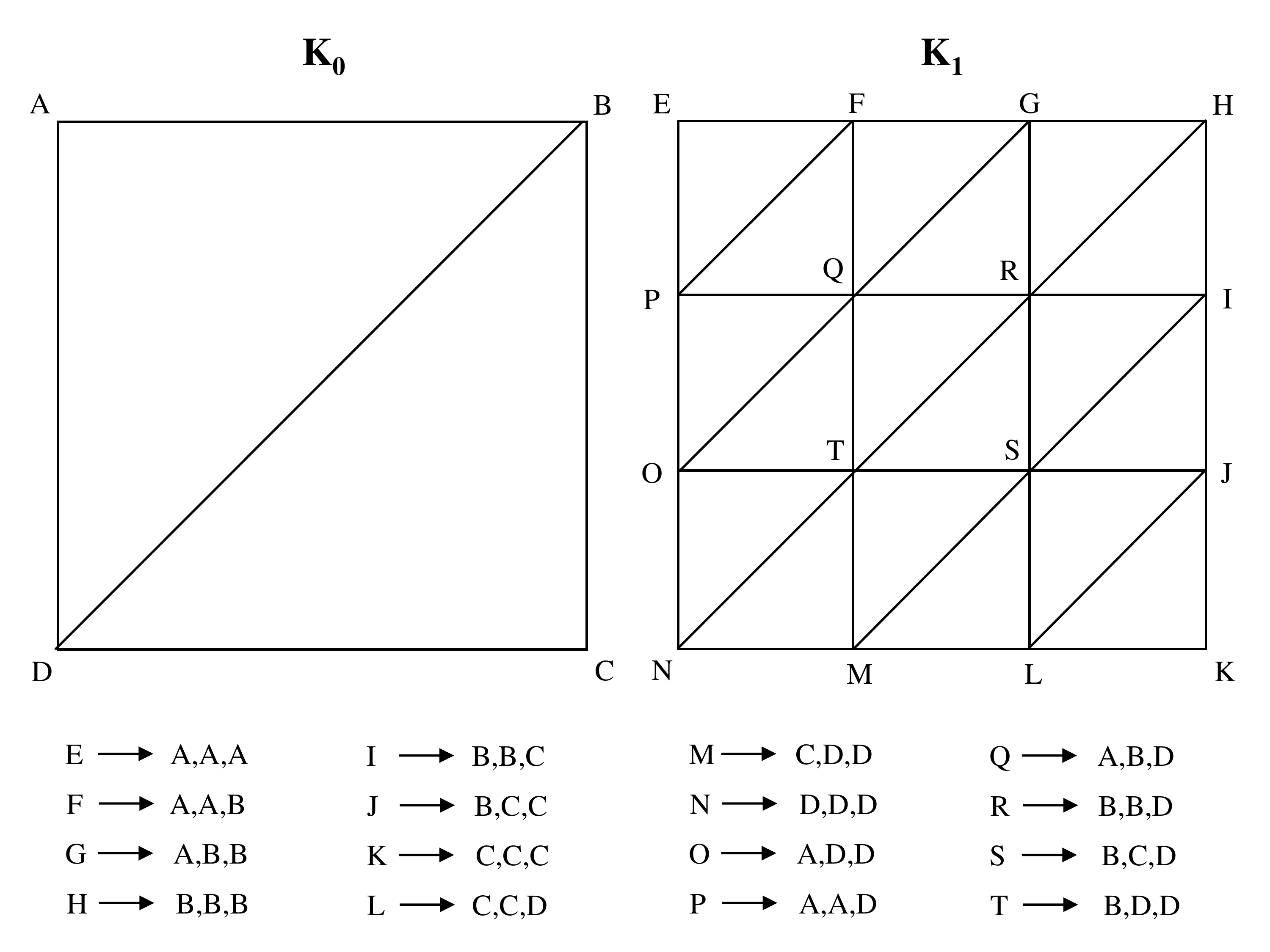}
\caption{Above: Complexes $K_0,K_1$ of the $(\beta,k)=(3,2)$ Kuhn grid $\{K_n:n\geq 0\}$ of Ex.\ 1.4. 
Below: Subgraph $(V_0\cup V_1,E_0)$ of the induced $\beta$-regular Bratteli diagram $(V,E)$.}
\end{figure}
\par

\subsection{B-V Source Entropy Rate and its Interpretation}

{\bf Lemma 1.4.} Let $D=(V,E)$ be a $\beta$-regular Bratteli diagram, where $\beta\geq 2$.
If $n\geq 0$ and $\lambda$ is a PMF on $V_{n+1}$, then
\begin{equation}
H(\lambda) \leq \beta H([\lambda]).
\label{5feb2016eq3}\end{equation}
\par
{\it Proof.} Via Lemma 1.2, it suffices to prove the result assuming that $D$ is canonical.
Let $X$ be a $V_{n+1}$-valued random object whose PMF is $\lambda$. Since $X$
and its $\beta$-decomposition $(X[0],X[1],\cdots,X[\beta-1])$ are functions of each other,
$$H(\lambda) = H(X) = H(X[0],X[1],\cdots,X[\beta-1]).$$
By sub-additivity of the entropy function,
$$H(X[0],X[1],\cdots,X[\beta-1]) \leq \sum_{i=0}^{\beta-1}H(X[i]).$$
Each $X[i]$ is $V_n$-valued. Let
$p_i$ be the PMF of $X[i]$ on $V_n$. We have
$$[\lambda] = \beta^{-1}\sum_{i=0}^{\beta-1}p_i.$$
By concavity of the entropy function and the preceding relations, 
$$H([\lambda]) \geq \beta^{-1}\sum_{i=0}^{\beta-1}H(p_i) \geq \beta^{-1}H(\lambda),$$
and our result is proved. 
\par

{\it Definition: Entropy Rate of a Bratteli-Vershik Source.} Let $\beta\geq 2$, let $D$ be any $\beta$-regular Bratteli diagram, and let $\mu$ be any source
in ${\cal S}(D)$. By Lemma 1.4, the sequence $\{\beta^{-n}H(\mu_n):n\geq 0\}$ is
non-increasing and therefore it possesses a limit as $n\to\infty$. This limit is defined
as the {\it entropy rate} $H_{\infty}(\mu)$ of $\mu$, that is,
$$H_{\infty}(\mu) \define \lim_{n\to\infty}\beta^{-n}H(\mu_n) = \inf_n\beta^{-n}H(\mu_n).$$
We have the upper bounds
$$H_{\infty}(\mu) \leq H(\mu_0) \leq \log_2|V_0|,\;\;\mu\in{\cal S}(D).$$
\par

{\it Coding Interpretation of Entropy Rate.} 
We discuss the operational significance of entropy rate in the lossless
encoding of Bratteli-Vershik sources. If $S$ is a finite set, a one-to-one function $\phi:S\to \{0,1\}^*$
satisfying the prefix condition is called a {\it lossless prefix encoder} on $S$. (The
prefix condition means that the codeword $\phi(s_1)$ is not a prefix of codeword $\phi(s_2)$
for any two distinct $s_1,s_2$ in $S$.) Let $p$ be any PMF on $S$. For any lossless prefix
encoder $\phi$ on $S$, define
$${\overline L}(\phi,p) \define \sum_{s\in S}|\phi(s)|p(s),$$
the expected codeword length resulting from using $\phi$ to encode the members of $S$
distributed according to $p$. Basic information theory
gives us the following facts (e.1)-(e.2).
\begin{itemize}
\item {\bf (e.1):} For any lossless prefix encoder $\phi$ on $S$ and any PMF $p$ on $S$,
$$ {\overline L}(\phi,p) \geq H(p).$$
\item {\bf (e.2):} For any PMF $p$ on $S$, there exists a lossless prefix encoder $\phi$ on $S$ 
such that
$${\overline L}(\phi,p) \leq H(p)+1.$$
\end{itemize}
Let $D=(V,E)$ be any $\beta$-regular Bratteli diagram, where $\beta\geq 2$. Let $n\geq 0$.
If $\phi_n$ is any lossless prefix encoder on $V_n$ and $\lambda$ is any PMF on $V_n$, 
we define the number
$$R(\phi_n,\lambda) \define \beta^{-n}{\overline L}(\phi_n,\lambda).$$
We discuss the interpretation of  $R(\phi_n,\lambda)$. Since $D$ is isomorphic to
a canonical diagram, we have
$$|V_n| \leq |V_0|^{\beta^n},\;\;n\geq 0.$$
Thus, there exists a default lossless prefix encoder
on $V_n$ employing fixed-length binary codewords of length $\lceil \beta^n\log_2|V_0|\rceil$. 
$R(\phi_n,\lambda)/\log_2|V_0|$ is
roughly (for large $n$) the ratio between 
the expected codeword length  ${\overline L}(\phi_n,\lambda)$ afforded by encoder $\phi_n$
and the codeword length afforded by the default encoder. 
It is thus sensible for us
to interpret the quantity $R(\phi_n,\lambda)$ as a measure of
{\it compression rate}. A lossless encoding scheme on $D$ 
is defined to be any sequence $\{\phi_n:n\geq 0\}$ in which $\phi_n$ is a lossless prefix
encoder on $V_n$ ($n\geq 0$). Moreover, if $\{\phi_n\}$ is such a scheme and 
$\mu\in {\cal S}(D)$, by (e.1) we have
\begin{equation}
\varliminf_{n\to\infty}R(\phi_n,\mu_n) \geq H_{\infty}(\mu).
\label{17feb2016eq1}\end{equation}
Thus, a scheme $\{\phi_n\}$ which yields the best asymptotic compression rate in encoding
the source $\mu$ would be one for which
\begin{equation}
\lim_{n\to\infty}R(\phi_n,\mu_n) = H_{\infty}(\mu).
\label{3feb2016eq1}\end{equation}
Such a scheme exists by (e.2). 
In summary, $H_{\infty}(\mu)$ has operational significance
of being the minimum asymptotic compression rate afforded by lossless
encoding schemes used to compress $\mu$.

\subsection{Organization of Paper}

In the rest of the paper,  we prove the six theorems for B-V sources that are stated below. Each of  them has an  obvious analogue for
the stationary sequential sources in ${\cal P}(A^{\infty})$. The first two theorems are proved in Section II and
are as follows.
\par

{\bf Theorem 1.5: Weak Universal Encoding Theorem.} {\it Let $D=(V,E)$ be any regular Bratteli diagram.  For each $n\geq 0$, there exists a
lossless prefix encoder $\phi_n$ on $V_n$ such that}
$$\lim_{n\to\infty}R(\phi_n,\mu_n) = H_{\infty}(\mu),\;\;\mu\in{\cal S}(D).$$
\par

{\bf Theorem 1.6: Strong Universal Encoding Theorem.} {\it Let $D=(V,E)$ be any regular Bratteli diagram. Let $\Lambda$ be a closed subset of 
topological space ${\cal S}(D)$ such that
the function $\mu\to H_{\infty}(\mu)$ on $\Lambda$ is a continuous function. For each $n\geq 0$, there exists a
lossless prefix encoder $\phi_n$ on $V_n$ such that}
$$\lim_{n\to\infty}\left\{\sup_{\mu\in\Lambda}\,\frac{{\overline L}(\phi_n,\mu_n)-H(\mu_n)}{\beta^n}\right\} = 0.$$
\par

{\bf Remarks.}
\begin{itemize}
\item There are well known weak universal and strong universal encoding theorems for sequential 
stationary sources \cite{davisson} that are analogues of these two results. We show in Section II that the sequential source 
results follow from the above B-V source results. 
\item In Section II, we show how one may use a simplicial grid to construct a family of B-V sources  satisfying the
hypotheses of the Strong Universal Lossless Encoding Theorem. The family will be in one-to-one correspondence
with the underlying space of the grid.
\end{itemize}
\par

Section III presents a sufficient amount of the theory of Bratteli-Vershik systems that will allow us to prove the
remaining results. Section IV presents the follow two decomposition theorems.
\par

{\bf Theorem 1.7: Ergodic Decomposition Theorem.} {\it Let $D=(V,E)$ be any regular Bratteli diagram. 
For each $\mu\in{\cal S}(D)$, there exists a unique probability measure $\lambda_{\mu}$ on
${\cal S}_e(D)$ such that}
\begin{equation}
\mu(v) = \int_{{\cal S}_e(D)}\sigma(v)d\lambda_{\mu}(\sigma),\;\;v\in V.
\label{7mar2016eq1}\end{equation}

\par

{\bf Theorem 1.8: Entropy Rate Decomposition Theorem.} {\it Let $D=(V,E)$ be any regular Bratteli diagram. Then}
$$H_{\infty}(\mu) = \int_{{\cal S}_e(D)}H_{\infty}(\sigma)d\lambda_{\mu}(\sigma),\;\;\mu\in{\cal S}(D).$$
\par

 There is a both a strong
and weak form of the Shannon-McMillan-Breiman theorem (SMB theorem) for B-V sources. The strong form gives a limit theorem in which 
there is both almost everywhere convergence
and $L^1$ convergence. The weak form gives convergence in distribution, and is a consequence of the strong form.
Section V establishes the strong form of the SMB theorem for B-V sources. However, the statement of the strong form
requires the Section III background material concerning Bratteli-Vershik systems in order for the statement to make sense.
Here, we state the weak form and refer the reader to Section V for the strong form.
\par

{\bf Theorem 1.9: SMB Theorem (Weak Form).} {\it Let $D=(V,E)$ be a $\beta$-regular Bratteli diagram, where $\beta\geq 2$. Let $\mu\in{\cal S}(D)$, and let $F_{\mu}:{\mathbb R}\to[0,1]$ be defined by}
\begin{equation}
F_{\mu}(x) \define \lambda_{\mu}(\{\sigma\in{\cal S}_e(D):H_{\infty}(\sigma)\leq x\}),\;\;x\in{\mathbb R}.
\label{3may2016eq2}\end{equation}
{\it Then}
$$\lim_{n\to\infty}\mu_n(\{v\in V_n:-\beta^{-n}\log_2\mu_n(v) \leq x\}) = F_{\mu}(x)$$
{\it for every $x\in{\mathbb R}$ at which $F_{\mu}$ is continuous.} 
\par

In Section VI, our final section, the following result on fixed-length lossy encoding of B-V sources is established, which
requires the SMB theorem for its proof. It is an analogue
of a result of Parthasarathy \cite{partha} on fixed-length lossy encoding of stationary sequential sources.
\par

{\bf Theorem 1.10: Fixed-Length Lossy Encoding Theorem.} {\it Let $D=(V,E)$ be any $\beta$-regular Bratteli diagram, where $\beta\geq 2$.
Let $\mu \in {\cal S}(D)$. For each $\delta\in(0,1)$, define}
\begin{equation}
M_n(\delta,\mu) \define \min\{|S|: S \subset V_n,\;\;\mu_n(S)\geq 1-\delta\},\;\;n\geq 0.
\label{3may2016eq1}\end{equation}
{\it For any $\delta\in(0,1)$,} 
$$A(\delta)\leq \varliminf_{n\to\infty} \frac{\log_2M_n(\delta,\mu)}{\beta^n} \leq 
\varlimsup_{n\to\infty}\frac{\log_2M_n(\delta,\mu)}{\beta^n} \leq B(\delta),$$
where
\begin{eqnarray*}
B(\delta) &\define& \inf\{x\in{\mathbb R}: F_{\mu}(x)>1-\delta\},\\
A(\delta) &\define& \sup\{x\in{\mathbb R}: F_{\mu}(x)<1-\delta\}.
\end{eqnarray*}
{\it Moreover, the set of $\delta$'s for which $A(\delta)\not = B(\delta)$ is countable.}

\section{Universal Lossless B-V Source Encoding}

We consider two types of universal encoding for families of Bratteli-Vershik sources, namely,
weak universal encoding and strong universal encoding. As corollaries of our 
universal encoding results for B-V sources, we will then obtain some previously known universal lossless encoding theorems
for stationary sequential source families, thereby
illustrating that some parts of universal encoding theory for stationary sequential information sources are special
cases of universal encoding theory for Bratteli-Vershik sources.

\subsection{Weak Universal Encoding Theory}

If $\psi$ is a lossless prefix
encoder on finite set $S$, and $\tau$ is a PMF on $S$, the {\it redundancy of $\psi$ with respect
to $\tau$} is the non-negative real number defined by
$${\rm RED}(\psi,\tau) \define {\overline L}(\psi,\tau)-H(\tau).$$
\par
Let $\beta\geq 2$ and let
$D=(V,E)$ be a $\beta$-regular Bratteli diagram. 
A lossless encoding scheme  $\{\phi_n\}$ on $D$ is defined to be  {\it weak universal for
${\cal S}(D)$} if
\begin{equation}
\lim_{n\to\infty}\beta^{-n}{\rm RED}(\phi_n,\mu_n) = 0,\;\;\mu\in{\cal S}(D),
\label{22feb2016eq5}\end{equation}
or equivalently, if
\begin{equation}
\lim_{n\to\infty} R(\phi_n,\mu_n) = H_{\infty}(\mu),\;\;\mu\in{\cal S}(D).
\label{19feb2016eq2}\end{equation}
\par

The goal of this subsection is to prove the following theorem.
\par

{\bf Theorem 2.1}. {\it Let $D=(V,E)$ be any $\beta$-regular Bratteli diagram, where $\beta\geq 2$. 
Then there exists a lossless encoding scheme on $D$
which is weak universal for ${\cal S}(D)$.}

\par

For the rest of this section, we fix $\beta$-regular Bratteli diagram $D=(V,E)$ and an indexing
$I$ of $D$. For each $x\in V^+$ and $i\in \{0,1,\cdots,\beta-1\}$, let $e_i(x)$ denote
the edge in $E(x)$ such that $I(e_i(x))=i$ and let $x[i]=s(e_i(x))$.
 If $n\geq 0$ and $\phi$ 
is a lossless prefix encoder on $V_n$, 
let $[\phi]:V_{n+1}\to \{0,1\}^*$ be the mapping defined by
$$[\phi](x) \define \phi(x[0])\phi(x[1])\cdots\phi(x[\beta-1]),\;\;x\in V_{n+1}.$$
Since the mapping $x\to (x[0],x[1],\cdots,x[\beta-1])$ is a one-to-one mapping 
of $V^+$ into $V^{\beta}$, $[\phi]$ is a lossless prefix encoder on $V_{n+1}$.
\par
{\bf Lemma 2.2.} {\it Let $n\geq 0$ and let $\phi$ be a lossless prefix encoder on $V_n$.
Then 
\begin{equation}
R([\phi],\lambda) =  R(\phi,[\lambda])
\label{11feb2016eq1}\end{equation}
for every PMF $\lambda$ on $V_{n+1}$.}
\par
{\it Proof.} Let $X$ be a $V_{n+1}$-valued random object whose PMF is $\lambda$.  
We have
$$\beta^{n+1}R([\phi],\lambda) = E\left[\sum_{i=0}^{\beta-1}|\phi(X[i])|\right] = 
\sum_{i=0}^{\beta-1}E[|\phi(X[i])|].$$
Each $X[i]$ is a $V_n$-valued random object. 
The average of the PMF's of the $X[i]$'s is the PMF $[\lambda]$ on $V_n$, that is,
$$\beta^{-1}\sum_{i=0}^{\beta-1}\Pr[X[i]=u] = [\lambda](u),\;\;u\in V_n.$$
Thus, for any real-valued function $f$ on $V_n$,
$$\beta^{-1}\sum_{i=0}^{\beta-1}E[f(X[i])] = \sum_{u\in V_n}f(u)[\lambda](u).$$
In the preceding equation, choose $f$ to be the function $u\to |\phi(u)|$, giving us
$$\beta^{-1} \sum_{i=0}^{\beta-1}E[|\phi(X[i])|] = \beta^{n}R(\phi,[\lambda]).$$
Equation (\ref{11feb2016eq1}) is now apparent.
\par\medskip

For $n\geq 0$, we define ${\cal E}(V_n)$ to be the set of all lossless prefix encoders on $V_n$.
An encoder $\phi$ in ${\cal E}(V_n)$ is said to be a {\it proper encoder} if
$$\sum_{v\in V_n}2^{-|\phi(v)|} = 1.$$
We define ${\cal E}^*(V_n)$ to be the set of all proper encoders in ${\cal E}(V_n)$. 
The concept of proper encoder will be useful to us in the following respect: given any
encoder $\phi\in {\cal E}(V_n)$, there exists an encoder $\phi^* \in {\cal E}^*(V_n)$
such that
$$|\phi^*(v)| \leq |\phi(v)|,\;\;v\in V_n.$$
Define
\begin{eqnarray*}
{\cal E}(D) &\define& \bigcup_{n=0}^{\infty}{\cal E}(V_n)\\
{\cal E}^*(D) &\define& \bigcup_{n=0}^{\infty}{\cal E}^*(V_n)
\end{eqnarray*}
If $\phi\in {\cal E}(D)$, we define its order ORD($\phi$) to be the integer $n\geq 0$ such that
$\phi \in {\cal E}(V_n)$. Given $\phi\in {\cal E}(D)$ and $\mu\in {\cal S}(D)$, we define
$$R(\phi,\mu) \define R(\phi,\mu_n),$$
where $n={\rm ORD}(\phi)$. Given $\phi\in{\cal E}(D)$, let $\{\phi^{(n)}:n\geq 0\}$ be the sequence
in ${\cal E}(D)$ defined recursively by
\begin{eqnarray*}
\phi^{(0)} &=& \phi\\
\phi^{(n)} &=& [\phi^{(n-1)}],\;\;n\geq 1
\end{eqnarray*} 
From Lemma 2.2, we have the properties
$${\rm ORD}(\phi^{(n)}) = {\rm ORD}(\phi) + n,\;\;\phi\in{\cal E}(D),\;\;n\geq 0.$$
$$R(\phi^{(n)},\mu) = R(\phi,\mu),\;\;\phi\in{\cal E}(D),\;\;\mu\in{\cal S}(D),\;\;n\geq 0.$$
\par

{\bf Lemma 2.3}. {\it There is an upper triangular array
$$\{\phi_{i,j}:j=i,i+1,i+2,\cdots; i=1,2,3,\cdots\} \subset {\cal E}(D)$$
such that}
\begin{itemize}
\item {\bf (a):} {\it ${\rm ORD}(\phi_{i,j}) = j-1$ for any $\phi_{i,j}$ in the array.}
\item {\bf (b):} {\it $\phi_{i,j} = \phi_{i,i}^{(j-i)}$ for any $\phi_{i,j}$ in the array.}
\item {\bf (c):} {\it If $\phi\in{\cal E}^*(D)$, then there exists $i\geq 1$ and $k\geq 0$ such that
$\phi^{(k)} = \phi_{i,i}$.}
\end{itemize}
\par
{\it Proof.} Since each set ${\cal E}^*(V_n)$ is finite, we may fix an
enumeration $\{\psi_i:i\geq 0\}$ of ${\cal E}^*(D)$ such that
${\rm ORD}(\psi_i) \leq {\rm ORD}(\psi_{i'})$ holds whenever $i<i'$.
Then we have ${\rm ORD}(\psi_i) \leq i$ for all $i\geq 0$. 
Define
$$\phi_{i,i} \define \psi_{i-1}^{(i-1-{\rm ORD}(\psi_{i-1}))},\;\;i\geq 1.$$
Define $\sigma_{i,j}$ for $j>i$ so that property (b) holds. Then, properties (a) and (c) also hold.

\par\medskip

Suppose $S$ is a finite set and $\{\phi_j:1\leq j \leq J\}$ is a finite sequence of
lossless prefix encoders on
$S$, where $J\geq 2$. For each $j\in\{1,2,\cdots,J\}$, let $B_j$ be the binary string of length $\lceil\log_2J\rceil$
which is the binary expansion of integer $j-1$ (most significant bit on the left). For each $s\in S$,
let $j(s)$ be the smallest $j\in\{1,\cdots,J\}$ such that 
$$|\phi_j(s)| = \min\{|\phi_{j'}(s)|:1\leq j'\leq J\}.$$
Then we have the lossless prefix encoder $\phi$ on $S$ defined by
$$\phi(s) \define B(s)\phi_{j(s)}(s),\;\;s\in S.$$
This encoder $\phi$ constructed from $\{\phi_j:1\leq j\leq J\}$ shall henceforth be denoted by
$\wedge_{j=1}^J\phi_j$. We have the property
\begin{equation}
|(\bigwedge_{j=1}^J\phi_j)(s)| \leq \lceil\log_2J\rceil + \min_{1\leq j\leq J}|\phi_j(s)|,\;\;s\in S.
\label{18feb2016eq1}\end{equation}

\par

{\it Proof of Theorem 2.1.} Let $\{\phi_{i,j}\}$ be any triangular array of encoders in ${\cal E}(D)$ chosen according to Lemma 2.3.
Let  $\{\tau_n:n\geq 0\}$ be the lossless encoding scheme on $D$ defined by
$$\tau_n \define \bigwedge_{i=1}^{n+1}\phi_{i,n+1},\;\;n\geq 0.$$
Let $\mu\in{\cal S}(D)$ be arbitrary. We show that
\begin{equation}
\varlimsup_{n\to\infty}R(\tau_n,\mu) \leq H_{\infty}(\mu),
\label{4dec2015eq1}\end{equation}
which will imply that $\{\tau_n\}$ is weak universal for ${\cal S}(D)$. 
Let $\epsilon>0$ be arbitrary. Pick $N\geq 0$ large enough so that
$$\beta^{-N}H(\mu_N) \leq H_{\infty}(\mu) + \epsilon,$$
$$\beta^{-N} < \epsilon$$
both hold. There exists $\phi \in {\cal E}_N^*(D)$  such that
$$R(\phi,\mu) \leq \beta^{-N}(H(\mu_N) + 1) < H_{\infty}(\mu) + 2\epsilon.$$
By property (c) of Lemma 2.3, there exists $M\geq 1$ such that $\phi_{M,M} \in \{\phi^{(k)}:k\geq 0\}$.  
We have
\begin{equation}
R(\phi_{M,M},\mu) = R(\phi,\mu) < H_{\infty}(\mu) + 2\epsilon.
\label{4dec2015eq2}\end{equation}
Let $m\geq M$. By (\ref{18feb2016eq1}), 
$$|\tau_{m-1}| \leq |\phi_{M,m}| + \lceil\log_2m\rceil,$$
from which it follows that
$$R(\tau_{m-1},\mu) \leq R(\phi_{M,m},\mu) + \beta^{-(m-1)}\lceil\log_2m\rceil.$$
But
$$R(\phi_{M,m},\mu) = R(\phi_{M,M},\mu).$$
In view of (\ref{4dec2015eq2}), we have proved that
$$R(\tau_{m-1},\mu) \leq \beta^{-(m-1)}\lceil\log_2m\rceil + H_{\infty}(\mu) + 2\epsilon,\;\;m\geq M,$$
from which it follows that
$$\varlimsup_{n\to\infty}R(\tau_n,\mu) \leq H_{\infty}(\mu) + 2\epsilon.$$
Since this statement must be true for every $\epsilon>0$, (\ref{4dec2015eq1}) follows,
completing the proof that $\{\tau_n\}$ is weak universal for ${\cal S}(D)$.

\subsection{Strong Universal Encoding Theory}

Let $\beta\geq 2$ and let
$D=(V,E)$ be a $\beta$-regular Bratteli diagram. Let $\Lambda$ be a subfamily of
the Bratteli-Vershik source family ${\cal S}(D)$. Let $\{\phi_n:n\geq 0\}$ be a lossless
encoding scheme on $D$. $\{\phi_n\}$ is defined to be  {\it strong universal for
$\Lambda$} if
$$\lim_{n\to\infty}\left\{\sup_{\mu\in\Lambda}\beta^{-n}{\rm RED}(\phi_n,\mu_n)\right\} = 0.$$
\par

{\bf Theorem 2.4.} {\it Let $D=(V,E)$ be any $\beta$-regular Bratteli diagram, where $\beta\geq 2$. 
Let $\Lambda$ be a closed subset of ${\cal S}(D)$. Let $H_{\Lambda}:\Lambda\to[0,\infty)$ be the function
$$H_{\Lambda}(\mu) \define H_{\infty}(\mu),\;\;\mu\in\Lambda.$$
If $H_{\Lambda}$ is continuous, then there exists a lossless encoding scheme on $D$ which
is strong universal for $\Lambda$.}
\par

{\it Proof.} Let $\{\phi_{i,j}\}$ be any triangular array of encoders in ${\cal E}(D)$ chosen according to Lemma 2.3.
Let  $\{\tau_n:n\geq 0\}$ be the lossless encoding scheme on $D$ defined by
$$\tau_n \define \bigwedge_{i=1}^{n+1}\phi_{i,n+1},\;\;n\geq 0.$$
For each $j\geq  1$, let $Q_j:{\cal S}(D)\to [0,\infty)$ be the continuous function
$$Q_j(\mu) \define \min_{1\leq i\leq j}R(\phi_{i,j},\mu),\;\;\mu\in {\cal S}(D).$$
Then we have the monotonicity property
$$Q_1(\mu) \geq Q_2(\mu) \geq Q_3(\mu) \geq \cdots,\;\;\mu\in{\cal S}(D).$$
Also,
$$\lim_{j\to\infty}Q_j(\mu) = H_{\infty}(\mu),\;\;\mu\in{\cal S}(D).$$
For each $j\geq  1$, let $H_j:{\cal S}(D)\to[0,\infty)$ be the continuous function
$$H_j(\mu) \define \beta^{-(j-1)}H(\mu_{j-1}),\;\;\mu\in {\cal S}(D).$$
We have
$$H_1(\mu) \geq H_2(\mu) \geq H_3(\mu) \geq \cdots,\;\;\mu\in{\cal S}(D),$$
and
$$\lim_{j\to\infty}H_j(\mu) = H_{\infty}(\mu),\;\;\mu\in{\cal S}(D).$$
Since the function $H_{\Lambda}$ is continuous and $\Lambda$ is compact,  
Dini's theorem \cite[Thm. 7.2.5]{ashmetric} tells us that each of the two monotone sequences of functions $\{Q_j\}$ and $\{H_j\}$
converges uniformly on $\Lambda$ to the function $H_{\Lambda}$. 
Therefore, the sequence $\{Q_j-H_j\}$ converges uniformly to $0$ on $\Lambda$, that is,
\begin{equation}
\lim_{n\to\infty}\left[\sup_{\mu\in{\Lambda}}\{Q_{n+1}(\mu)-H_{n+1}(\mu)\}\right] = 0.
\label{15dec2015eq1}\end{equation}
Let $\mu\in\Lambda$. The relationships
$$\beta^{-n}{\rm RED}(\tau_n,\mu_n) = R(\tau_n,\mu_n)-H_{n+1}(\mu),$$
$$R(\tau_n,\mu_n) \leq \beta^{-n}\lceil \log_2(n+1)\rceil + Q_{n+1}(\mu)$$
hold for $n\geq 0$. Hence,
$$0\leq \sup_{\mu\in\Lambda}\beta^{-n}{\rm RED}(\tau_n,\mu_n) \leq$$
$$ \beta^{-n}\lceil \log_2(n+1)\rceil + \sup_{\mu\in\Lambda}\{Q_{n+1}(\mu)-H_{n+1}(\mu)\}.$$
Since the left side converges to zero as $n\to\infty$, 
 $\{\tau_n\}$ is strong universal for $\Lambda$.

\par

{\it Definition.} A family $\Lambda$ of Bratteli-Vershik sources on a regular Bratteli diagram $D$ 
is said to be {\it strongly universally encodable} if there exists a lossless
encoding scheme on $D$ which is strong universal for $\Lambda$.  
We present three examples of B-V source families which can be seen to be
strongly universally encodable via Theorem 2.4. The third of these examples
is of most interest; it shows how a simplicial grid can give rise to a strongly universally encodable
family of B-V sources.
\par

{\it Example 2.1.} Suppose regular Bratteli diagram $D$ is such that every source in ${\cal S}(D)$ has entropy rate $0$. 
Then ${\cal S}(D)$ is strongly universally encodable.
\par

{\it Example 2.2.} Suppose regular Bratteli diagram $D$ is such that ${\cal S}_e(D)$ is finite.  Then
${\cal S}(D)$ is strongly universally encodable.
\par

{\it Example 2.3.} Let $\beta$ be an integer $\geq 2$. Let $\{K_n:n\geq 0\}$ be a $\beta$-admissible
simplicial grid 
in Euclidean space $\mathbb E$. In Section I, for each $n\geq 1$ and
$x\in V(K_n)$, we defined the unique multiset
$M(x) = \{x_1,x_2,\cdots,x_{\beta}\}$ whose distinct elements are the vertices of
a simplex in $K_{n-1}$  and $x=\beta^{-1}(x_1+\cdots+x_{\beta})$. 
Let $S(x)$ be the set of all $\beta$-tuples $(u_1,u_2,\cdots,u_{\beta})$ such that
$M(x) = \{u_1,u_2,\cdots,u_{\beta}\}$. 
For each $n\geq 0$ and $x\in V(K_n)$, we define
a subset $C_n(x)$ of $V(K_0)^{\beta^n}$ recursively as follows.
The initial sets for the recursion are
$$C_0(x) \define \{x\},\;\; x\in V(K_0).$$
For $n\geq 1$ and $x\in V(K_n)$, we recursively define
$$C_n(x) \define \bigcup_{(u_1,u_2,\cdots,u_{\beta})\in S(x)}C_{n-1}(u_1)\times C_{n-1}(u_2)\times\cdots\times C_{n-1}(u_{\beta}).$$
Note that if $(z_1,z_2,\cdots,z_{\beta^n})$ is any of the $\beta^n$-tuples in  $C_n(x)$, then
$x=\beta^{-n}(z_1+z_2+\cdots+z_{\beta^n})$. Thus, the collection of sets $\{C_n(x):x\in V(K_n)\}$ satisfies
the property that its members are pairwise disjoint.  Let $D=(V,E)$ be the $\beta$-regular canonical Bratteli diagram
for which
$$V_n = \bigcup_{x\in V(K_n)}C_n(x),\;\;n\geq 0.$$
Let $\Theta$ be the underlying space of the grid $\{K_n\}$.
For each $\theta\in \Theta$, let $\tau^{\theta}:V\to[0,1]$ be the mapping defined by
$$\tau^{\theta}(u) \define \frac{p_{\theta|K_n}(x)}{|C_n(x)|},\;\;u\in C_n(x),\;\;x\in V(K_n),\;\;n\geq 0.$$
It can be checked that $\tau^{\theta}$ is a source in ${\cal S}(D)$. 
Let $\Lambda$ be the subfamily of ${\cal S}(D)$ defined by 
\begin{equation}
\Lambda \define \{\tau^{\theta}:\theta\in\Theta\}.
\label{26apr2016eq1}\end{equation}
Let $H_{\Theta}: \Theta\to[0,\infty)$ be the function
\begin{equation}
H_{\Theta}(\theta) \define H_{\infty}(\tau^{\theta}),\;\;\theta\in\Theta.
\label{26apr2016eq2}\end{equation}
$\Lambda$  is a closed subset of topological space ${\cal S}(D)$ and the
function $H_{\Theta}$ is continuous.
Thus, the family of B-V sources $\Lambda$ is strongly universally encodable by Theorem 2.4.
To illustrate, let $\{K_n:n\geq 0\}$ be the $(\beta,k)=(2,1)$ Kuhn grid, whose
underlying space is $\Theta=[0,1]$. 
The resulting entropy rate function $H_{\Theta}:[0,1]\to[0,\infty)$ is plotted in 
Figure 4. The paper \cite{kief2011} explains how to employ iterated function system theory
to quickly generate millions of points on the $H_{\Theta}$ curve.
\par

{\bf Remarks.} 
\begin{itemize}
\item One can construct a specific strong universal encoding scheme for $\Lambda$ of (\ref{26apr2016eq1}) by adapting
the tree-based compression scheme put forth in \cite[Fig.\ 1]{kief2011}. (Letting
$t(n)$ be the finite rooted ordered tree with $\beta^n$ leaves and $\beta$ children for
each non-leaf, for each $x\in V_n$, we will label the vertices of $t(n)$ in a certain
way in our Section V proof of the SMB theorem, starting with label $x$ on the root. 
The scheme of \cite{kief2011} compresses $x$ by compressing this labeled tree.)
\item With some further assumptions on the $\beta$-admissible grid $\{K_n\}$, the entropy rate
function $H_{\Theta}$ defined in (\ref{26apr2016eq2}) will be a self-affine function
(a function whose graph is the attractor of an iterated function system consisting
of affine functions \cite[Chap. 11]{falconer}).  An infinite collection of 
self-affine entropy rate functions arising from simplicial grids are catalogued in \cite{kief2011}.
including the function $H_{\Theta}:[0,1]\to[0,\infty)$ given at the end of Example 2.3.
\end{itemize}
\par

\begin{figure}[htb]
\centering
\mbox{\includegraphics[scale=0.45]{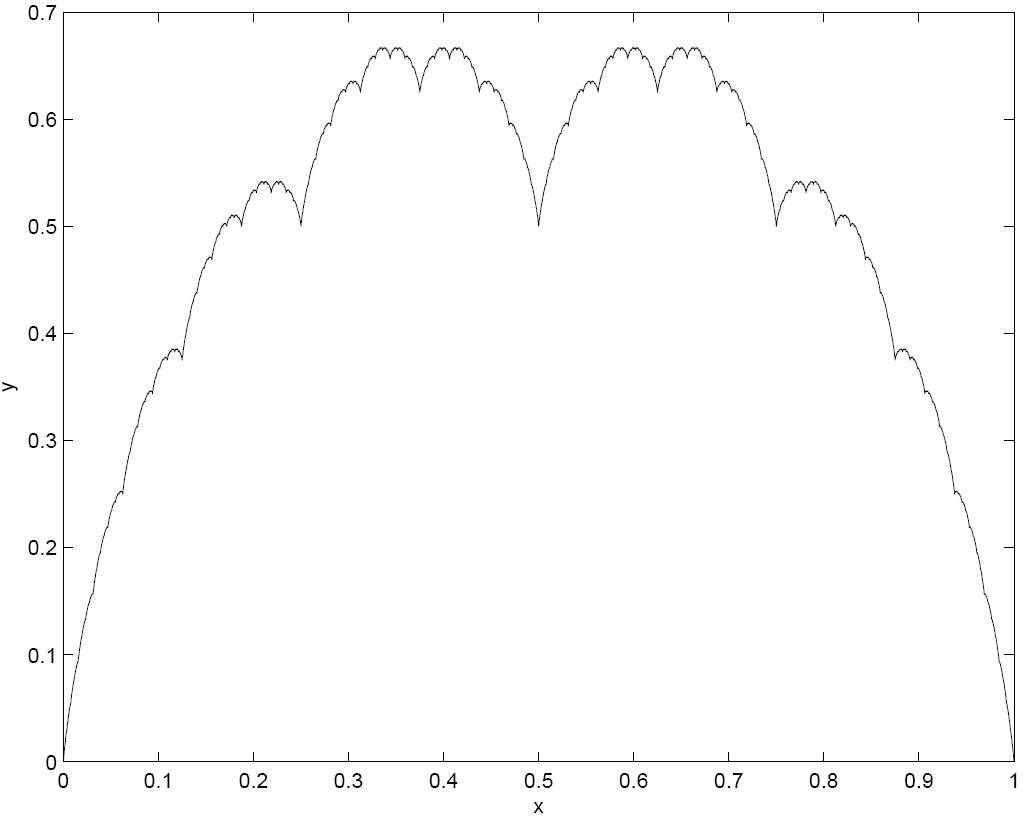}}
\caption{Plot of self affine entropy rate function $H_{\Theta}:[0,1]\to[0,\infty)$ given in
Example 2.3.}
\end{figure}

\subsection{Universal Sequential Source Encoding Schemes} 
Let $A$ be a finite set with at least two elements, and let $D_2(A)=(V,E)$ be the 
$2$-canonical Bratteli diagram $D_2(A)$ of Example 1.3. This is the diagram such that
$V_n = A^{2^n}$ ($n\geq 0$), and such that the
two edges with range vertex $x\in V_n$ ($n\geq 1$) have source
vertices $x[0],x[1]$ in $V_{n-1}$, where $(x[0],x[1])$ is the $2$-decomposition of $x$. 
This subsection shows how we may use universal lossless encoding schemes
for subfamilies of the B-V source family ${\cal S}(D_2(A))$ to construct 
universal lossless encoding schemes for subfamilies
of the sequential source family ${\cal P}(A^{\infty})$. 
\par

If $\alpha$ is a source
in ${\cal P}(A^\infty)$ and $k\geq 1$, then $\alpha^{(k)}$ shall denote the marginal PMF
of $\alpha$ on $A^k$. Furthermore, $H_{\infty}(\alpha)$ shall denote the entropy rate of $\alpha$,
defined by
$$H_{\infty}(\alpha) \define \lim_{k\to\infty}k^{-1}H(\alpha^{(k)}).$$
This limit always exists and is equal to $\inf_kk^{-1}H(\alpha^{(k)})$.
\par

We review universal sequential source encoding concepts put forth in Davisson's ground-breaking
paper \cite{davisson}. A sequence $\{\psi_k:k\geq 0\}$ in which 
$\psi_k$ is a lossless prefix encoder on $A^k$ ($k\geq 1$)
is called a {\it lossless encoding scheme on $A^{\infty}$}.  
A lossless encoding scheme $\{\psi_k\}$ on $A^{\infty}$ is defined to be {\it weak universal
for ${\cal P}(A^{\infty})$} \cite{davisson} if
\begin{equation}
\lim_{k\to\infty}k^{-1}{\rm RED}(\psi_k,\alpha^{(k)}) = 0,\;\;\alpha\in{\cal P}(A^{\infty}),
\label{22feb2016eq1}\end{equation}
or equivalently, if 
\begin{equation}
\lim_{k\to\infty}k^{-1}{\overline L}(\psi_k,\alpha^{(k)}) = H_{\infty}(\alpha),\;\;\alpha\in{\cal P}(A^{\infty}).
\label{19feb2016eq1}\end{equation}
Let ${\cal P}^*(A^{\infty})$ be a subfamily of ${\cal P}(A^{\infty})$. 
A lossless encoding scheme $\{\psi_k\}$ on $A^{\infty}$ is defined to be {\it strong universal
for ${\cal P}^*(A^{\infty})$} \cite{davisson} if
$$\lim_{k\to\infty}\,\left\{\sup_{\alpha\in{\cal P}^*(A^{\infty})}k^{-1}{\rm RED}(\psi_k,\alpha^{(k)})\right\} = 0.$$
\par 

As discussed in  Ex.\ 1.3,
each sequential source $\alpha\in {\cal P}(A^{\infty})$ gives rise
to the source $\alpha^{\dagger}\in {\cal S}(D_2(A))$ whose restriction to $V_n$ is $\alpha^{(2^n)}$ ($n\geq 0$).
 The sources $\alpha$ and $\alpha^{\dagger}$ have the same entropy rate.
The family of sequential sources ${\cal P}(A^{\infty})$
is thus in one-to-one
correspondence with the family of Bratteli-Vershik sources 
$${\cal P}(A^{\infty})^{\dagger} = \{\alpha^{\dagger}:\alpha\in{\cal P}(A^{\infty})\} \subset {\cal S}(D_2(A)).$$
\par

{\it Construction of Schemes on $A^{\infty}$ from schemes on $D_2(A)$}. Let $\{\phi_n:n\geq 0\}$ be any lossless
encoding scheme on $D_2(A)$. Using $\{\phi_n\}$, we show how to construct
a lossless encoding scheme $\{\psi_k:k\geq 0\}$ on $A^{\infty}$. 
For $k\geq 1$, let $S_k$ be the set of positive integers
$$S_k = \{m(k,1),m(k,2),\cdots,m(k,J_k)\}$$
whose elements left to right are the decreasing powers of two which sum to $k$. Factor each $x\in A^k$
as 
$$x=x(1)x(2)\cdots x(J_k),$$
where $x(j)$ is of length $m(k,j)$ ($j=1,2,\cdots,J_k$). Define
$$n(k,j) \define \log_2m(k,j).$$
For each $k\geq 1$, define the mapping 
$\psi_k:A^k\to\{0,1\}^*$ by
\begin{equation}
\mbox{\small $\psi_k(x) = \phi_{n(k,1)}(x(1))\phi_{n(k,2)}(x(2))\cdots \phi_{n(k,J_k)}(x(J_k)),\;\;x\in A^k.$}
\label{25feb2016eq1}\end{equation}
It is straightforward to check that $\psi_k$ is a lossless prefix encoder. Therefore, $\{\psi_k:k\geq 1\}$ is a lossless encoding scheme on $A^{\infty}$. If the scheme $\{\phi_n:n\geq 0\}$ on $D_2(A)$ and
the scheme  $\{\psi_k:k\geq 1\}$ on $A^{\infty}$ are related in this way,
we henceforth state that $\{\phi_n\}$ {\it induces} $\{\psi_k\}$ or that
 $\{\psi_k\}$ {\it is induced by} $\{\phi_n\}$. 

\par
The following two results, which are consequences of Theorems 2.1 and 2.4,  give circumstances
under which a universal scheme on $D_2(A)$ induces a universal scheme on $A^{\infty}$.
They imply results of Davisson \cite[Thm. 7]{davisson} 
on the existence of 
weak and strong universal encoding schemes for sequential source families.
This shows that parts of universal encoding theory for sequential sources are
special cases of universal encoding theory for   Bratteli-Vershik sources.

\par

{\bf Theorem 2.5.} {\it Let $\{\phi_n:n\geq 0\}$ be any lossless encoding scheme on  $D_2(A)$ which is weak universal for ${\cal S}(D_2(A))$. Then the lossless encoding scheme on $A^{\infty}$ induced by $\{\phi_n\}$ is weak universal for ${\cal P}(A^{\infty})$.}
\par

{\bf Theorem 2.6.} {\it Let $\Gamma$ be a subfamily of ${\cal P}(A^{\infty})$ closed with respect to the weak topology on ${\cal P}(A^{\infty})$. Let $H_{\Gamma}:\Lambda\to[0,\infty)$ be the function
$$H_{\Gamma}(\alpha) \define H_{\infty}(\alpha),\;\;\alpha\in\Gamma.$$
Suppose that $H_{\Gamma}$ is a continuous function. Then there exists a lossless encoding scheme on  $D_2(A)$ which is strong universal for  $\Gamma^{\dagger}$. Furthermore, any such scheme induces a
lossless encoding scheme on $A^{\infty}$ which is  strong universal for $\Gamma$.}
\par

{\it Proof of Theorem 2.5.} Let $\{\phi_n:n\geq 0\}$ be a scheme on ${\cal S}(D_2(A))$ which is weak
universal for ${\cal S}(D_2(A))$. Let $\{\psi_k:k\geq 1\}$ be the scheme on $A^{\infty}$ induced by
$\{\phi_n\}$. Let $\alpha\in {\cal P}(A^{\infty})$ be arbitrary and let  $\epsilon>0$ be arbitrary.
The proof is completed by showing that
\begin{equation}
\varlimsup_{k\to\infty}k^{-1}{\overline L}(\psi_k,\alpha^{(k)}) \leq H_{\infty}(\alpha)+\epsilon.
\label{9dec2015eq2}\end{equation}
Since
the scheme $\{\phi_n\}$ is weak universal for ${\cal S}(D_2(A))$, 
\begin{equation}
\lim_{n\to\infty}R(\phi_n,\alpha^{\dagger}) = H_{\infty}(\alpha^{\dagger}) = H_{\infty}(\alpha).
\label{18feb2016eq5}\end{equation}
As a consequence, there is a positive real number $C$ and a positive integer $N$ such that
$$R(\phi_n,\alpha^{\dagger}) \leq C,\;\;n\geq 0,$$
$$R(\phi_n,\alpha^{\dagger}) \leq H_{\infty}(\alpha) + \epsilon,\;\;n\geq N.$$
By (\ref{25feb2016eq1}), 
\begin{equation}
k^{-1}{\overline L}(\psi_k,\alpha^{(k)}) = k^{-1}\sum_{m\in S_k}m\,R(\phi_{\log_2m},\alpha^{\dagger}),\;\;k\geq 1.
\label{18feb2016eq6}\end{equation}
For each $k\geq 1$, we partition $S_k$ into the two sets
$$S_k(1) = \{m\in S_k: m \leq 2^N\}$$
$$S_k(2) = \{m\in S_k: m > 2^N\}$$
We have the bounds
$$\sum_{m\in S_k(1)}m \leq 2^{N+1},$$
$$R(\phi_{\log_2m},\alpha^{\dagger}) \leq H_{\infty}(\alpha) + \epsilon,\;\;m\in S_k(2).$$
Applying these bounds to the right side of equation (\ref{18feb2016eq6}),
$$k^{-1}{\overline L}(\psi_k,\alpha^{(k)}) \leq k^{-1}2^{N+1}C + H_{\infty}(\alpha)+\epsilon,\;\;k\geq 1,$$
from which (\ref{9dec2015eq2}) follows by letting $k\to\infty$.
\par

{\it Proof of Theorem 2.6.} Let $\Lambda=\Gamma^{\dagger}$. Then $\Lambda$ is a subfamily of ${\cal S}(D_2(A))$
which satisfies the assumptions of Theorem 2.4. Thus, there exists a 
lossless encoding scheme on $D_2(A)$ which is strong universal for $\Gamma^{\dagger}$. 
Let $\{\phi_n:n\geq 0\}$ be any such scheme and let $\{\psi_k:k\geq 0\}$ be the lossless encoding
scheme on $A^{\infty}$ induced by $\{\phi_n\}$. Let $\epsilon>0$ be arbitrary. The proof is completed
by showing that
\begin{equation}
\varlimsup_{k\to\infty}\left\{\sup_{\alpha\in\Gamma}\,k^{-1}{\rm RED}(\psi_k,\alpha^{(k)})\right\} \leq\epsilon.
\label{25feb2016eq2}\end{equation}
We have
\begin{equation}
k^{-1}\,{\rm RED}(\psi_k,\alpha^{(k)}) \leq k^{-1}{\overline L}(\psi_k,\alpha^{(k)}) - H_{\infty}(\alpha),\;\;\alpha\in\Gamma.
\label{25feb2016eq5}\end{equation}
By the Dini theorem argument used in the proof of Theorem 2.4, there is a positive integer $N_1$
such that
$$2^{-n}H(\alpha^{(2^n)}) \leq H_{\infty}(\alpha) + \epsilon/2,\;\;n\geq N_1,\;\;\alpha\in\Gamma.$$
Since $\{\phi_n\}$ is strong universal for $\Gamma^{\dagger}$, there is a positive integer
$N_2$ such that
$$R(\phi_n,\alpha^{\dagger}) \leq 2^{-n}H(\alpha^{(2^n)}) + \epsilon/2,\;\;n\geq N_2,\;\;\alpha\in\Gamma.$$
Let $N=\max(N_1,N_2)$. Then we have
$$R(\phi_n,\alpha^{\dagger}) \leq H_{\infty}(\alpha)+\epsilon,\;\;n\geq N,\;\;\alpha\in\Gamma.$$
The mapping $\alpha\to \max_{0\leq n<N}R(\phi_n,\alpha)$ is a continuous mapping on $\Gamma$ and
$\Gamma$ is compact; therefore this mapping is bounded on $\Gamma$.  There thus exists a positive real number $C$ such that
$$R(\phi_n,\alpha^{\dagger}) \leq C,\;\;n\geq 0,\;\;\alpha\in\Gamma.$$
As argued in the proof of Theorem 2.5,
$$k^{-1}{\overline L}(\psi_k,\alpha^{(k)}) \leq k^{-1}2^{N+1}C + H_{\infty}(\alpha)+\epsilon,\;\;k\geq 1,\;\;\alpha\in\Gamma.$$
Applying inequality (\ref{25feb2016eq5}), it follows that
$$\sup_{\alpha\in\Gamma}\,k^{-1}{\rm RED}(\psi_k,\alpha^{(k)}) \leq k^{-1}2^{N+1}C + \epsilon,\;\;k\geq 1.$$
Letting $k\to\infty$, (\ref{25feb2016eq2}) holds and our proof is complete.

\section{Some Bratteli-Vershik System Theory}

In this section, we explain how a regular Bratteli diagram $D$ induces 
the Bratteli-Vershik dynamical systems that were alluded to in Section I. We develop some theory for these systems, 
including the fact that there is a  one-to-one correspondence between
the sources in ${\cal S}(D)$ and the Bratteli-Vershik systems induced by $D$. In later sections, we will be able to establish
results about a Bratteli-Vershik source $\mu$ by exploiting dynamical properties of the Bratteli-Vershik system corresponding
to $\mu$.
\par

Let $\beta$ be an integer $\geq 2$ and let $S_{\beta}$ be the set $\{0,1,\cdots,\beta-1\}$.
Throughout this section, $D=(V,E)$ is a fixed $\beta$-regular Bratteli diagram and 
$I:E\to S_{\beta}$ is any fixed embedding. For each vertex $x\in V^+$ and $i\in S_{\beta}$,
$e_i(x)$ denotes the unique edge in $E(x)$ such that $I(e_i(x))=i$ and $x[i]$ denotes
the vertex $s(e_i(x))$. The $\beta$-tuple $(x[0],x[1],\cdots,x[\beta-1])$ is called
the $\beta$-decomposition of $x$; we have $M_D(x)=\{x[0],\cdots,x[\beta-1]\}$.
\par

In the following, if $(U_0,U_1,\cdots,U_n)$ is a deterministic or random sequence of finite
length and $0\leq i\leq j\leq n$, then $U_i^j\define (U_i,\cdots,U_j)$. If
$(U_0,U_1,U_2,\cdots)$ is an infinite sequence and $i\geq 0$, then $U_i^{\infty}\define (U_i,U_{i+1},\cdots)$
and $U_0^i = (U_0,\cdots,U_i)$.

\subsection{$\beta$-Expansions and $\beta$-adic Arithmetic}

Let $n\geq 1$. Define
$$S_{\beta,n} \define \{0,1,\cdots,\beta^n-1\}.$$
The set of integers $S_{\beta,n}$ and the set of $n$-tuples $S_{\beta}^n$ are both of cardinality $\beta^n$.
Each integer  $i\in S_{\beta,n}$ has a unique expansion 
$$[i]_{\beta,n} = (i_0,i_1,\cdots,i_{n-1}) \in S_{\beta}^n$$
into digits from $S_{\beta}$ in which
\begin{equation}
i = i_0 + i_1\beta + \cdots + i_{n-1}\beta^{n-1}.
\label{24mar2016eq1}\end{equation}
We call $[i]_{\beta,n}$ the $\beta$-expansion of $i$. 
Note that in going from left to right in the expansion $(i_0,\cdots,i_{n-1})$, we are going
from least significant digit $i_0$ to most significant digit $i_{n-1}$.
(Thus, for example, we write $[35]_{2,4} = 110001$
instead of reversed as $100011$.) The mapping $i\to [i]_{\beta,n}$ is a one-to-one mapping
of set $S_{\beta,n}$ onto set $S_{\beta}^n$. 
\par

Listing  the $n$-tuples in $S_{\beta}^n$ in lexicographical order and then reversing
the order of the entries in each $n$-tuple, the list
$$L(n) \define ([0]_{\beta,n},[1]_{\beta,n},\cdots,[\beta^n-1]_{\beta,n})$$
is obtained. For example, if $\beta=3$ and $n=2$, the lexicographical
ordering of $S_{\beta}^n$ is 
$$00,01,02,10,11,12,20,21,22.$$
Reversing these, we obtain the list
$$L(2) = (00,10,20,01,11,21,02,12,22)$$
giving the expansions of $0,1,\cdots,8$, respectively. 
\par
Suppose that $i,i+1$ belong to $S_{\beta,n}$ and we are given the expansion $[i]_{\beta,n}$. It is desired
to compute the expansion $[i+1]_{\beta,n}$ directly from the expansion $[i]_{\beta,n}$. In the rest of this subsection, we develop an efficient method for accomplishing this computation
which exploits $\beta$-adic arithmetic.  
\par

Let $({\mathbb Z}_{\beta},\oplus)$ be the additive group of
the $\beta$-adic integers \cite[Chap.\ 1]{robert}.  ${\mathbb Z}_{\beta}$ consists of all unilateral
infinite sequences ${\bf z}=(z_0,z_1,z_2,\cdots)$ in which all $z_i\in S_{\beta}$. The entries
$z_i$ are called the $\beta$-adic digits of $z$. The $\beta$-adic addition operation $\oplus$
on ${\mathbb Z}_{\beta}$ 
operates similarly
to the usual addition algorithm for integers expanded into decimal digits,  
except our convention is to carry $\beta$-adic digits from left to right 
rather than right to left. For example, if $\beta=3$,
$$(1,0,1,2,1,\cdots) \oplus (2,1,0,1,1,\cdots) = (0,2,1,0,0,\cdots).$$
If $\beta=7$, we have
$$(3,3,2,5,1,\cdots) \oplus (6,2,0,2,3,\cdots) = (2,6,2,0,5,\cdots).$$
Let ${\bf 1}\in{\mathbb Z}_{\beta}$ be the $\beta$-adic integer $(z_0,z_1,z_2,\cdots)$ in which $z_0=1$ and all
other entries are $0$. (There is also a $\beta$-adic multiplication operation $\otimes$ on ${\mathbb Z}_{\beta}$
such that $({\mathbb Z}_{\beta},\oplus,\otimes)$ is a ring, 
and ${\bf 1}$ is the identity element with respect to $\otimes$.) 
The {\it adding machine transformation} on ${\mathbb Z}_{\beta}$ \cite[Chap.\ 7]{pollicott}
is the one-to-one mapping of ${\mathbb Z}_{\beta}$ onto itself in which ${\mathbf z}\in {\mathbb Z}_{\beta}$
is mapped into ${\bf 1}\oplus{\bf z}$. 
If ${\bf z}=(z_i:i\geq 0)$ contains at least one
entry $z_i<\beta-1$, it is easy to compute ${\bf 1}\oplus {\bf z}$ via the following three-step procedure: (i) find the first entry $z_i$ 
of ${\bf z}$ which is $<\beta-1$; (ii) set all entries preceding $z_i$ equal to
zero; (iii)  increase $z_i$ by $1$ and keep all subsequent entries unchanged. To illustrate, for $\beta=2$, we have
$${\bf 1}\oplus (0,1,0,1,1,z_5,z_6,\cdots) = (1,1,0,1,1,z_5,z_6,\cdots),$$
$${\bf 1} \oplus (1,1,0,0,1,z_5,z_6,\cdots) = (0,0,1,0,1,z_5,z_6,\cdots).$$
\par

Let $n\geq 1$. If $\alpha\in S_{\beta}^n$, let $\alpha 0^{\infty}$ be the element of ${\mathbb Z}_{\beta}$
starting with the entries of $\alpha$, followed by infinitely many zeroes. The mapping
$\alpha\to \alpha 0^{\infty}$ is an embedding of $S_{\beta}^n$ into ${\mathbb Z}_{\beta}$.
Let ${\tilde S}_{\beta}^n$ be the set consisting of the $\beta^n-1$ 
$n$-tuples in $S_{\beta}^n$ which are not equal to $(\beta-1,\beta-1,\cdots,\beta-1)$.
If $\alpha\in {\tilde S}_{\beta}^n$, we define ${\mathbf 1}\oplus_n \alpha$ to be the element
$\alpha' \in S_{\beta}^n$ such that 
$${\bf 1}\oplus \alpha 0^{\infty} = \alpha'0^{\infty}.$$
We have
$${\bf 1} \oplus_n [i]_{\beta,n} = [i+1]_{\beta,n},\;\;0\leq i < \beta^n.$$
This formula gives us the efficient method alluded to previously for directly computing $[i+1]_{\beta,n}$ from $[i]_{\beta,n}$. The list $L(n)$ of all $n$-tuples in $S_{\beta}^n$ referred to earlier is then
$$L(n) = (\alpha_0,\alpha_1,\cdots,\alpha_{\beta^n-1}),$$
where the entries of the list are generated via the recursion
$$\alpha_0 = (0,0,\cdots,0),$$
$$\alpha_{i+1} = {\bf 1}\oplus_n\alpha_{i},\;\;0\leq i < \beta^n-1.$$
\par

For example, let $\beta=3$ and $n=3$. Since $\beta^n=3^3=27$, list
$L(3)$ takes the form
$$L(3) = (\alpha_0,\alpha_1,\cdots,\alpha_{26}).$$ The first ten $3$-tuples in 
the list $L(3)$ are generated as follows:
$$\alpha_0 = (000) = [0]_{3,3},$$
$${\bf 1}\oplus_3(000) = (100) = \alpha_1 = [1]_{3,3}$$
$${\bf 1}\oplus_3(100) = (200) = \alpha_2 = [2]_{3,3}$$
$${\bf 1}\oplus_3(200) = (010) = \alpha_3 = [3]_{3,3}$$
$${\bf 1}\oplus_3(010) = (110) = \alpha_4 = [4]_{3,3}$$
$${\bf 1}\oplus_3(110) = (210) = \alpha_5 = [5]_{3,3}$$
$${\bf 1}\oplus_3(210) = (020) = \alpha_6 = [6]_{3,3}$$
$${\bf 1}\oplus_3(020) = (120) = \alpha_7 = [7]_{3,3}$$
$${\bf 1}\oplus_3(120) = (220) = \alpha_8 = [8]_{3,3}$$
$${\bf 1}\oplus_3(220) = (001) = \alpha_9 = [9]_{3,3}$$
The remaining $17$ members of the list $L(3)$ are similarly generated. 
\par

\subsection{Dynamics of Finite Paths in $D$}

A finite path in $D=(V,E)$ is a sequence of finitely many connected edges from $E$. 
We shall be interested in finite paths in $D$ which start at level $0$ (that is, at some vertex in $V_0$) 
and then ascend level-by-level, successively visiting vertices in $V_1,V_2,\cdots$ until the
terminating vertex is reached. If such a path consists of $n$ edges, then the terminating vertex of the path will lie in vertex set $V_n$.
\par

Let $n\geq 1$. $\Pi_D(0,n)$ is the set of all paths in $D$ that start at a vertex in $V_0$
and terminate at a vertex in $V_n$. That is, the paths in $\Pi_D(0,n)$ are the $n$-tuples 
$(\omega_0,\omega_1,\cdots,\omega_{n-1})$
such that $\omega_i\in E_i$ ($0\leq i \leq n-1$) and $r(\omega_i)=s(\omega_{i+1})$ ($0 \leq i\leq n-2$).
The {\it address} $\alpha(y)$ of path $y=(\omega_0,\omega_1,\cdots,\omega_{n-1})\in \Pi_D(0,n)$ is
the $n$-tuple in $S_{\beta}^n$  defined by
$$\alpha(y) \define (I(\omega_0),I(\omega_1),\cdots,I(\omega_{n-1})).$$
The {\it index} $i(y)$ of this path $y$ is defined
to be the integer
computed from the address $\alpha(y)$ via
 $$i(y) \define I(\omega_0) + \beta I(\omega_1) + \cdots + \beta^{n-1} I(\omega_{n-1}).$$
The index $i(y)$ belongs to the set $S_{\beta,n}$. We have
$$\alpha(y) = [i(y)]_{\beta,n},$$
that is, the address of path $y$ is the $\beta$-expansion of the index of $y$. 
We can thus compute the index of
a path in $\Pi_D(0,n)$ from its address and vice-versa.

Let $n\geq 1$ and $x\in V_n$. $\Pi_D(0,n,x)$ is the set of all paths in $D$ that start at a vertex in $V_0$
and end at vertex $x$. That is, $\Pi_D(0,n,x)$ consists of all paths $(\omega_0,\cdots,\omega_{n-1})$
in $\Pi_D(0,n)$ such that $r(\omega_{n-1})=x$. Each path in $\Pi_D(0,n,x)$ is uniquely determined
by its address as follows: If path $(\omega_0,\cdots,\omega_{n-1})\in \Pi_D(0,n,x)$
has address $(i_0,\cdots,i_{n-1})$,  the path is computed via the
backward recursion
\begin{eqnarray}
\omega_{n-1} &=& e_{i_{n-1}}(x),\label{21mar2016eq2}\\
\omega_j &=& e_{i_j}(s(\omega_{j+1})),\;\;0 \leq j \leq n-2.\label{21mar2016eq3}
\end{eqnarray}
Alternatively, each path in $\Pi_D(0,n,x)$ is uniquely determined by its index as follows: If
path $y\in \Pi_D(0,n,x)$ has index $i$, the address of $y$ is then $[i]_{\beta,n}$, from which
$y$ is determined by the above backward recursion. 
The mapping which maps each path in  $\Pi_D(0,n,x)$ into its
 address is a one-to-one
mapping of $\Pi_D(0,n,x)$ onto $S_{\beta}^n$. For each $\alpha\in S_{\beta}^n$,
we let $y[\alpha,x]$ denote the path in $\Pi_D(0,n,x)$ whose address is $\alpha$. 
For each $i\in S_{\beta,n}$, we let $y(i,x)$ denote the path in $\Pi_D(0,n,x)$ whose index is $i$.
\par

The path sets $\{\Pi_D(0,n,x): x\in V_{n}\}$ are pairwise disjoint and their union is $\Pi_D(0,n)$.
Since each path set $\Pi_D(0,n,x)$ has cardinality $\beta^n$, 
the cardinality of $\Pi_D(0,n)$ is $\beta^{n}|V_{n}|$.
\par

Let $n\geq 1$. A path in $\Pi_D(0,n)$ whose address is $(\beta-1,\beta-1,\cdots,\beta-1)$
is called a {\it final path}; equivalently, a path is final if its index is $\beta^n-1$.
A path in $\Pi_D(0,n)$ whose address is $(0,0,\cdots,0)$
is called an {\it initial path}; equivalently, a path is initial if its index is $0$.
Define ${\tilde\Pi}_D(0,n)$ to be the set of paths in $\Pi_D(0,n)$ which are not final.
That is,
$${\tilde\Pi}_D(0,n) \define \{y[\alpha,x]: \alpha\in {\tilde S}_{\beta}^n,\;\;x\in V_n\}.$$
Let $T_n:{\tilde\Pi}_D(0,n)\to\Pi_D(0,n)$
be the one-to-one mapping defined by
$$T_{n}(y(i,x)) \define y(i+1,x),\;\;0\leq i<\beta^n-1,\;\;x\in V_n.$$
Equivalently,
$$T_{n}(y[\alpha,x]) = y[{\bf 1}\oplus_n\alpha,x],\;\;\alpha\in{\tilde S}_{\beta}^n,\;\;x\in V_n.$$
Thus, if $y\in {\tilde \Pi}_D(0,n,x)$ for some $x\in V_n$, then ${ T}_{n}(y)$ is the path in $\Pi_D(0,n,x)$
whose address is ${\bf 1}\oplus_n\alpha(y)$ (which is also the path whose index is  $i(y)+1)$. 
Consequently, given $x\in V_n$, we may start with the initial path in $\Pi_D(0,n,x)$
and iteratively apply the function ${ T}_{n}$ to
dynamically generate the entire set of paths $\{y(i,x):0\leq i\leq \beta^n-1\}$ comprising $\Pi_D(0,n,x)$.
\par

In the next subsection, we see how to splice together the mappings
$T_1, T_2, T_3,\cdots$ to obtain 
 a unique transformation mapping a set of infinite paths in $D$ into itself; this transformation is
the Vershik transformation.

\subsection{Vershik Transformation on Aperiodic Infinite Paths in $D$}

We now consider infinite paths in $D$ that start at vertices in $V_0$; these paths
ascend level-by-level, visiting vertices at  every level $V_n$ ($n\geq 1$) along the way.
Certain of these infinite paths are called aperiodic paths, and it is these aperiodic
paths that will be our focus in this subsection. 
\par

$\Pi_D(0,\infty)$  denotes the path space consisting of all infinite paths in $D$ which
start at a vertex in $V_0$. $\Pi_D(0,\infty)$ thus consists 
of all infinite sequences $\omega=(\omega_0,\omega_1,\omega_2,\cdots)$ 
such that $\omega_0^{n-1} \in \Pi_D(0,n)$ for every $n\geq 1$. 
The Cartesian product space 
$$E(0,\infty) \define E_0\times E_1\times E_2\times\cdots$$
is a compact space under the Cartesian product topology, 
and $\Pi_D(0,\infty)$ is a closed subset of $E(0,\infty)$. Therefore, $\Pi_D(0,\infty)$ is a compact topological space under the topology it inherits from $E(0,\infty)$. The address $\alpha(\omega)$ of $\omega\in \Pi_D(0,\infty)$ is defined to be the
$\beta$-adic integer $(I(\omega_n):n\geq 0) \in {\mathbb Z}_{\beta}$. 
Equivalently, $\beta$-adic integer ${\bf z}=(z_n:n\geq 0)$
is the address of $\omega\in\Omega_D(0,\infty)$ if and only if the path $\omega_0^{n-1}\in\Pi_D(0,n)$
has address $z_0^{n-1}$ for every $n\geq 1$.
 We thus have
$$\{\omega\in\Pi_D(0,\infty): \alpha(\omega)={\bf z}\} =$$
$$ \bigcap_{n=1}^{\infty}\{\omega\in\Pi_D(0,\infty):\alpha(\omega_0^{n-1})=z_0^{n-1}\}.$$
The sets on the right in the preceding equation are non-increasing and non-empty subsets of $\Pi_D(0,\infty)$;
they are also compact sets because they are closed subsets of $\Pi_D(0,\infty)$. 
Thus, the set of paths in $\Pi_D(0,\infty)$ with address ${\bf z}$ is a non-empty compact set
for every ${\bf z}\in{\mathbb Z}_{\beta}$.
\par

A sequence ${\bf z}=(z_n:n\geq 0) \in {\mathbb Z}_{\beta}$ is 
defined to be {\it aperiodic} if it is not eventually
periodic, that is, there exists no $N\geq 0$ such that the sequence $(z_n:n\geq N)$ is periodic.
A path in  $\Pi_D(0,\infty)$ is defined to
be an {\it aperiodic path} if its address is aperiodic.
$\Omega_D$ is defined to
be the set of all aperiodic paths in $\Pi_D(0,\infty)$. $\Omega_D$ is uncountable
because $\alpha(\Omega_D)$ is the uncountable set of  aperiodic sequences in $ {\mathbb Z}_{\beta}$.
$\Omega_D$ is a topological space under the topology it inherits from $\Pi_D(0,\infty)$.
\par

For $n\geq 1$, $i\in S_{\beta,n}$, $\alpha\in S_{\beta}^n$, and $x\in V_n$, we define cylinder sets
\begin{eqnarray*}
C_n(i) &\define& \{\omega\in\Omega_D: \alpha(\omega_0^{n-1})=[i]_{\beta,n}\}\\
C_n(i,x) &\define& \{\omega\in\Omega_D:\omega_0^{n-1}=y(i,x)\}\\
C_n[\alpha,x] &\define& \{\omega\in\Omega_D:\omega_0^{n-1}=y[\alpha,x]\}
\end{eqnarray*}
Each of these cylinder sets is a clopen subset of $\Omega_D$ (that is, it is both closed and open). The
countable collection 
$$\{C_n(i,x):i\in S_{\beta,n},\;\;x\in V_n,\;\;n\geq 1\}$$
of cylinder sets
is a basis for the topology of $\Omega_D$. We let ${\cal F}(\Omega_D)$ denote the sigma-field of subsets
of $\Omega_D$ generated by this collection of cylinder sets.   This gives us
the measurable space $(\Omega_D,{\cal F}(\Omega_D))$.
\par

If $\omega=(\omega_n:n\geq 0)$ is a path in $\Omega_D$, we define the positive integer $N(\omega)$ by
$$N(\omega)\define\min\{n\geq 1:I(\omega_{n-1})<\beta-1\}.$$
$N(\omega)$ has the following meaning: If $n$ is a positive integer,
the path $\omega_0^{n-1}\in \Pi_D(0,n)$ is not a final path if and only if $n\geq N(\omega)$.
Thus, $T_n(\omega_0^{n-1})$ is defined as a path in $\Pi_D(0,n)$ if and only if $n\geq N(\omega)$.

\par

{\bf Lemma 3.1.} $D=(V,E)$ is a $\beta$-regular Bratteli diagram. Let $\omega\in\Omega_D$. For $n\geq N(\omega)$, 
$$T_{n+1}(\omega_0^{n}) = (T_n(\omega_0^{n-1}),\omega_n),$$
the path in $\Pi_D(0,n+1)$ obtained by appending
edge $\omega_n\in E_n$ to the end of path $T_{n}(\omega_0^{n-1})\in\Pi_D(0,n)$. 
\par

{\it Proof.} Let $\omega\in\Omega_D$, let $n\geq N(\omega)$, and let
$$T_{n+1}(\omega_0^n) =  {\tilde\omega} = ({\tilde\omega}_0,\cdots,{\tilde\omega}_n),$$
$$T_n(\omega_0^{n-1}) =  {\hat\omega} = ({\hat\omega}_0,\cdots,{\hat\omega}_{n-1}).$$
We have to show that ${\tilde\omega}_n=\omega_n$ and ${\tilde\omega}_0^{n-1} = {\hat\omega}$. We have
\begin{equation}
\alpha({\tilde\omega}) = {\bf 1}\oplus_{n+1} \alpha(\omega_0^n),
\label{31mar2016eq2}\end{equation}
$$\alpha({\hat\omega}) = {\bf 1}\oplus_n \alpha(\omega_0^{n-1}).$$
Thus, $\alpha({\hat\omega})$ must be a prefix of $\alpha({\tilde\omega})$ since $\alpha(\omega_0^{n-1})$
is a prefix of $\alpha(\omega_0^n)$. That is, letting
$$\alpha({\tilde\omega}) = {\tilde\alpha} = ({\tilde\alpha}_0,\cdots,{\tilde\alpha}_n)$$
$$\alpha({\hat\omega}) = {\hat\alpha} = ({\hat\alpha}_0,\cdots,{\hat\alpha}_{n-1}),$$
we must have
\begin{equation}
{\hat\alpha} = {\tilde\alpha}_0^{n-1}.
\label{31mar2016eq1}\end{equation}
Equation (\ref{31mar2016eq1}) tells us that the paths ${\hat\omega}$ and ${\tilde\omega}_0^{n-1}$
have the same address $\hat\alpha$. Thus, the two paths ${\hat\omega}$ and ${\tilde\omega}_0^{n-1}$
will be identical if we can show that they end at the same vertex in $V_n$.
Let $x\in V_{n+1}$ be the vertex at which path ${\tilde\omega}$ ends. Since $T_{n+1}(\omega_0^n)={\tilde\omega}$, path $\omega_0^n$ also
ends at $x$, and we thus have
\begin{equation}
r(\omega_n) = r({\tilde\omega}_n) = x.
\label{31mar2016eq3}\end{equation}
Let 
$$\alpha(\omega_0^{n}) = (\alpha_0,\cdots,\alpha_n).$$ 
The address of path ${\tilde\omega}$ is computed from the address of path $\omega_0^n$
via the computation on the right side of equation (\ref{31mar2016eq2}). This computation can only
change at most the first $n$ coordinates of the address of $\omega_0^n$ since one of these coordinates
is $<\beta-1$. Thus, $\alpha_n = {\tilde\alpha}_n$, and we have
$$I(\omega_n) = \alpha_n = {\tilde\alpha}_n = I({\tilde\omega}_n).$$
But by (\ref{31mar2016eq3}), both edges $\omega_n$ and ${\tilde\omega}_n$ belong to $E(x)$, so we
must have $\omega_n={\tilde\omega}_n$. Since ${\hat\omega}=T_n(\omega_0^{n-1})$, paths ${\hat\omega}$ 
and $\omega_0^{n-1}$ end at the
same vertex in $V_n$. Since $\omega_n={\tilde\omega}_n$, paths $\omega_0^{n-1}$ and
${\tilde\omega}_0^{n-1}$ end at the same vertex in $V_n$. Thus, paths
${\hat\omega}$  and ${\tilde\omega}_0^{n-1}$ end at the same vertex in $V_n$, allowing
us to conclude that ${\hat\omega} = {\tilde\omega}_0^{n-1}$ as pointed out earlier.
\par

{\it Definition.} The {\it Vershik transformation} is the mapping $T:\Omega_D\to\Omega_D$ defined
as follows. Let $\omega\in\Omega_D$ and let $N=N(\omega)$. Then
$$T(\omega) \define (T_N(\omega_0^{N-1}),\omega_N^{\infty}).$$
By Lemma 3.1, we have the property
$$T(\omega) = (T_n(\omega_0^{n-1}),\omega_n^{\infty}),\;\;n\geq N(\omega),\;\;\omega\in\Omega_D.$$
Thus, for $n$ sufficiently large, the path formed by the first $n$ components of $T(\omega)$ is
$T_n(\omega_{0}^{n-1})$, and the remaining components of $T(\omega)$ coincide with the 
corresponding components of $\omega$. That is,
$$T(\omega)_0^{n-1} = T_n(\omega_0^{n-1}),\;\;n\geq N(\omega),\;\;\omega\in\Omega_D.$$
$$T(\omega)_n^{\infty} = \omega_n^{\infty},\;\;n\geq N(\omega),\;\;\omega\in\Omega_D.$$
From this, it is clear that the definition of the Vershik transformation that was used represents a reasonable
way to splice together the collection of finite path maps $\{T_n:n\geq 1\}$ to obtain a mapping
on infinite paths.  
\par

$T$ is a one-to-one mapping of $\Omega_D$ onto itself and thus the inverse transformation $T^{-1}$ exists.
 Here is a simple argument to establish this.
Let us write $T(I)$ for $T$ to denote its dependence upon the embedding $I$. Let $J:E\to S_{\beta}$
be the embedding $J = \beta-1-I$. Then we have another Vershik transformation $T(J):\Omega_D\to\Omega_D$.
 It is straightforward to argue that both $T(I)\circ T(J)$ and $T(J)\circ T(I)$ are the identity transformation on $\Omega_D$. Thus, $T(J)=T^{-1}$.
\par
{\bf Properties of the Vershik Transformation.} We list some easily established properties of $T$.
\begin{itemize}
\item {\bf (a):} Both $T$ and $T^{-1}$ are continuous, and thus
$T$ is a homeomorphism of the topological space $\Omega_D$. 
\item {\bf (b):} $T$ is bimeasurable (that is, both
$T$ and $T^{-1}$ are measurable mappings), and thus $T$ is an automorphism 
of the measurable space $(\Omega_D,{\cal F}(\Omega_D))$. 
\item {\bf (c):} Let $\sim$ be the equivalence relation on $\Omega_D$ such that
$\omega \sim {\hat\omega}$ if and only if 
 there exists $n\geq 0$ such that $\omega_n^{\infty} = {\hat\omega}_n^{\infty}$. Then
$$\omega \sim T(\omega),\;\;\omega\in\Omega_D,$$
and 
$T$ preserves the relation $\sim$, that is, if $\omega\sim \omega'$ then $T(\omega)\sim T(\omega')$.
\item {\bf (d):} For every $\omega\in\Omega_D$,
$$\alpha(T(\omega)) = {\bf 1}\oplus\alpha(\omega),\;\;\omega\in\Omega_D.$$
\item {\bf (e):} For every $n\geq 1$ and $x\in V_n$,
$$T(C_n(i,x)) = C_n(i+1,x),\;\;x\in V_n,\;\;0\leq i<\beta^n-1.$$
\item {\bf (f):} For every $n\geq 1$,
\begin{eqnarray*}
T(C_n(i)) &=& C_n(i+1),\;\;0\leq i<\beta^n-1\\
T(C_n(\beta^n-1)) &=& C_n(0)
\end{eqnarray*}
\end{itemize}
\par
{\bf Remark.} It can be shown that $T$ is the only mapping on $\Omega_D$ satisfying the
above properties.
\par

\subsection{Characterization of Bratteli-Vershik Dynamical Systems} 

A dynamical system is a quadruple $(\Lambda,{\cal F},P,U)$ in which
$(\Lambda,{\cal F},P)$ is a probability space and $U$ is a one-to-one bimeasurable transformation of
 $\Lambda$ onto itself. Dynamical system $(\Lambda,{\cal F},P,U)$ is measure-preserving 
if $U$ preserves the measure $P$, meaning that
$P(E)=P(U(E))$ for every measurable subset $E$ of $\Lambda$. Measure-preserving
dynamical system $(\Lambda,{\cal F},P,U)$ is ergodic if $P(E)\in\{0,1\}$ for
every event $E\in{\cal F}$ in which $U(E)=E$ (such events $E$ are called the
invariant events of the system).
\par
 
For the given $\beta$-regular Bratteli diagram $D$, let ${\cal P}(\Omega_D,T)$ be the set of all probability measures $P$ on $\Omega_D$ such that $T$ preserves $P$. For each $P\in {\cal P}(\Omega_D,T)$, we have
a measure-preserving dynamical system $(\Omega_D,{\cal F}(\Omega_D),P,T)$, which is called
a {\it Bratteli-Vershik system} induced by $D$. We
let ${\cal P}_e(\Omega_D,T)$ be the collection of all $P\in {\cal P}(\Omega_D,T)$
such that the Bratteli-Vershik system $(\Omega_D,{\cal F}(\Omega_D),P,T)$ is ergodic. 
Theorem 3.2 which follows gives  (a) a one-to-one
correspondence between ${\cal S}(D)$ and  ${\cal P}(\Omega_D,T)$, and (b)
a one-to-one correspondence between ${\cal S}_e(D)$ and  ${\cal P}_e(\Omega_D,T)$. 
In this way, all Bratteli-Vershik systems are characterized in terms of Bratteli-Vershik
sources which give rise to them. Theorem 3.2 will be employed in Sec.\ IV to obtain
the ergodic decomposition theorem for Bratteli-Vershik sources.

\par

{\bf Theorem 3.2.} Let $D$ be the arbitrary $\beta$-regular Bratteli diagram fixed
at the beginning of this section. For each $\mu\in{\cal S}(D)$, there is a unique probability measure
$P_{\mu}$ on $\Omega_D$ such that
\begin{equation}
P_{\mu}(C_n[\alpha,x]) = \beta^{-n}\mu_n(x),\;\;x\in V_n,\;\;\alpha\in S_{\beta}^n,\;\;n\geq 1.
\label{5apr2016eq1}\end{equation}
The following properties hold:
\begin{itemize} 
\item {\bf (a):} If $\mu\in{\cal S}(D)$,
\begin{equation}
P_{\mu}(\{\omega\in\Omega_D:s(\omega_n)=x\}) = \mu_n(x),\;\;x\in V_n,\;\;n\geq 0.
\label{6apr2016eq1}\end{equation}
\item {\bf (b):} $\{P_{\mu}:\mu\in{\cal S}(D)\} = {\cal P}(\Omega_D,T)$.
\item {\bf (c):} $\{P_{\mu}:\mu\in{\cal S}_e(D)\} = {\cal P}_e(\Omega_D,T)$.
\end{itemize}
\par

We present notation and background needed for our proof of Theorem 3.2. 
For  $n\geq 0$, let
 $Y_n:\Omega_D\to E_n$ be the measurable mapping defined by
$$Y_n(\omega) \define \omega_n,\;\;\omega\in\Omega_D.$$
For  $n\geq 1$, the
measurable mapping $Y^{(n)}\define (Y_0,Y_1,\cdots,Y_{n-1})$ on $\Omega_D$ takes its
values in the set $\Pi_D(0,n)$. 
For $n\geq 0$, the measurable mapping $X_n\define s(Y_n)$
on $\Omega_D$ takes its values in $V_n$. Note that $X_n=r(Y_{n-1})$ for $n\geq 1$. 
If $n\geq 0$, the measurable mapping $Z_n\define I(Y_n)$ on $\Omega_D$ takes its values in $S_{\beta}$.
If $n\geq 1$, the measurable mapping 
$$Z^{(n)} \define i(Y^{(n)}) = Z_0 + \beta Z_1 + \cdots + \beta^{n-1}Z_{n-1}$$
on $\Omega_D$ takes its values in $S_{\beta,n}$. 
For $n\geq 1$, 
$$Y^{(n)} = y(Z^{(n)},X_n)$$
and thus $Y^{(n)}$ and $(Z^{(n)},X_n)$ are functions of each other.
 Also,
$$X_{n} = X_{n+1}[Z_n],\;\;n\geq 0,$$
and $(Z^{(n)},X_n)$ is a function of $(Z^{(n+1)},X_{n+1})$ for $n\geq 1$. 
All of the types of cylinder sets defined earlier can be viewed as measurable events
involving the measurable functions just defined. In particular, we have
$$C_n(i) = \{Z^{(n)}=i\}$$
$$C_n(i,x) = \{Z^{(n)}=i,\;\;X_n=x\} = \{Y^{(n)}=y(i,x)\}$$
$$C_n[\alpha,x] = \{(Z_0,\cdots,Z_{n-1})=\alpha,\;\;X_n=x\} = \{Y^{(n)}=y[\alpha,x]\}.$$
Let $Y,X,Z$ be the measurable functions on $\Omega_D$ defined by
\begin{eqnarray*}
Y &\define& (Y_0,Y_1,Y_2,\cdots)\\
X &\define& (X_0,X_1,X_2,\cdots)\\
Z &\define& (Z_0,Z_1,Z_2,\cdots)
\end{eqnarray*}
$Y$ is the identity transformation on $\Omega_D$. 
$X$ takes its values in the Cartesian product measurable space $V_0\times V_1\times V_2\cdots$; its components
track the vertices visited along infinite path $Y$. $Z$ takes its values in ${\mathbb Z}_{\beta}$, and
is the address of path $Y$, that is, $Z = \alpha(Y)$.
Later on, placing a certain type of probability measure $P$ on ${\cal F}(\Omega_D)$, we will
view $Y,Z,X$ as random sequences on probability space $(\Omega_D,{\cal F}(\Omega_D),P)$. 
\par

We now present two lemmas needed in the proof of Theorem 3.2. 
We omit the simple proof of the first of these two lemmas.\par

{\bf Lemma 3.3.} Let $\mu\in{\cal S}(D)$ and let $n\geq 1$. On some probability space,
suppose we have a pair $({\tilde X}_n,{\tilde Z}_{n-1})$ of random objects such that
\begin{itemize}
\item ${\tilde X}_n$ is $V_n$-valued and has probability distribution $\mu_n$. 
\item $Z_{n-1}$ is $S_{\beta}$-valued and its probability distribution is the
uniform distribution on $S_{\beta}$.
\item ${\tilde X}_n$ and ${\tilde Z}_{n-1}$ are statisically independent.
\end{itemize}
Then the $V_{n-1}$-valued random object $X_n[Z_{n-1}]$ has probability distribution $\mu_{n-1}$.
\par

{\bf Lemma 3.4.} Let $\mu\in{\cal S}(D)$. On some probability space, there exists a random pair $({\tilde X},{\tilde Z})$ such that
\begin{itemize}
\item {\bf (a):} ${\tilde X}$ is a random sequence $({\tilde X}_0,{\tilde X}_1,\cdots)$, where ${\tilde X}_n$ is $V_n$-valued and has probability
distribution $\mu_n$ ($n\geq 0$).
\item {\bf (b):} ${\tilde Z}$ is a random sequence $({\tilde Z}_0,{\tilde Z}_1,\cdots)$, where the ${\tilde Z}_n$'s
are independent and $S_{\beta}$-valued, each uniformly distributed over $S_{\beta}$.
\item {\bf (c):} For $n\geq 1$, ${\tilde X}_n$ and $({\tilde Z}_0,{\tilde Z}_1,\cdots,{\tilde Z}_{n-1})$ are independent.
\item {\bf (d):} For $n\geq 1$, 
$$\Pr({\tilde X}_{n-1}={\tilde X}_{n}[{\tilde Z}_{n-1}]) = 1.$$
\end{itemize}
\par

{\it Proof.} On some probability space, there exists random pair $(X^0,Z^0)$ such that
\begin{itemize}
\item $X^0$ is a random sequence $(X^0_0,X^0_1,X^0_2,\cdots)$ with independent components in which $X_n^0$
is  $V_n$-valued and has probability
distribution $\mu_n$ ($n\geq 0$).
\item $Z^0$ is a random sequence $(Z^0_0,Z^0_1,Z^0_2,\cdots)$ with independent components in which $Z_n^0$
is $S_{\beta}$-valued and uniformly distributed over $S_{\beta}$ ($n\geq 0$).
\item $X^0, Z^0$ are independent.
\end{itemize}
For each $N\geq 1$, let $X^N=(X^N_0,X^N_1,X^N_2,\cdots)$ be the random sequence defined recursively by
$$X^N_n = X^0_n,\;\;n\geq N,$$
$$X^N_n = X^N_{n+1}[Z^0_n],\;\;0 \leq n < N.$$
Exploiting Lemma 3.3, we have the properties
\begin{itemize}
\item {\bf (e):} For every $N\geq 1$, random pair $(X^N,Z^0)$ obeys the property that
$X^N_n$ and $(Z_0^0,Z_1^0,\cdots,Z_{n-1}^0)$ are independent for every $n\geq 1$.
\item {\bf (f):} For every $N\geq 1$, random sequence $X^N$ obeys the property that
component $X^N_n$ of $X^N$ is $V_n$-valued with probability distribution $\mu_n$ for every $n\geq 0$.
\item {\bf (g):} For $N>n\geq 0$, $\Pr(X_n^N = X_{n+1}^N[Z^0_n]) = 1$.
\item {\bf (h):} For each $n\geq 1$, the random $2n$-tuple 
$$(X^N_0,X^N_1,\cdots,X^N_{n-1},Z^0_0,Z^0_1,\cdots,Z^0_{n-1})$$
has probability distribution which does not depend on $N$ for $N\geq n$.
\end{itemize}
Let measurable space $\Lambda_1$ be the Cartesian product space
$$\Lambda_1 = V_0\times V_1\times V_2\times\cdots,$$
and let measurable space $\Lambda_2$ be the Cartesian product space $\Lambda_1\times{\mathbb Z}_{\beta}$.
$Z^0$ is ${\mathbb Z}_{\beta}$-valued, and $X^N$ is $\Lambda_1$-valued for each $N\geq 1$. Thus,
the random pair $(X^N,Z^0)$ is $\Lambda_2$-valued for each $N\geq 1$. For each $N\geq 1$, let
$P_N$ be the probability measure on $\Lambda_2$ which is the probability distribution of $(X^N,Z^0)$.
By property (h), the sequence $\{P_N:N\geq 1\}$ converges weakly to a probability measure $P$
on $\Lambda_2$. Let $({\tilde X},{\tilde Z})$ be any $\Lambda_2$-valued random
pair whose probability distribution is $P$. The sequence
of random pairs $\{(X^N,Z^0):N\geq 1\}$ then converges in distribution to the random pair
$({\tilde X},{\tilde Z})$. As a consequence, ${\tilde Z}$ and $Z^0$ have the same probability
distribution, giving us property (b), and then 
properties (a),(c),(d) follow by letting $N\to\infty$
in properties (f),(e),(g), respectively.
\par

{\it Proof of Theorem 3.2 Part 1.} Fix $\mu\in {\cal S}(D)$. We prove there exists unique probability
measure $P_{\mu}$ on $\Omega_D$ such that (\ref{5apr2016eq1}) holds, and we also prove that
(\ref{6apr2016eq1}) holds. 
On some probability space $\Lambda$, we may define processes
$${\tilde X} = ({\tilde X}_0,{\tilde X}_1,{\tilde X}_2,\cdots)$$
$${\tilde Z} = ({\tilde Z}_0,{\tilde Z}_1,{\tilde Z}_2,\cdots)$$
according to Lemma 3.4. As we may throw away a set of measure zero if necessary, we may assume that 
${\tilde X}_n={\tilde X}_{n+1}[{\tilde Z}_n]$ holds everywhere on $\Lambda$ for every $n\geq 0$.
For each $n\geq 0$, let ${\tilde Y}_n$ be the $E_n$-valued random object 
${\tilde Y}_n = e_{{\tilde Z}_n}({\tilde X}_{n+1})$. This gives us random sequence
$${\tilde Y} \define ({\tilde Y}_0,{\tilde Y}_1,{\tilde Y}_2,\cdots).$$
We have:
\begin{itemize}
\item For every $n\geq 0$, 
$$r({\tilde Y}_n) = {\tilde X}_{n+1},\;\;s({\tilde Y}_n)={\tilde X}_n,\;\;{\tilde Z}_n=I({\tilde Y}_n)$$
hold everywhere on $\Lambda$.
\item ${\tilde Y}$ takes its values in $\Pi_D(0,\infty)$.
\item The address of ${\tilde Y}$ is ${\tilde Z}$.
\end{itemize}
Since ${\tilde Z}$ is a non-degenerate IID process,
with probability $1$ it takes its values in the set of aperiodic sequences in ${\mathbb Z}_{\beta}$.
Therefore, random path ${\tilde Y}$ lies in $\Omega_D$
with probability $1$. Let $P_{\mu}$ be the probability measure on ${\cal F}(\Omega_D)$ such that
$$P_{\mu}(E) = \Pr\{ {\tilde Y} \in E\},\;\;E\in{\cal F}(\Omega_D).$$
$(Y,X,Z)$, regarded as random triple on probability space $(\Omega_D,{\cal F}(\Omega_D),P_{\mu})$,
has the same probability distribution as the random triple $({\tilde Y},{\tilde X},{\tilde Z})$.
Evaluating the left side of equation (\ref{5apr2016eq1}), we have
$$P_{\mu}(C_n[\alpha,x]) = P_{\mu}\{(Z_0,\cdots,Z_{n-1})=\alpha,\;X_n=x\} =$$
$$ \Pr\{({\tilde Z}_0,\cdots,{\tilde Z}_{n-1})=\alpha,\;{\tilde X}_n=x\}.$$
By independence of $({\tilde Z}_0,\cdots,{\tilde Z}_{n-1})$ and ${\tilde X}_n$, this last expression
factors as
$$\Pr\{ ({\tilde Z}_0,\cdots,{\tilde Z}_{n-1})=\alpha\}\Pr\{{\tilde X}_n=x\} = \beta^{-n}\mu_n(x).$$
Thus, (\ref{5apr2016eq1}) holds. $P_{\mu}$ is unique because the cylinder sets generate
${\cal F}(\Omega_D)$. Evaluating the left side of (\ref{6apr2016eq1}),
$$P_{\mu}(\{\omega\in\Omega_D:s(\omega_n)=x\}) = P_{\mu}\{X_n=x\} = \Pr\{{\tilde X}_n=x\} = \mu_n(x),$$
and thus equation (\ref{6apr2016eq1}) holds.
\par

{\it Proof of Theorem 3.2 Part 2.} We prove statement(b) of the theorem. Let $P\in {\cal P}(\Omega_D,T)$.
For each $n\geq 0$, let $\mu_n$ be the PMF on $V_n$ defined by
$$\mu_n(x) \define P\{X_n=x\},\;\;x\in V_n.$$
Let $n\geq 1$ and $x\in V_n$. The cylinder sets $\{C_n(i,x):0\leq i\leq \beta^n-1\}$ all have
the same probability under $P$ because $T$ preserves $P$ and $T^i(C_n(0,x))=C_n(i,x)$. There are $\beta^n$ of these cylinder
sets and their union is the event $\{X_n=x\}$, so each one must have probability equal to
$\beta^{-n}\mu_n(x)$. We have shown
$$P\{Y^{(n)}=y,\;\;X_n=x\} = \beta^{-n}\mu_n(x),\;\;x\in V_n,\;\;y\in\Pi_D(0,n,x),\;\;n\geq 1.$$
We prove that
\begin{equation}
\mu_n(x) = \beta^{-1}\sum_{i=0}^{\beta-1}\mu_{n+1}\{w\in V_{n+1}:w[i]=x\},\;\;x\in V_n,\;\;n\geq 0.
\label{6apr2016eq2}\end{equation}
Fix $n\geq 0$ and  $x\in V_n$. Let $F_n$ be any event in ${\cal F}(\Omega_D)$ (a particular
$F_n$ will be chosen later on). 
Let
$$G = \{(e,w): e\in E_n,\;\;w\in V_{n+1},\;\;s(e)=x,\;\;r(e)=w\}.$$
We have
$$P(F_n\cap\{X_n=x\}) = P(F_n\cap\{(Y_n,X_{n+1})\in G\}),$$
since the events $\{X_n=x\}$ and $\{(Y_n,X_{n+1})\in G\}$ are identical. 
For $i\in S_{\beta}$, let $G_i=\{(e,w)\in G: e=e_i(w)\}$.
The $G_i$'s form
a partition of $G$, and so
$$P(F_n\cap\{(Y_n,X_{n+1})\in G\}) =$$
$$ \sum_{i=0}^{\beta-1}\sum_{(e,w)\in G_i}P(F_n\cap\{Y_n=e,\;X_{n+1}=w\}) =$$
$$\sum_{i=0}^{\beta-1}\sum_{w\in V_{n+1}}\sum_{e\in G_i[w]}P(F_n\cap\{Y_n=e,\;X_{n+1}=w\}),$$
where $G_i[w]$ is the section of $G_i$ at $w$, namely,
$$G_i[w] = \{e\in E_n:(e,w)\in G_i\}.$$
We have $G_i[w] = \{e_i(w)\}$ if $w\in V_{n+1}$ is such that $s(e_i(w))=x$,
whereas $G_i[w]$ is the empty set if $s(e_i(w))\not = x$. This fact, coupled
with the fact that $s(e_i(w))=w[i]$, gives us
$$P(F_n\cap\{X_n=x\}) =$$
\begin{equation}
 \sum_{i=0}^{\beta-1}\left\{\sum_{w\in V_{n+1}:w[i]=x}P(F_n\cap\{Y_n=e_i(w),\;X_{n+1}=w\})\right\}.
\label{13oct2015eq1}\end{equation}
If $n=0$, choose $F_n = \Omega_D$, and if $n>0$, choose $F_n$ to be any cylinder set
of form $\{Y^{(n)}=y\}$, where $y\in \Omega_D(0,n,x)$. Whether $n=0$ or $n>0$, we have
$$P(F_n\cap \{X_n=x\}) = \beta^{-n}P(\{X_n=x\}) = \beta^{-n}\mu_n(x),$$
$$P(F_n\cap\{Y_n=e_i(w),\;X_{n+1}=w\}) = \beta^{-(n+1)}\mu_{n+1}(w).$$
Substituting into (\ref{13oct2015eq1}), we have
$$\beta^{-n}\mu_n(x) = \beta^{-n}\beta^{-1}\sum_{i=0}^{\beta-1}\left\{\sum_{w\in V_{n+1}:w[i]=x}\mu_{n+1}(w)\right\}.$$
 Cancelling $\beta^{-n}$ from both sides, equation (\ref{6apr2016eq2}) has been established.
Let $\mu:V\to[0,1]$ be the mapping whose restriction to $V_n$ is $\mu_n$ for every $n\geq 0$.
From (\ref{6apr2016eq2}), we conclude that $\mu\in{\cal S}(D)$ and therefore $P=P_{\mu}$.
To finish the proof of Theorem 3.2(b), let $\mu\in{\cal S}(D)$ and
let $P=P_{\mu}$; we show $P\in{\cal P}(\Omega_D,T)$.
Let $Q$ be the probability measure on ${\cal F}(\Omega_D)$
such that $Q(F)=P(T(F))$ for $F\in{\cal F}(\Omega_D)$. We have to show $P=Q$.
For each $n\geq 1$, let ${ {\cal F}}_n$
be the finite sub sigma-field of ${\cal F}({\Omega}_D)$ whose atoms are
the $|V_{n}|(\beta^{n}-1)$ cylinder sets
$$C_n(i,x),\;\;0\leq i < \beta^n-1,\;\;x\in V_n$$
together with the event $C_n(\beta^n-1)$. 
For each atom of ${\cal F}_n$ of the form $C_n(i,x)$,
$$Q(C_n(i,x)) = P(T(C_n(i,x)) = P(C_n(i+1,x)) = P(C_n(i,x)),$$
where the rightmost equality is due to the fact that both $P(C_n(i+1,x))$ and $P(C_n(i,x))$
are equal to $\beta^{-n}\mu_n(x)$. Thus, $P,Q$ coincide on each sigma-field ${\cal F}_n$ ($n\geq 1$).
Let ${ {\cal F}}_{\infty} \define \cup_n{ {\cal F}}_n$. 
Since ${\cal F}_n \subset {\cal F}_{n+1}$ for $n\geq 1$, 
${ {\cal F}}_{\infty}$
is a field of subsets of ${\Omega}_D$. $P,Q$ coincide on ${ {\cal F}}_{\infty}$,
and therefore $P=Q$ because the field ${\cal F}_{\infty}$ generates ${\cal F}(\Omega_D)$.
\par

{\it Proof of Theorem 3.2 Part 3.} We prove statement(c) of the theorem. 
Let ${\cal P}^{\dagger} = \{P_{\mu}: \mu\in {\cal S}_e(D)\}$.  The mapping $\mu\to P_{\mu}$ is
a one-to-one affine mapping of convex set ${\cal S}(D)$ onto convex set ${\cal P}(\Omega_D,T)$.
Therefore, it maps the set of extreme points of ${\cal S}(D)$ onto the set of extreme points
of ${\cal P}(\Omega_D,T)$. Thus, ${\cal P}^{\dagger}$ is the set of extreme points of ${\cal P}(\Omega_D,T)$.
By \cite[Thm.\ 2]{blum}, the set of extreme points of ${\cal P}(\Omega_D,T)$ coincides with
${\cal P}_e(\Omega_D,T)$. Therefore, $\{P_{\mu}: \mu\in {\cal S}_e(D)\} = {\cal P}_e(\Omega_D,T)$
and statement(c) holds.
\par

{\bf Remarks.} As a by-product of the earlier proofs, we see that
a probability measure $P$ on $\Omega_D$ belongs to  ${\cal P}(\Omega_D,T)$ if and only if
the pair  $(X,Z)$, regarded as a pair of random sequences on probability
space $(\Omega_D,{\cal F}(\Omega_D),P)$, satisfies the properties:
\begin{itemize}
\item {\bf (a):} ${ Z}$ is an IID random sequence, with each component $Z_n$ uniformly distributed over 
$S_{\beta}$.  
\item {\bf (b):} For each  $n\geq 1$, ${ X}_n$ and $({ Z}_0,{ Z}_1,\cdots,{ Z}_{n-1})$ are independent.
\end{itemize}
Given probability measure $P$ on $\Omega_D$ satisfying (a) and (b), then the source $\mu\in{\cal S}(D)$
such that $P=P_{\mu}$ is given by
$$\mu(x) = P\{X_n=x\},\;\;x\in V_n,\;\;n\geq 0.$$
\par

Each measure-preserving dynamical system has an entropy, whose definition is reviewed
 in the next paragraph. Theorem 3.5 below relates the entropies of  Bratteli-Vershik
dynamical systems to the entropy rates of the Bratteli-Vershik sources that give rise to them.
This relationship is needed later on in Sec.\ V to establish the Shannon-McMillan-Breiman
theorem for Bratteli-Vershik sources. 
\par
Suppose $(\Lambda,{\cal F},P,U)$ is a
measure preserving dynamical system. If $\cal W$ is a collection of random objects
on probability space $(\Lambda,{\cal F},P)$, we let $\sigma({\cal W})$ denote the sub sigma-field
of $\cal F$ generated by $\cal W$, that is, $\sigma({\cal W})$ is the smallest sub sigma-field
of $\cal F$ with respect to which all of the RVs in $\cal W$ are measurable. 
Let $W$ be an arbitrary random variable on $(\Lambda,{\cal F},P)$ taking finitely many values.
For each integer $i$, let $W_i$ be the random variable $W\circ T^i$. Define
$$H_P(\infty,W) \define \lim_{n\to\infty}n^{-1}H_P(W_0,\cdots,W_{n-1}).$$
(In information-theoretic terms, $H_P(\infty,W)$ is the entropy rate of the random
sequence $(W_0,W_1,W_2,\cdots)$.) The entropy of dynamical system $(\Lambda,{\cal F},P,U)$ is defined as
$$H(\Lambda,{\cal F},P,T) \define \sup_WH_P(\infty,W).$$
Furthermore, suppose that $W$ is a generator of the system $(\Lambda,{\cal F},P,U)$, meaning that
$\sigma(\{W_i:i\in{\mathbb Z}\})={\cal F}$. A well known result \cite[Cor.\ 3.12]{parry} tells us that
$$H(\Lambda,{\cal F},P,T) = H_P(\infty,W).$$
\par

{\bf Theorem 3.5.} Let $D$ be the arbitrary $\beta$-regular Bratteli diagram fixed
at the beginning of this section. For each $\mu\in{\cal S}(D)$, the entropy of the
Bratteli-Vershik system $(\Omega_D,{\cal F}(\Omega_D),P_{\mu},T)$
is $H_{\infty}(\mu).$

{\it Proof of Theorem 3.5 Part 1.} Fix $\mu\in{\cal S}(D)$ and let $P=P_{\mu}$. We 
show that $H(\Omega_D,{\cal F}(\Omega_D),P,T) \leq H_{\infty}(\mu)$. Let $n\geq 1$.
Choose $W^n$ to be a random variable 
on probability space $(\Omega_D,{\cal F}(\Omega_D),P)$ taking $|V_n|+\beta^n-1$ values such that the atoms of $\sigma(\{W^n\})$
are the  $|V_{n}|$ cylinder sets
$$C_n(\beta^n-1,x) = \{Z^{(n)}=\beta^n-1,\;X_n=x\},\;\;x\in V_n$$
together with the $\beta^{n}-1$ cylinder sets
$$C_n(i) = \{Z^{(n)}=i\},\;\;0\leq i < \beta^{n}-1.$$
For each $i\in {\mathbb Z}$, let $W^n_i$ be the random variable $W^n\circ T^i$. 
We have measure-preserving
dynamical system $(\Omega_D,{\cal F}_n,P,T)$ in which ${\cal F}_n = \sigma(\{W^n_i:i\in{\mathbb Z}\})$.
$W^n$ is a generator for this system, and so
$$H(\Omega_D,{\cal F}_n,P,T) = \lim_{m\to\infty}m^{-1}H_P(W^n_0,W^n_1,\cdots,W^n_{m-1}).$$
The terms on the right side of this equation are non-increasing as $m$ increases, and so
\begin{equation}
H(\Omega_D,{\cal F}_n,P,T) \leq \beta^{-n}H_P(W^n_0,W^n_1,\cdots,W^n_{\beta^n-1}).
\label{10apr2016eq1}\end{equation}
By properties (e) and (f) of the Vershik transformation, one sees that the atoms of
the sigma-field $\sigma(\{W^n_i: i\in S_{n,\beta}\})$ are the $\beta^n|V_n|$ cylinder
sets
$$C_n(i,x)=\{Z^{(n)}=i,\;X_n=x\},\;\;i\in S_{\beta,n},\;\;x\in V_n.$$
The random vector $(W^n_0,W^n_1,\cdots,W^n_{\beta^n-1})$ and
the random pair $(Z^{(n)},X_n)$ are thus functions of each other.
We then have
\begin{eqnarray*}
H_P(W^n_0,W^n_1,\cdots,W^n_{\beta^n-1}) &=& H_P(Z^{(n)},X_n)\\
&=& H_P(Z^{(n)}) + H_P(X_{n})\\
&=& \log_2(\beta^{n}) + H(\mu_{n})
\end{eqnarray*}
Substituting into the right side of (\ref{10apr2016eq1}) and letting $n\to\infty$, we obtain
$$\varlimsup_n\;H(\Omega_D,{\cal F}_n,P,T) \leq \varlimsup_{n}\;\beta^{-n}[\log_2(\beta^{n}) + H(\mu_{n})] = H_{\infty}(\mu).$$
We have the monotonicity property
$${\cal F}_1 \subset {\cal F}_2 \subset {\cal F}_3 \subset \cdots$$
since $(Z^{(n)},X_n)$ is a function of $(Z^{(n+1)},X_{n+1})$ for every $n\geq 1$.
In addition to this property, we have the 
property that the field of sets $\cup_n{\cal F}_n$ generates ${\cal F}({\Omega}_D)$. 
From these properties, \cite[Thm.\ 5.10]{parry} tells us that
the sequence $\{H(\Omega_D,{\cal F}_n,P,T):n\geq 1\}$ is non-decreasing and
$$H(\Omega_D,{\cal F}(\Omega_D),P,T) =  \lim_{n\to\infty}H(\Omega_D,{\cal F}_n,P,T).$$
We conclude that $H(\Omega_D,{\cal F}(\Omega_D),P,T)\leq H_{\infty}(\mu)$.
\par

{\it Proof of Theorem 3.5 Part 2.} We show that
$$H_{\infty}(\mu) \leq H(\Omega_D,{\cal F}(\Omega_D),P_{\mu},T),\;\;\mu\in{\cal S}(D).$$
For convenience, we assume that the diagram $D=(V,E)$
is canonical in this proof. (There is no loss of generality in assuming this, since
a regular diagram is isomorphic to a canonical one.) Thus, $V_n$ is a subset of
$V_0^{\beta^n}$ for every $n\geq 1$. Letting $n\geq 1$ and $x\in V_n$, we have
the factorization
$$x=x[0]x[1]\cdots x[\beta-1],$$
in which the $x[i]$'s belong to $V_{n-1}$. (The indexing $I:E\to S_{\beta}$
is assumed to be the  natural one in which 
the edge $e_i(x)$ such that $I(e_i(x))=i$ has source vertex $x[i]\in V_{n-1}$.)
Fix $\mu\in{\cal S}(D)$ and let $P=P_{\mu}$.
The sequences $X=(X_n:n\geq 0)$ and $Z=(Z_n:n\geq 0)$ of measurable functions
on $\Omega_D$ defined earlier are regarded as random sequences on probability space
$(\Omega_D,{\cal F}(\Omega_D),P)$ in this proof. 
For each integer $i$, let $W_i$ be the $V_0$-valued random object
$W_i = X_0\circ T^i$. 
We have
$$\lim_{m\to\infty}m^{-1}H(W_0,W_1,\cdots,W_{m-1}) \leq H(\Omega_D,{\cal F}(\Omega_D),P,T).$$
Fix arbitrary positive integer $m$.  Fix arbitrary positive integer $n$ such that $\beta^{n}>m$.
Since $V_n \subset V_0^{\beta^n}$ and $X_n$ is $V_n$-valued, $X_n$ is a random $\beta^n$-tuple, which we write as
$${ X}_{n} = (U_0,U_1,\cdots,U_{\beta^{n}-1}),$$
where coordinate $U_i$ of $X_n$ is a $V_0$-valued random object.
Using the formula
$$X_{j-1} = X_j[Z_{j-1}],\;\;1\leq j \leq n,$$
it follows that
\begin{equation}
\{{ Z}^{(n)}=i\} \subset \{{ X}_0 = U_i\},\;\;0\leq i\leq \beta^{n}-1.
\label{3nov2015eq1}\end{equation}
Let
$$R_j=\{0\leq i \leq \beta^{n}-m: {\rm mod}(i,m)=j\},\;\;j\in \{0,1,\cdots,m-1\}.$$
 Fix $j\in \{0,1,\cdots,m-1\}$.
The blocks $\{U_i^{i+m-1}:i\in R_j\}$ are adjacent non-overlapping blocks of length $m$
in ${ X}_{n}$ which together constitute all of $X_n$ except for at most $m-1$
coordinates of $X_n$ at the beginning and at most $m-1$ coordinates of
$X_n$ at the end. Each coordinate $U_i$ of $X_n$ has entropy at most $\log_2|V_0|$. Thus, by subadditivity of entropy,
$$H(\mu_n) = H_P({ X}_{n}) \leq 2(m-1)\log_2|V_0| + \sum_{i\in R_j}H_P(U_i^{i+m-1}).$$
Summing over $j$, we obtain
\begin{equation}
mH(\mu_n)  \leq 2m(m-1)\log_2|V_0| + \sum_{i=0}^{\beta^{n}-m}H_P(U_i^{i+m-1}).
\label{4nov2015eq1}\end{equation}
Suppose $Z^{(n)}=i$, where $0 \leq i \leq \beta^{n}-m$. Let $j\in\{0,\cdots,m-1\}$.
Then $0\leq i+j\leq \beta^n-1$ and so by properties of the Vershik transformation, we
have $Z^{(n)}\circ T^j=i+j$ and $X_n\circ T^j=X_n$. By (\ref{3nov2015eq1}), 
$W_j=X_0\circ T^j$ is equal to coordinate $i+j$ of $X_n\circ T^j$, which is $U_{i+j}$,
coordinate $i+j$ of $X_n$. We have shown that
$$\{{ Z}^{(n)}=i\} \subset \{W_0^{m-1}=U_i^{i+m-1}\},\;0\leq i\leq \beta^n-m,$$
which implies
$$H_P(W_0^{m-1}|Z^{(n)}=i) = H_P(U_i^{i+m-1}|{ Z}^{(n)}=i),\;\;0\leq i\leq\beta^n-m.$$
For each such $i$, the random $m$-tuple $U_i^{i+m-1}$
and ${ Z}^{(n)}$ are statistically independent, because $U_i^{i+m-1}$ is
a function of ${ X}_{n}$ and $ X_{n}, { Z}^{(n)}$ are independent.
Thus,
$$H_P(U_i^{i+m-1}|{ Z}^{(n)}=i) = H_P(U_i^{i+m-1}),\;\;0\leq i\leq\beta^n-m.$$
Employing (\ref{4nov2015eq1}), 
\begin{equation}
\sum_{i=0}^{\beta^{n}-m}H_P(W_0^{m-1}|{ Z}^{(n)}=i) = \sum_{i=0}^{\beta^{n}-m}H_P(U_i^{i+m-1})
\label{4nov2015eq2}\end{equation}
$$\geq  mH(\mu_n) - 2m(m-1)\log_2|V_0|.$$
We also have
$$H_P(W_0^{m-1}) \geq H_P(W_0^{m-1}|{ Z}^{(n)})$$
$$ = \sum_{i=0}^{\beta^{n}-1}{ P}\{{ Z}^{(n)}=i\}H_P(W_0^{m-1}|{Z}^{(n)}=i)$$
$$\geq \beta^{-n}\sum_{i=0}^{\beta^{n}-m}H_P(W_0^{m-1}|{ Z}^{(n)}=i),$$
using the fact that ${ Z}^{(n)}$ is equiprobable over the set $\{0,1,\cdots,\beta^{n}-1\}$.
Combining with (\ref{4nov2015eq2}), we have shown that
$$m^{-1}H_P(W_0^{m-1}) \geq \beta^{-n}H(\mu_n) -2\beta^{-n}(m-1)\log_2|V_0|$$
holds for any pair $(m,n)$ of positive integers for which $\beta^{n}>m$. Holding $m$
fixed and letting $n\to\infty$
on both sides of this statement,
$$m^{-1}H_P(W_0^{m-1}) \geq H_{\infty}(\mu),\;\;m\geq 1.$$
Letting $m\to\infty$,
\begin{equation}
H(\Omega_D,{\cal F}(\Omega_D),P,T) \geq \lim_{m\to\infty}m^{-1}H_P(W_0^{m-1}) \geq H_{\infty}(\mu),
\label{12apr2016eq1}\end{equation}
and our proof is complete.
\par
{\bf Remark.} Letting $W_j=X_0\circ T^j$ for $j\in{\mathbb Z}$, we see 
that
$$H(\Omega_D,{\cal F}(\Omega_D),P_{\mu},T) = H_{\infty}(\mu) = \lim_{m\to\infty}m^{-1}H_{P_{\mu}}(W_0^{m-1})$$
holds for every $\mu\in{\cal P}(D)$,
because by Theorem 3.5 the rightmost and leftmost quantities in (\ref{12apr2016eq1}) are equal.

\section{Decomposition Theorems}

In this section, we prove the
Ergodic Decomposition Theorem (Theorem 1.7) and the Entropy
Rate Decomposition Theorem (Theorem 1.8). We prove Theorem 1.7 first. 
Our proof of Theorem 1.7 employs the following representation theorem, proved in \cite{blum}.
 \par

{\bf Blum-Hanson Representation Theorem.} Let $(\Gamma,{\cal F})$ be a measurable space, and let $U$ be a one-to-one
bimeasurable transformation of $\Gamma$ onto itself. Let ${\cal P}$ be the set of
probability measures on $\Gamma$ which are preserved by $U$,
let ${\cal F}_U$ be the set of events in $\cal F$ which are $U$-invariant,
and let ${\cal P}_e=\{P\in{\cal P}:P({\cal F}_U)=\{0,1\}\}$ (the set of measures
in ${\cal P}$ which are ergodic with respect to $U$). 
Suppose that
\begin{eqnarray}
\{F\in {\cal F}_U:P(F)=0\;{\rm for}\;{\rm all}\;P\in{\cal P}_e\} &=&\nonumber\\
\{F\in {\cal F}_U:P(F)=0\;{\rm for}\;{\rm all}\;P\in{\cal P}\}&&\label{11nov2015eq1}
\end{eqnarray}
Let ${\cal F}({\cal P}_e)$ be the smallest sigma-field of subsets of ${\cal P}_e$
such that for each $F\in{\cal F}$, the mapping $P\to P(F)$ is a measurable mapping
from ${\cal P}_e$ into ${\mathbb R}$. Then for each $P\in {\cal P}$, there exists
a unique probability measure $\lambda_P$ on ${\cal F}({\cal P}_e)$ such that
$$P(F) = \int_{{\cal P}_e}Q(F)d\lambda_P(Q),\;\;F\in {\cal F}.$$
\par

We proceed with the proof of Theorem 1.7 after establishing Lemmas 4.1-4.2 below. Throughout
the rest of this section, $D=(V,E)$ is a fixed $\beta$-regular Bratteli diagram.
We need the following definitions.\par

{\it Definitions.}
\begin{itemize}
\item For each $v\in V$, $\Phi_v:{\cal S}_e(D)\to{\mathbb R}$ is the mapping
defined by
$$\Phi_v(\mu) \define \mu(v),\;\;\mu\in{\cal S}_e(D).$$
\item For each $G\in {\cal F}({\Omega}_D)$, $\Phi_G:{\cal P}_e({\Omega}_D,T)\to {\mathbb R}$
is the mapping defined by
$$\Phi_G(P) \define P(G),\;\;P\in{\cal P}_e({\Omega}_D,T).$$
\item Define ${\cal F}({\cal P}_e({\Omega}_D,T))$ to
be the smallest of all sigma-fields $\cal F$ of subsets of ${\cal P}_e({\Omega}_D,T)$ such that
$\Phi_G$ is ${\cal F}$-measurable for each $G\in{\cal F}({\Omega}_D)$.
\item Define the mapping $\Psi:{\cal S}_e(D)\to{\cal P}_e({\Omega}_D,T)$ by
$$\Psi(\mu) \define {P}_{\mu},\;\;\mu\in{\cal S}_e(D).$$
\end{itemize}
\par

Recall that in Sec.\ I, ${\cal S}_e(D)$ was taken to be the measurable space with
sigma-field the family of all Borel subsets with respect to the topology on ${\cal S}_e(D)$.
Let ${\cal F}({\cal S}_e(D))$ denote this sigma-field of Borel sets. ${\cal F}({\cal S}_e(D))$
is then the smallest sigma-field of subsets of ${\cal S}_e(D)$ containing all ${\cal S}_e(D)$-open sets.  
\par

{\bf Lemma 4.1.} ${\cal F}({\cal S}_e(D))$ is the smallest of all sigma-fields $\cal F$ of subsets 
of ${\cal S}_e(D)$ such that the mapping $\Phi_v$ is ${\cal F}$-measurable for every $v\in V$.
\par

{\it Proof.} Let ${\cal F}_1$ be the smallest of all sigma-fields $\cal F$ of subsets of ${\cal S}_e(D)$
such that the mapping $\Phi_v$ is ${\cal F}$-measurable for every $v\in V$.
Part 1 of the proof is to show that ${\cal F}_1 \subset {\cal F}({\cal S}_e(D))$.
Part 2 of the proof is to show that ${\cal F}({\cal S}_e(D)) \subset {\cal F}_1$.
\par
{\it Part 1 of Proof.}  ${\cal F}_1 \subset {\cal F}({\cal S}_e(D))$ is established
by showing that every $\Phi_v$ is ${\cal F}({\cal S}_e(D))$-measurable. 
Fix $\Phi_v$ and any real number $x$. We have to show that the set $S=\Phi_v^{-1}((-\infty,x))$
belongs to ${\cal F}({\cal S}_e(D))$. Since $\Phi_v$ is continuous with respect to
the topology on ${\cal S}_e(D)$, the set $S$ is ${\cal S}_e(D)$-open. But then
$S\in {\cal F}({\cal S}_e(D))$ because this sigma-field contains all ${\cal S}_e(D)$-open sets.

\par

{\it Part 2 of Proof.} ${\cal F}({\cal S}_e(D)) \subset {\cal F}_1$ is established
by showing that ${\cal F}_1$ contains every ${\cal S}_e(D)$-open set. Let
$\cal B$ be the collection of all subsets of ${\cal S}_e(D)$ of the form
$$\bigcap_{v\in S}\Phi_v^{-1}(I_v),$$
where $S$ is a finite subset of $V$ and each $I_v$
is a non-empty bounded open sub-interval of the real line having rational endpoints.
${\cal S}_e(D)$'s topology is the set of all unions of members of $\cal B$.
Every member of $\cal B$
belongs to ${\cal F}_1$, and $\cal B$ is countable.
If
$O$ is an arbitrary ${\cal S}_e(D)$-open set, we argue that $O\in {\cal F}_1$
as follows: let $\{O_i\}$
be a subset of $\cal B$ such that $O=\cup_iO_i$, which is a countable union of members of ${\cal F}_1$
since $\cal B$ is countable. $O$ must therefore belong to ${\cal F}_1$, since any sigma-field is closed with respect to countable unions.

{\bf Lemma 4.2.} The mapping $\Psi$ is a one-to-one bimeasurable mapping
of measurable space $({\cal S}_e(D),{\cal F}({\cal S}_e(D)))$ onto measurable
space $({\cal P}_e({\Omega}_D,T),{\cal F}({\cal P}_e({\Omega}_D,T)))$.
\par

{\it Proof.} Let ${\cal F}_1 = {\cal F}({\cal S}_e(D))$ and let ${\cal F}_2 = {\cal F}({\cal P}_e(\Omega_D,T))$.
From our previous results, $\Psi$ is a one-to-one
mapping of ${\cal S}_e(D)$ onto ${\cal P}_e(\Omega_D,T)$. What remains is to prove that $\Psi$ is
bimeasurable. That is, we have to prove the two statements
\begin{eqnarray}
\Psi^{-1}({\cal F}_2) &\subset& {\cal F}_1\label{12nov2015eq1}\\
\Psi({\cal F}_1) &\subset& {\cal F}_2\label{12nov2015eq2}
\end{eqnarray}
This gives us two parts of the proof.
\par
{\it Part 1: Proof of (\ref{12nov2015eq1}).} Define the sigma-field ${\cal F}_2^*$ of subsets of ${\cal P}_e(\Omega_D,T)$ by
$${\cal F}_2^* \define \{F\in {\cal F}_2: \Psi^{-1}(F) \in {\cal F}_1\}.$$
If we can show that ${\cal F}_2 \subset {\cal F}_2^*$, then we will have 
${\cal F}_2 = {\cal F}_2^*$, establishing (\ref{12nov2015eq1}).
By definition of ${\cal F}_2$, it will follow that
${\cal F}_2 \subset {\cal F}_2^*$ if we can show that the mapping $\Phi_G$ is
${\cal F}_2^*$-measurable for every $G\in {\cal F}({\Omega}_D)$. First, we consider
the case when $G$ is the cylinder set $G = C_n(i,v)$,  
where $n\geq 1$, $i\in S_{\beta,n}$, and $v\in V_{n}$ are arbitrary. 
To show that $\Phi_G$ is ${\cal F}_2^*$-measurable, we must show that
the set $S_x=\Phi_G^{-1}((-\infty,x)) \in {\cal F}_2^*$ for every $x\in {\mathbb R}$,
which by definition of ${\cal F}_2^*$ means we must show that $\Psi^{-1}(S_x)\in {\cal F}_1$. We have
$$\Psi^{-1}(S_x) = \{\mu\in{\cal S}_e(D):{P}_{\mu}(G)< x\} = \{\mu\in{\cal S}_e(D):\mu(v)< \beta^{n}x\},$$
using the fact that $P_{\mu}(C_n(i,v)) = \beta^{-n}\mu(v)$. 
Thus, $\Psi^{-1}(S_x) \in {\cal F}_1$ because it is an ${\cal S}_e(D)$-open set.
Let
$${\cal M} = \{G\in {\cal F}({\Omega}_D):\Phi_G\;{\rm is}\;{\cal F}_2^*{\rm -measurable}\}.$$
${\cal M}$ is a monotone class containing the field generated by the cylinder sets. By the monotone class 
theorem \cite[Thm.\ 1.3.9]{ashdade}, ${\cal M}={\cal F}({\Omega}_D)$. We conclude that
$\Phi_G$ is ${\cal F}_2^*$-measurable for every $G \in {\cal F}({\Omega}_D)$, completing
Part 1.

\par

{\it Part 2: Proof of (\ref{12nov2015eq2}).} Define the sigma-field ${\cal F}_1^*$ of subsets of ${\cal S}_e(D)$ by
$${\cal F}_1^* \define \{F\in {\cal F}_1: \Psi(F) \in {\cal F}_2\}.$$
If we can show that ${\cal F}_1 \subset {\cal F}_1^*$, then we will have 
${\cal F}_1 = {\cal F}_1^*$, establishing (\ref{12nov2015eq2}). By Lemma 4.1, it will follow that
${\cal F}_1 \subset {\cal F}_1^*$ if we can show that the mapping $\Phi_v$ is
${\cal F}_1^*$-measurable for every $v\in V$. Fix $v\in V$. Then $v\in V_n$ for some $n\geq 0$.
Let $G=\{{X}_n=v\}$. Let $x\in{\mathbb R}$ be arbitrary. Then 
$$\Psi(\Phi_v^{-1}((-\infty,x))) = \Phi_G^{-1}((-\infty,x))\in{\cal F}_2,$$
and so the event $\Phi_v^{-1}((-\infty,x))$ belongs to ${\cal F}_1^*$. 
We conclude that $\Phi_v$ is ${\cal F}_1^*$-measurable, completing Part 2.

\subsection{Proof of Theorem 1.7}

 In this proof, we use the notations ${\cal P}(T)$ and
${\cal P}_e(T)$ to denote the spaces ${\cal P}({\Omega}_D,T)$ and ${\cal P}_e({\Omega}_D,T)$, respectively. 
As before, let ${\cal F}_1 = {\cal F}({\cal S}_e(D))$ and let ${\cal F}_2 = {\cal F}({\cal P}_e(\Omega_D,T))$.
Recall that  ${\cal F}_T$ is the sub sigma-field 
of ${\cal F}(\Omega_D)$ consisting of the $T$-invariant sets. We have broken down the proof of Theorem 1.7 into
three parts. 

\par

{\it Part 1.} Let $P\in {\cal P}(T)$. We show there exists probability measure $\tau_P$ on 
measurable space $({\cal P}_e(T),{\cal F}_2)$ which represents $P$
in the sense that
 \begin{equation}
P(F) = \int_{{\cal P}_e(T)}Q(F)d\tau_P(Q),\;\;F\in {\cal F}({\Omega}_D).
\label{11nov2015eq30}\end{equation}
Let $\mu\in {\cal S}(D)$ be the measure such that
${P}_{\mu}=P$. By Lemma 1.1, there exists probability measure $\lambda$ on
measurable space $({\cal S}_e(D),{\cal F}_1)$ such that
\begin{equation}
\mu(v) = \int_{{\cal S}_e(D)}\sigma(v)d\lambda(\sigma),\;\;v\in V.
\label{11nov2015eq10}\end{equation}
From this it follows that
\begin{equation}
P(F) = \int_{{\cal S}_e(D)}{ P}_{\sigma}(F)d\lambda(\sigma)
\label{11nov2015eq20}\end{equation}
for every cylinder set 
$$F = C_n(i,v),\;\;i\in S_{\beta,n},\;\;v\in V_{n},\;\;n\geq 1.$$ 
(Multiply both sides
of (\ref{11nov2015eq10}) by $\beta^{-n}$ and then use the facts that $P(F)=\beta^{-n}\mu(v)$ and ${P}_{\sigma}(F) = \beta^{-n}\sigma(v)$.)  
The collection of events $F\in {\cal F}({\Omega}_D)$
for which  equation (\ref{11nov2015eq20}) holds is a monotone class containing the field
generated by the cylinder sets, and thus (\ref{11nov2015eq20}) holds for every $F\in {\cal F}({\Omega}_D)$
by the monotone class theorem \cite[Thm.\ 1.3.9]{ashdade}. By Lemma 4.2, we have 
probability measure $\tau_P$ on $({\cal P}_e(T),{\cal F}_2)$ defined by
$$\tau_P(G) \define \lambda(\Psi^{-1}(G)) = \lambda(\{\mu\in{\cal S}_e(D): {P}_{\mu} \in G\}),\;\;G\in {\cal F}_2.$$
Making a change of variable in (\ref{11nov2015eq20}), we obtain (\ref{11nov2015eq30}).

\par
{\it Part 2.} For each $P\in {\cal P}(T)$,
we show that $\tau_P$ representing $P$ in the sense of (\ref{11nov2015eq30}) is unique.
Suppose $G\in {\cal F}_T$
and $Q(G)=0$ for all $Q\in {\cal P}_e(T)$. Fix arbitrary $P\in {\cal P}(T)$. By Part 1, 
measure $\tau_P$ on ${\cal P}_e(T)$ exists such that (\ref{11nov2015eq30}) holds.
Setting $F=G$ in this equation, we conclude that $P(G)=0$. We have verified that
property (\ref{11nov2015eq1}) holds for the dynamical system $(\Lambda,{\cal F},U) = 
({\Omega}_D, {\cal F}({\Omega}_D),T)$. Thus, 
for each $P\in {\cal P}(T)$, the representing measure $\tau_P$ in (\ref{11nov2015eq30})
is unique by the Blum-Hanson Representation Theorem.
\par

{\it Part 3.} Let $\mu\in{\cal S}(D)$. To complete the proof of Theorem 1.7, 
we have to show that the measure $\lambda$ on $({\cal S}_e(D),{\cal F}_1)$ satisfying
 (\ref{11nov2015eq10}) is unique. That is, we must show $\lambda_1=\lambda_2$,
where $\lambda_1$ and $\lambda_2$ are any two 
probability measures on $({\cal S}_e(D),{\cal F}_1)$ for which
\begin{equation}
\mu(v) = \int_{{\cal S}_e(D)}\sigma(v)d\lambda_1(\sigma) = \int_{{\cal S}_e(D)}\sigma(v)d\lambda_2(\sigma),\;\;v\in V.
\label{5nov2015eq2}\end{equation}
Let $P={ P}_{\mu}$. Let $\tau_1,\tau_2$
be the probability measures on $({\cal P}_e(T),{\cal F}_2)$ such that
$$\tau_i(G) \define \lambda_i(\Psi^{-1}(G)),\;\;G\in {\cal F}_2,\;\;i=1,2.$$
As argued in Part 1,
$$P(F) = \int_{{\cal P}_e(T)}Q(F)d\tau_1(Q) = \int_{{\cal P}_e(T)}Q(F)d\tau_2(Q),\;\;F\in{\cal F}({\Omega}_D).$$
By Part 2, $\tau_1=\tau_2$.  Let $G\in {\cal F}_1$ be arbitrary. Let $G'=\Psi(G)$, which
is an event in ${\cal F}_2$ by Lemma 4.2. Since $G=\Psi^{-1}(G')$, we have 
$$\tau_i(G') = \lambda_i(\Psi^{-1}(G')) = \lambda_i(G),\;\;i=1,2.$$
Therefore, $\lambda_1(G)=\lambda_2(G)$ since $\tau_1(G')=\tau_2(G')$. We conclude that $\lambda_1=\lambda_2$,
completing the proof of Theorem 1.7.

\subsection{Proof of Theorem 1.8}

Throughout this subsection, we have a fixed $\beta$-regular Bratteli diagram $D=(V,E)$.
By Theorem 1.5, there exists for each $n\geq 0$ a lossless prefix encoder $\phi_n:V_n\to\{0,1\}^*$
such that
$$\lim_{n\to\infty}\beta^{-n}{\overline L}(\phi_n,\sigma_n) = H_{\infty}(\sigma),\;\;\sigma\in{\cal S}(D).$$
For each $n\geq 0$, let $\psi_n:V_n\to\{0,1\}^*$ be a lossless prefix encoder employing fixed-length codewords of length
$\lceil\log_2|V_n|\rceil$. 
For each $n\geq 0$, let $\phi_n^*:V_n\to\{0,1\}^*$ be the lossless prefix encoder $\phi_n^* = \phi_n \wedge \psi_n$.
For $\sigma\in{\cal S}(D)$ and $n\geq 0$, we have 
$$\beta^{-n}H(\sigma_n)\leq \beta^{-n}{\overline L}(\phi_n^*,\sigma_n) \leq \beta^{-n}({\overline L}(\phi_n,\sigma_n) + 1).$$
Therefore,
$$\lim_{n\to\infty}\beta^{-n}{\overline L}(\phi_n^*,\sigma_n) = H_{\infty}(\sigma),\;\;\sigma\in{\cal S}(D).$$
Since $|V_n| \leq |V_0|^{\beta^n}$, 
$$\beta^{-n}{\overline L}(\phi_n^*,\sigma_n) \leq \beta^{-n}(\lceil\log_2|V_n|\rceil+1) \leq \log_2|V_0| + 2$$
holds for all $\sigma\in{\cal S}(D)$ and $n\geq 0$. 
\par

Fix $\mu\in {\cal S}(D)$. Let $\lambda_{\mu}$ be the unique probability measure on ${\cal S}_e(D)$ such that
$$\mu(x) = \int_{{\cal S}_e(D)}\sigma(x)d\lambda_{\mu}(\sigma),\;\;x\in V.$$
Then
\begin{equation}
\beta^{-n}{\overline L}(\phi_n^*,\mu_n) =  \int_{{\cal S}_e(D)}\beta^{-n}{\overline L}(\phi_n^*,\sigma_n)d\lambda_\mu(\sigma),\;\;n\geq 0. \label{15apr2016eq1}\end{equation}
By the bounded convergence theorem of probability theory,
$$\lim_{n\to\infty}\int_{{\cal S}_e(D)}\beta^{-n}{\overline L}(\phi_n^*,\sigma_n)d\lambda_{\mu}(\sigma) =$$
$$\int_{{\cal S}_e(D)}\left[\lim_{n\to\infty}\beta^{-n}{\overline L}(\phi_n^*,\sigma_n)\right]d\lambda_{\mu}(\sigma)\\
 = \int_{{\cal S}_e(D)}H_{\infty}(\sigma)d\lambda_{\mu}(\sigma).$$
Thus, letting $n\to\infty$ on both sides of equation (\ref{15apr2016eq1}), we obtain
$$H_{\infty}(\mu) = \int_{{\cal S}_e(D)}H_{\infty}(\sigma)d\lambda_{\mu}(\sigma),$$
the desired conclusion of Theorem 1.8.

\section{Shannon-McMillan-Breiman Theorem}

This section proves the Shannon-McMillan-Breiman theorem (SMB theorem) for Bratteli-Vershik information sources.
We first review the SMB theorem for
stationary finite-alphabet sequential sources. Let $A$ be a finite non-empty set.
Let $\{U_i:i\geq 0\}$ be a stationary random sequence of $A$-valued random objects
defined on a probability space $\Lambda$.
For $n\geq 1$, let $p_n$ be the PMF on $A^n$ which is the probability distribution of 
the $A^n$-valued random vector $(U_1,\cdots,U_n)$. The SMB theorem for $\{U_i\}$
states that the sequence of random variables $\{-n^{-1}\log_2p_n(U_1,\cdots,U_n):n\geq 1\}$
converges in $L^1[\Lambda]$ space norm and almost surely to a random variable on $\Lambda$
belonging to $L^1[\Lambda]$. $L^1[\Lambda]$ convergence in the SMB theorem is due to
Shannon and McMillan \cite{mcmillan}. Almost sure convergence in the SMB theorem is
due to Breiman \cite{breiman}.
\par

Let $D=(V,E)$ be a $\beta$-regular Bratteli diagram, fixed throughout
this section. Recall the collections $\{Y_n:n\geq 0\}$, $\{X_n:n\geq 0\}$, and $\{Y^{(n)}:n\geq 1\}$
of measurable functions on $\Omega_D$ defined in Section III. 
Let $\mu\in{\cal S}(D)$ be a fixed Bratteli-Vershik information  source on $D$ 
and let $P=P_{\mu}$, which is the probability measure on ${\cal F}(\Omega_D)$ such that
$$P\{Y^{(n)}=y,\;\;X_n=x\} = \beta^{-n}\mu_n(x),\;\;x\in V_n,\;\;y\in \Pi_D(0,n,x),\;\;n\geq 1.$$
This gives us the probability space 
$\Lambda_P =  ({\Omega}_D,{\cal F}({\Omega}_D),P)$. 
The expected value of a random variable $U$ on $\Lambda_P$ shall be denoted $E_P[U]$.
Let $L^1[\Lambda_P]$ be the space of random variables $U$ on $\Lambda_P$ for which $E_P[|U|]<\infty$.
 For each $n\geq 0$,
the random object $X_n:\Lambda_P\to V_n$ has probability distribution $\mu_n$. 
Furthermore, the random variable $-\log_2\mu_n(X_n)$ belongs to $L^1[P]$ and its expected value is $H(\mu_n)$. 
The SMB theorem for Bratteli-Vershik sources (Theorem 5.3 below) tells us that
the sequence of random variables $\{-\beta^{-n}\log_2\mu_n(X_n):n\geq 0\}$ converges
in $L^1[\Lambda_P]$ and almost surely $[P]$ to
some limit function in $L^1[\Lambda_P]$, and moreover, the limit function and its distribution
are identified. To give the full statement of this theorem, we need to lay some
groundwork below.
\par

Let us recall some of the machinery from Section III. Fix an indexing $I:E\to S_{\beta}$. 
For each $n\geq 1$ and $x\in V_n$ and $i\in S_{\beta}$, we have the
vertex $x[i] \in V_{n-1}$ which is the source vertex of the edge in $E(x)$ whose
$I$-value is $i$.  Letting $Z_n\define I(Y_n)$, we obtain the
IID sequence $Z=\{Z_n:n\geq 0\}$ of random variables on $\Lambda_P$ in which each
$Z_n$ is $S_{\beta}$-valued and is uniformly distributed over $S_{\beta}$. 
For $n\geq 1$, $(Z_0,Z_1,\cdots,Z_{n-1})$ and $X_n$ are statistically independent random
objects on $\Lambda_P$ and $X_{n-1}=X_n[Z_{n-1}]$ holds everywhere on $\Omega_D$. Furthermore,
letting 
$$Z^{(n)} = Z_0 + Z_1\beta + \cdots + Z_{n-1}\beta^{n-1},\;\;n\geq 1,$$
$Z^{(n)}$ is uniformly distributed over $S_{\beta,n}$. 
\par

$N$ is the positive integer valued
random variable on $\Lambda_P$ defined in Sec.\ III by
$$N \define \min\{n\geq 1: Z_{n-1}<\beta-1\}.$$
We have $N=n$ if and only if the path $Y^{(n)}\in \Pi_D(0,n)$ is not a final path and no
prefix of $Y^{(n)}$ is a final path. The random variable $N$ gave rise to the Vershik
transformation in Sec.\ III: if $N(\omega)=n$, then $T(\omega)_0^{n-1} = T_n(\omega_0^{n-1})$
and $T(\omega)_n^{\infty}=\omega_n^{\infty}$. $N$ has a geometric distribution:
$$P\{N=n\} = (\beta^{-1})^{n-1}(1-\beta^{-1}),\;\;n\geq 1.$$
Another useful fact about random variable $N$ is that it is a stopping time relative to the
random sequence $Z=(Z_i:i\geq 0)$, that is, the event $\{N=n\}$ is $(Z_0,\cdots,Z_{n-1})$-measurable
for every $n\geq 1$. 
\par

{\bf Lemma 5.1.} Let $\mu\in {\cal S}(D)$, let $P=P_{\mu}$, let
$\Lambda_P$ be the probability space $(\Omega_D,{\cal F}(\Omega_D),P)$, and let $n\geq 1$.
Then 
\begin{equation}
\{x\in V_n:\mu(x)>0\} = \bigcap_{i=0}^{\beta-1}\{x\in V_n:\mu(x)>0,\;\mu(x[i])>0\},
\label{21apr2016eq3}\end{equation}
and for the resulting $L^1[\Lambda_P]$ random variable
\begin{equation}
W_n \define \log_2\frac{\mu_{n}(X_{n})}{\prod_{i=0}^{\beta-1}\mu_{n-1}(X_{n}[i])},
\label{19apr2016eq3}\end{equation}
we have 
\begin{eqnarray*}
E_P[W_n] &=& \beta H(\mu_{n-1}) - H(\mu_n)\\
E_P[|W_n|] &\leq & 2e^{-1}\log_2e + \beta H(\mu_{n-1}) - H(\mu_n)
\end{eqnarray*}
\par

{\it Proof.} Fix $\mu$ and let $P=P_{\mu}$. This gives us probability space $\Lambda_P$.
Let $n\geq 1$. Let $x\in V_n$ satisfy $\mu_n(x)>0$. Let $i\in S_{\beta}$ and let $u=x[i]$.
We show $\mu_{n-1}(u)>0$, establishing (\ref{21apr2016eq3}). We have
\begin{equation}
\mu_{n-1}(u) = \beta^{-1}\sum_{j=0}^{\beta-1}\mu_n(\{x'\in V_n:x'[j]=u\}).
\label{21apr2016eq4}\end{equation}
Since $x\in \{x'\in V_n:x'[i]=u\}$, the $\mu_n$-measure of this set is $>0$
and hence $\mu_{n-1}(u)>0$. Recall that $X_m$ has PMF $\mu_m$ for every $m\geq 0$. Hence,
$$P\{\mu_n(X_n)>0\} = \mu_n(\{x\in V_n:\mu_n(x)>0\}) = 1,$$
and so $\log_2\mu_n(X_n)$ belongs to $L^1[\Lambda_P]$. 
By (\ref{21apr2016eq3}),
$$P\{\mu_{n-1}(X_n[i])>0\} = \mu_n(\{x\in V_n:\mu_{n-1}(x[i])>0\}) = 1,\;\;i\in S_{\beta},$$
and so $\log_2\mu_{n-1}(X_n[i])$ belongs to $L^1[\Lambda_P]$ for $i\in S_{\beta}$. It follows
that $W_n$ belongs to $L^1[\Lambda_P]$ because
$$W_n = \log_2\mu_n(X_n) - \sum_{i=0}^{\beta-1}\log_2(X_n[i]).$$
Since (\ref{21apr2016eq4}) holds for every $u\in V_{n-1}$, it follows that 
$$E_P[f(X_{n-1})] = \beta^{-1}\sum_{i=0}^{\beta-1}E_P[f(X_n[i])]$$
for every extended real-valued function $f:V_{n-1}\to[0,\infty]$.
Taking $f$ to be the function 
$$f(x) \define -\log_2\mu_{n-1}(x),\;\;x\in V_{n-1},$$
we have
\begin{eqnarray*}
E_P[W_n] &=& E_P[\log_2\mu_n(X_n)] + \sum_{i=0}^{\beta-1} E_P[f(X_n[i])]\\
&=& E_P[\log_2\mu_n(X_n)] + \beta E_P[f(X_{n-1})]\\
&=& -H(\mu_n) + \beta H(\mu_{n-1})
\end{eqnarray*}
The proof is completed by showing that
\begin{equation}
E_P[|W_n|] \leq 2e^{-1}\log_2e + E_P[W_n]
\label{19apr2016eq4}\end{equation}
As discussed in Sec.\ I, there is an isomorphism carrying $D=(V,E)$
into a canonical diagram ${\tilde D}=({\tilde V},{\tilde E})$ in which every ${\tilde V}_n$ is a subset of $V_0^{\beta^n}$. Under this isomorphism, for each $n\geq 0$ and $x\in V_n$, there corresponds 
a string $\eta(x)$ in $V_0^{\beta^n}$. Furthermore, for $n\geq 1$ and $x\in V_n$,
$$\eta(x) = \eta(x[0])\eta(x[1])\cdots\eta(x[\beta-1]).$$
On probability space $\Lambda_P$, we have the collections of random objects $\{{\tilde X}_n:n\geq 0\}$ and
$\{{\tilde X}_n[i]:n\geq 1,\;i\in S_{\beta}\}$ such that
\begin{itemize}
\item ${\tilde X}_n$ is the $V_0^{\beta^n}$-valued random object $\eta(X_n)$.
\item ${\tilde X}_n[i]$ is the $V_0^{\beta^{n-1}}$-valued random object
$\eta(X_n[i])$.
\end{itemize}
For $n\geq 0$, let ${\tilde \mu}_n$ be the PMF of ${\tilde X}_n$ on $V_0^{\beta^n}$.
For $n\geq 1$, let $\lambda_n$ be the $\beta$-fold product PMF on $V_0^{\beta^n}$ defined by
$$\lambda_n \define {\tilde\mu}_{n-1}\times{\tilde\mu}_{n-1}\times\cdots\times{\tilde\mu}_{n-1}.$$
Fix $n\geq 1$. Since
$$\mu_n(X_n) = {\tilde \mu}_n({\tilde X}_n),$$
$$\mu_{n-1}(X_n[i]) =  {\tilde \mu}_{n-1}({\tilde X}_n[i]),\;\;i\in S_{\beta},$$
it follows that
$$W_n = \log_2\left[\frac{{\tilde\mu}_n({\tilde X}_n)}{\lambda_n({\tilde X}_n)}\right]$$
almost surely $[P]$. Let $d{\tilde \mu}_n/d\lambda_n$ be the Radon-Nikodym derivative of ${\tilde \mu}_n$ with respect to $\lambda_n$. The function $\log_2(d{\tilde \mu}_n/d\lambda_n)$ is called the information density
of ${\tilde \mu}_n$ with respect to $\lambda_n$. We have
\begin{eqnarray*}
E_P[W_n] &=& \int_{V_0^{\beta^n}}\log_2(d{\tilde \mu}_n/d\lambda_n)d{\tilde\mu}_n,\\
E_P[|W_n|]&=& \int_{V_0^{\beta^n}}|\log_2(d{\tilde \mu}_n/d\lambda_n)|d{\tilde\mu}_n.
\end{eqnarray*}
Inequality (\ref{19apr2016eq4}) then follows from the inequality
$$\int|\log_2(d{\tilde \mu}_n/d\lambda_n)|d{\tilde\mu}_n \leq 2e^{-1}\log_2e + \int\log_2(d{\tilde \mu}_n/d\lambda_n)d{\tilde\mu}_n,$$
which is well known in statistics and information theory \cite[Lemma 5.2.6]{grayEI}.

\par

{\it Definition: Random variable $h_{\mu}$ for B-V source $\mu$.} Let $\mu\in {\cal S}(D)$, 
let $P=P_{\mu}$, and let
$\Lambda_P$ be the probability space $(\Omega_D,{\cal F}(\Omega_D),P)$. 
If $E\in{\cal F}(\Omega_D)$, the characteristic function of event $E$ is the
random variable $\chi_E:\Lambda_P\to\{0,1\}$ such that $\{\chi_E=1\}=E$.
With $W_n$ as defined in Lemma 5.1, a random variable $h_{\mu}$ is then defined on $\Lambda_P$ such that
\begin{equation}
h_{\mu} = -(\beta-1)^{-1}\left[\chi_{\{N=1\}}\log_2\mu_1(X_1) +  \sum_{n=2}^{\infty}\chi_{\{N=n\}}W_n\right]
\label{19apr2016eq6}\end{equation}
holds almost surely $[P]$. Define
$$V_n(\mu) \define \{x\in V_n: \mu(x)>0\},\;\;n\geq 0.$$
If $\omega\in\Omega_D$, then $h_{\mu}(\omega)$ is evaluated as follows:
\begin{itemize}
\item If $N(\omega)=1$ and $X_1(\omega)=x \in V_1(\mu)$, then
\begin{equation}
h_{\mu}(\omega) = -(\beta-1)^{-1}\log_2\mu(x).
\label{21apr2016eq1}\end{equation}
\item If $N(\omega)=n>1$ and $X_n(\omega)=x\in V_n(\mu)$, then
\begin{equation}
h_{\mu}(\omega) = -(\beta-1)^{-1}\log_2\frac{\mu(x)}{\prod_{i=0}^{\beta-1}\mu(x[i])}.
\label{21apr2016eq2}\end{equation}
Note that $\prod_{i=0}^{\beta-1}\mu(x[i])>0$ for $x\in V_n(\mu)$ by Lemma 5.1.
\end{itemize}

\par

{\bf Lemma 5.2.} Let $\mu\in {\cal S}(D)$, let $P=P_{\mu}$, and let
$\Lambda_P$ be the probability space $(\Omega_D,{\cal F}(\Omega_D),P)$. The random
variable $h_{\mu}$ belongs $L^1[\Lambda_P]$ and its expected value is $H_{\infty}(\mu)$.
\par

{\it Proof.} Since $N$ is a stopping time relative to the random sequence
$(Z_i:i\geq 0)$, the random variables $\chi_{\{N=n\}}$ and $W_n$ on $\Lambda_P$ are statistically independent
for every $n\geq 1$. Thus,
\begin{eqnarray*}
E_P[\chi_{\{N=n\}}W_n] &=& P\{N=n\}E_P[W_n]\\
& =& (1-\beta^{-1})\beta^{-(n-1)}E_P[W_n]\\
&=& (1-\beta^{-1})\beta^{-(n-1)}(\beta H(\mu_{n-1})-H(\mu_{n}))
\end{eqnarray*}
and
\begin{eqnarray*}
E_P[\chi_{\{N=n\}}|W_n|] &=& P\{N=n\}E_P[|W_n|]\\
& =& (1-\beta^{-1})\beta^{-(n-1)}E_P[|W_n|]\\
&\leq & (1-\beta^{-1})\beta^{-(n-1)}(C + \beta H(\mu_{n-1})-H(\mu_{n})),
\end{eqnarray*}
where $C = 2e^{-1}\log_2e$.
From equation (\ref{19apr2016eq6}), we have
$$|h_{\mu}| = (\beta-1)^{-1}\left[-\chi_{\{N=1\}}\log_2\mu_1(X_1) +  \sum_{n=2}^{\infty}\chi_{\{N=n\}}|W_n|\right].$$
By the monotone convergence theorem, it is legitimate to integrate the right side of this equation
term by term. This gives us
$$E_P[|h_{\mu}|] = \beta^{-1}\left[E_P[-\log_2\mu_1(X_1)]+ \sum_{n=2}^{\infty}\beta^{-(n-1)}E_P[|W_n|]\right] \leq$$
$$ \beta^{-1}\left[H(\mu_1) + 
\sum_{n=2}^{\infty}\beta^{-(n-1)}\{C +\beta H(\mu_{n-1})-H(\mu_{n})\}\right] =$$
$$[\beta(\beta-1)]^{-1}C + 2\beta^{-1}H(\mu_1) -  H_{\infty}(\mu) < \infty.$$
The random variable $|h_{\mu}|$ consequently belongs to $L^1[\Lambda_P]$ and thus so does $h_{\mu}$. 
By the dominated convergence theorem, the right side of
equation (\ref{19apr2016eq6}) can be integrated term by term, giving us 
$$E_P[h_{\mu}] = \beta^{-1}\left[E_P[-\log_2\mu_1(X_1)] - \sum_{n=2}^{\infty}\beta^{-(n-1)}E_P[W_n]\right] =$$
$$ \beta^{-1}\left[H(\mu_1) -
\sum_{n=2}^{\infty}\beta^{-(n-1)}\{\beta H(\mu_{n-1})-H(\mu_{n})\}\right] =$$
$$\beta^{-1}[H(\mu_1) -  (H(\mu_1)-\beta H_{\infty}(\mu))] = H_{\infty}(\mu)$$
\par

{\it Definition: Random variable $h_{\mu}^*$ for B-V source $\mu$.} Let $\mu\in {\cal S}(D)$, 
let $P=P_{\mu}$, and let
$\Lambda_P$ be the probability space $(\Omega_D,{\cal F}(\Omega_D),P)$. Define
$$h_{\mu}^* \define  E_P[h_{\mu}|{\cal F}_T],$$
 the conditional expectation of $h_{\mu}$ with
respect to the sigma-field ${\cal F}_T$ of $T$-invariant events. Since $h_{\mu}$ belongs
to $L^1[\Lambda_P]$, $h_{\mu}^*$ also belongs to $L^1[\Lambda_P]$ and we have
$$E_P[h_{\mu}^*] = E_P[h_{\mu}] = H_{\infty}(\mu).$$

\par

{\bf Theorem 5.3 (SMB Theorem).} Let $D=(V,E)$ be a $\beta$-regular Bratteli diagram, where $\beta\geq 2$.
Let $\mu\in {\cal S}(D)$, let $P=P_{\mu}$, and let
$\Lambda_P$ be the probability space $(\Omega_D,{\cal F}(\Omega_D),P)$.  
The sequence $\{-\beta^{-n}\log_2\mu_n(X_n):n\geq 0\}$ converges to $h_{\mu}^*$ in  $L^1[\Lambda_P]$ norm.
It also converges to $h_{\mu}^*$ 
almost surely $[P]$. 
\par

The proof of the SMB theorem is deferred to subsection V-A. The following result follows from
Theorem 5.3 and is the asymptotic equipartition property for ergodic Bratteli-Vershik sources. 
\par

{\bf Asymptotic Equipartition Property}. Let $D=(V,E)$ be a $\beta$-regular Bratteli diagram, where $\beta\geq 2$.
Let $\mu\in {\cal S}_e(D)$, let $P=P_{\mu}$, and let
$\Lambda_P$ be the probability space $(\Omega_D,{\cal F}(\Omega_D),P)$.
Then
\begin{equation}
\lim_{n\to\infty} E_P[|-\beta^{-n}\log_2\mu_n(X_n)-H_{\infty}(\mu)|] = 0.\label{26nov2015eq10}
\end{equation}
Also, for every $\epsilon>0$, 
\begin{equation}
\lim_{n\to\infty}P\{\left|-\beta^{-n}\log_2\mu_n(X_n)-H_{\infty}(\mu)\right| > \epsilon\} = 0.
\label{26nov2015eq11}\end{equation}
\par
{\it Proof.} Since $P\in {\cal P}_e(\Omega_D,T)$, every ${\cal F}_T$-measurable random variable
on $\Lambda_P$ is constant almost surely $[P]$. Thus, 
 $$h_{\mu}^* = E_P[h_{\mu}^*] = H_{\infty}(\mu)\;\;{\rm a.s.}\;[P].$$
Statement (\ref{26nov2015eq10}) then follows from the convergence in $L^1[\Lambda_P]$ norm in the
SMB theorem. 
The Markov inequality gives us
$$P\{\left|-\beta^{-n}\log_2\mu_n(X_n)-H_{\infty}(\mu)\right| > \epsilon\} \leq$$
$$ \epsilon^{-1}E_P[|-\beta^{-n}\log_2\mu_n(X_n)-H_{\infty}(\mu)|],$$
and then statement (\ref{26nov2015eq11}) follows from (\ref{26nov2015eq10}).
\par

{\bf Remark.} From statement (\ref{26nov2015eq11}), it follows that 
$$\lim_{n\to\infty}\mu_n(\{x\in V_n:2^{-\beta^n(H_{\infty}(\mu)+\epsilon)}\leq\mu_n(x)\leq 2^{-\beta^n(H_{\infty}(\mu)-\epsilon)}\}) = 0$$
holds for every $\epsilon>0$. This tells us that for large $n$, $\mu_n$ is approximately a uniform distribution on a subset
of $V_n$ of cardinality roughly equal to $2^{\beta^nH_{\infty}(\mu)}$. This fact has implications for fixed-length lossy encoding of ergodic Bratteli-Vershik sources, as we shall see in Section VI.  
\par
The following result identifies the probability distribution of the SMB limit function $h_{\mu}^*$ in
Theorem 5.3. In this result and subsequently, 
${\cal F}({\mathbb R})$ denotes the sigma-field of Borel subsets of ${\mathbb R}$.

\par

{\bf Theorem 5.4.} Let $D=(V,E)$ be a $\beta$-regular Bratteli diagram, where $\beta\geq 2$.
Let $\mu\in {\cal S}(D)$, let $P=P_{\mu}$, and let
$\Lambda_P$ be the probability space $(\Omega_D,{\cal F}(\Omega_D),P)$.   Then
\begin{equation}
P\{h_{\mu}^*\in G\} = \lambda_{\mu}(\{\sigma\in{\cal S}_e(D):H_{\infty}(\sigma) \in G\}),\;\;G\in{\cal F}({\mathbb R}).
\label{23nov2015eq3}
\end{equation}
\par

{\bf Remarks.} 
\begin{itemize}
\item For $\mu\in{\cal S}(D)$, let $F_{\mu}:{\mathbb R}\to[0,1]$ be the cumulative distribution function defined in (\ref{3may2016eq2}). 
Since $L^1$ convergence implies convergence in distribution,  
Theorems 5.3-5.4 give us 
$$\lim_{n\to\infty}\mu_n(\{x\in V_n(\mu):-\beta^{-n}\log_2\mu_n(x) \leq h\}) = F_{\mu}(h)$$
at every point $h\in {\mathbb R}$ at which $F_{\mu}$ is continuous. This  is the weak form of the SMB theorem (Theorem 1.9 of Section I).
 \item Formula (\ref{23nov2015eq3}) is easily seen to be true if $\mu\in {\cal S}_e(D)$, as follows.
Using the fact that $h_{\mu}^*=H_{\infty}(\mu)$ almost surely $[P_{\mu}]$,  
$$P_{\mu}\{h_{\mu}^*\in G\} = P_{\mu}(\{\omega\in{\Omega}_D: H_{\infty}(\mu)\in G\}),$$
which is $1$ or $0$ depending on whether $H_{\infty}(\mu)$ belongs to $G$ or not. 
$\lambda_{\mu}$ is the point mass at $\mu$, so that $\lambda_{\mu}(F)$ for an event $F \in {\cal F}({\cal S}_e(D))$
is $1$ or $0$ depending on whether $\mu\in F$ or not. Thus, (\ref{23nov2015eq3}) clearly holds.
However, the proof of (\ref{23nov2015eq3}) for $\mu\not\in{\cal S}_e(D)$ is more involved, as we shall see below.
\item An analogue of Theorem 5.4 is known for stationary sequential sources \cite{illjournal}
or, more generally, for asymptotically mean stationary sources \cite[Theorem 3.1.1]{grayEI}.
\end{itemize} 

{\it Proof of Theorem 5.4.} Fix $\mu\in{\cal S}(D)$. Let $P=P_{\mu}$ and we have the probability space
$\Lambda_P=(\Omega_D,{\cal F}(\Omega_D),P)$. Note that $\sigma_n(V_n(\sigma))=1$ for each $\sigma\in {\cal S}(D)$. 
Let $S_{\mu}$ be the measurable subset of ${\cal S}_e(D)$ consisting of all $\sigma\in{\cal S}_e(D)$ 
such that $\sigma_n$ is absolutely continuous with respect to $\mu_n$ for every $n\geq 0$.
That is, $\sigma\in S_{\mu}$ if and only if 
$$V_n(\sigma) = V_n(\sigma)\cap V_n(\mu),\;\;n\geq 0.$$
By the ergodic decomposition theorem, $\lambda_{\mu}(S_{\mu})=1$. 
For $n\geq 0$, let
$\phi_n:S_{\mu}\to[0,\infty)$ be the mapping
$$\phi_n(\sigma) \define \beta^{-n}\sum_{x\in V_n(\sigma)}\sigma_n(x)\log_2\frac{\sigma_n(x)}{\mu_n(x)},\;\;\sigma\in S_{\mu}.$$
($\beta^n\phi_n(\sigma)$ is simply the Kullback-Leibler distance between PMF
$\sigma_n$ and PMF $\mu_n$ and is therefore non-negative.)
It is easy to verify that
$$\int_{S_{\mu}}\phi_nd\lambda_{\mu} = \beta^{-n}H_n(\mu) - \int_{S_{\mu}}\beta^{-n}H_n(\sigma)d\lambda_{\mu}(\sigma),\;\;n\geq 0.$$
By the dominated convergence theorem and the ergodic decomposition of entropy rate, we then have
$$\lim_{n\to\infty}\int_{S_{\mu}}\phi_nd\lambda_{\mu} = H_{\infty}(\mu) - \int_{S_{\mu}}H_{\infty}(\sigma)d\lambda_{\mu}(\sigma) = 0.$$
By a well-known property of Kullback-Leibler distance \cite[Lemma 5.2.6]{grayEI}, 
for every $n\geq 0$ we have 
$$\sum_{x\in V_n(\sigma)}\sigma_n(x)\left|\log_2\frac{\sigma_n(x)}{\mu_n(x)}\right| \leq (2e^{-1}\log_2e) + \beta^n\phi_n(\sigma),\;\;\sigma\in S_{\mu}.$$
For $\sigma\in S_{\mu}$, $\epsilon>0$, and $n\geq 0$, we define subsets of $V_n$ by
\begin{eqnarray*}
W_1(\sigma,\epsilon,n) &\define& \{x\in V_n(\sigma):\;\left|\beta^{-n}\log_2\frac{\sigma_n(x)}{\mu_n(x)}\right| > \epsilon\}\\
W_2(\sigma,\epsilon,n) &\define& \{x\in V_n(\sigma):\left|-\beta^{-n}\log_2\sigma_n(x) - H_{\infty}(\sigma)\right|>\epsilon\}\\
W(\sigma,\epsilon,n) &\define& W_1(\sigma,\epsilon,n) \cup W_2(\sigma,\epsilon,n)
\end{eqnarray*}
We have
$$\sigma_n(W_1(\sigma,\epsilon,n)) \leq \epsilon^{-1}\left[(2e^{-1}\log_2e)\beta^{-n} + \phi_n(\sigma)\right],$$
and hence
\begin{equation}
\lim_{n\to\infty}\int_{S_{\mu}}\sigma_n(W_1(\sigma,\epsilon,n))d\lambda_{\mu}(\sigma) = 0,\;\;\epsilon>0.
\label{26nov2015eq1}\end{equation}
We also have
\begin{equation}
\lim_{n\to\infty}\int_{S_{\mu}}\sigma_n(W_2(\sigma,\epsilon,n))d\lambda_{\mu}(\sigma) = 0,\;\;\epsilon>0.
\label{26nov2015eq2}\end{equation}
(The integrand converges to zero pointwise for $\sigma\in S_{\mu}$ by the asymptotic equipartition property,
and then integrate, applying the dominated convergence theorem.)
We conclude from (\ref{26nov2015eq1})-(\ref{26nov2015eq2}) that
\begin{equation}
\lim_{n\to\infty}\int_{S_{\mu}}\sigma_n(W(\sigma,\epsilon,n))d\lambda_{\mu}(\sigma) = 0,\;\;\epsilon>0.
\label{26nov2015eq3}\end{equation}
In the following, fix $y\in{\mathbb R}$ and $\epsilon>0$. For $\sigma\in S_{\mu}$ and $n\geq 0$, we have
$$\{x\in V_n(\sigma): \;-\beta^{-n}\log_2\mu_n(x)\leq y\} \subset$$
$$ \{x\in V_n(\sigma):\;H_{\infty}(\sigma)\leq y+2\epsilon\} \cup W(\sigma,\epsilon,n),$$
and therefore
$$\sigma_n(\{x\in V_n:-\beta^{-n}\log_2\mu_n(x)\leq y\}) \leq$$
$$\sigma_n(\{x\in V_n: H_{\infty}(\sigma)\leq y+2\epsilon\}) + \sigma_n(W(\sigma,\epsilon,n)) = $$
$$G(\sigma) + \sigma_n(W(\sigma,\epsilon,n)),$$
where $G:S_{\mu}\to{\mathbb R}$ is the function such that
\[ G(\sigma) \define \left \{ \begin{array} {r@{\quad}l}
1, & H_{\infty}(\sigma)\leq y+2\epsilon\\
0, & {\rm otherwise}
\end{array} \right. \]
Integrating,
$$\mu_n(\{x\in V_n:-\beta^{-n}\log_2\mu_n(x)\leq y\}) =$$
$$\int_{S_{\mu}}\sigma_n(\{x\in V_n:-\beta^{-n}\log_2\mu_n(x)\leq y\})d\lambda_{\mu}(\sigma) \leq$$
$$\int_{S_{\mu}}G(\sigma)d\lambda_{\mu}(\sigma) + \int_{S_{\mu}}\sigma_n(W(\sigma,\epsilon,n))d\lambda_{\mu}(\sigma).$$
Using (\ref{26nov2015eq3}) and the fact that
$$\int_{S_{\mu}}G(\sigma)d\lambda_{\mu}(\sigma) = \lambda_{\mu}(\{\sigma\in {\cal S}_e(D):H_{\infty}(\sigma)\leq y+2\epsilon\}),$$
we obtain the upper bound
$$\varlimsup_{n\to\infty} \mu_n(\{x\in V_n:-\beta^{-n}\log_2\mu_n(x)\leq y\})\leq$$
\begin{equation}
\lambda_{\mu}(\{\sigma\in {\cal S}_e(D):H_{\infty}(\sigma)\leq y+2\epsilon\}).
\label{26nov2015eq5}\end{equation}
Note that
$$\{x\in V_n(\sigma): \;H_{\infty}(\sigma)\leq y-2\epsilon\} \subset$$
$$ \{x\in V_n(\sigma):\;-\beta^{-n}\log_2\mu_n(x)\leq y\} \cup W(\sigma,\epsilon,n).$$
We obtain from this the lower bound 
$$\varliminf_{n\to\infty} \mu_n(\{x\in V_n:-\beta^{-n}\log_2\mu_n(x)\leq y\})\geq$$
\begin{equation}
\lambda_{\mu}(\{\sigma\in {\cal S}_e(D):H_{\infty}(\sigma)\leq y-2\epsilon\}),
\label{26nov2015eq6}\end{equation}
via reasoning steps similar to the steps used in obtaining (\ref{26nov2015eq5}). 
Let $F_1:{\mathbb R}\to[0,1]$ and $F_2:{\mathbb R}\to[0,1]$ be the cumulative distribution functions defined by
\begin{eqnarray*}
F_1(y) &\define& P\{h_{\mu}^*\leq y\},\;\;y\in{\mathbb R}\\
F_2(y) &\define& \lambda_{\mu}(\{\sigma\in{\cal S}_e(D): H_{\infty}(\sigma)\leq y\}),\;\;y\in{\mathbb R}
\end{eqnarray*}
We complete the proof by showing that $F_1=F_2$, which implies statement (\ref{23nov2015eq3}).
 Let $Q$ be the set of all real numbers at which 
$F_1,F_2$ are both continuous. Let $y\in Q$.
By almost sure convergence in the SMB theorem, we also have convergence
in distribution. Thus,
$$\lim_{n\to\infty}\mu_n(\{x\in V_n:-\beta^{-n}\log_2\mu_n(x)\leq y\}) = F_1(y).$$
By (\ref{26nov2015eq5})-(\ref{26nov2015eq6}), 
$$F_2(y-2\epsilon) \leq F_1(y) \leq F_2(y+2\epsilon),\;\;\epsilon>0.$$
Letting $\epsilon\to 0$, we obtain
$$F_2(y) \leq F_1(y) \leq F_2(y)$$
and hence
$$F_1(y) = F_2(y),\;\;y\in Q.$$
The complement of $Q$ is a countable set and therefore $Q$ is dense in $\mathbb R$. Fix
any real number $y$. Our proof is complete once we show $F_1(y)=F_2(y)$.
Let $\{y_n:n\geq 1\}$ be a sequence in $Q$
converging downward to $y$. We have
$$F_1(y_n) = F_2(y_n),\;\;n\geq 1,$$
and $F_1$ and $F_2$ are right continuous functions. Therefore,
$$F_1(y) = \lim_{n\to\infty}F_1(y_n) = \lim_{n\to\infty}F_2(y_n) = F_2(y).$$

\subsection{Proof of Theorem 5.3}

Fix $\beta$-regular Bratteli diagram $D=(V,E)$ and  $\mu \in {\cal S}(D)$. Let $P=P_{\mu}$. We have the probability space $\Lambda_P=(\Omega_D,{\cal F}(\Omega_D),P)$. 
Define random variable $U_i = h_{\mu}\circ T^i$ on $\Lambda_P$ ($i\in{\mathbb Z}$). 
The following result plays a key role in our proof of Theorem 5.3.\par

{\bf Lemma 5.5.} For $n\geq 1$,
\begin{equation}
-\log_2\mu_n(X_n) = \sum_{j=-Z^{(n)}}^{-Z^{(n)}+\beta^n-2}U_j\;\;{\rm a.s.}\;[P].
\label{23nov2015eq1}\end{equation}
\par

{\it Proof.} Fix $n\geq 1$ and $x\in V_n$ such that $\mu_n(x)>0$. For each 
$i\in S_{\beta,n}=\{0,1,\cdots,\beta^n-1\}$,
recall that we have the cylinder subset
$$C_n(i,x) = \{X_n=x,\;Z^{(n)}=i\}$$
of $\Omega_D$. We show that
\begin{equation}
-\log_2\mu_n(x) = \sum_{j=-i}^{-i+\beta^n-2}h_{\mu}(T^j\omega),\;\;\omega\in C_n(i,x),\;\;i\in S_{\beta,n},
\label{2dec2015eq1}\end{equation}
which establishes the statement (\ref{23nov2015eq1}).
Let $t(n)$ be the finite rooted labeled tree in which 
\begin{itemize}
\item Each non-leaf vertex $u$ has $\beta$ ordered child vertices, denoted by
$c[u,0], c[u,1), \cdots, c[u,\beta-1]$. 
\item There are $\beta^n$ leaves, and every root-to-leaf path consists of $n$ edges. 
\item Each vertex $u$ of $t(n)$ carries a label $x^u\in V$, determined as follows:
the root vertex of $t(n)$ carries the label $x$, and for each non-leaf vertex $u$ of $t(n)$,
child vertex $w=c[u,i]$ carries label $x^w=x^u[i]$ ($i=0,1,\cdots,\beta-1)$.
\end{itemize}
Let $u^*$ denote the root vertex of $t(n)$. 
Let $u$ be a leaf of $t(n)$. Starting at $u$ and following the leaf-to-root path,
we encounter vertices $u_0,u_1,\cdots,u_n$ in order, where $u_0=u$, $u_n=u^*$,
and $u_j$ is the parent vertex of $u_{j-1}$ ($j=1,\cdots,n$). The sequence $(i_0,i_1,\cdots,i_{n-1})\in\{0,1,\cdots,\beta-1\}^n$ such that $u_j=c[u_{j+1},i_j]$ ($j=0,\cdots,n-1$) is defined to be the {\it address}
of $u$. Also, the integer 
$$i_0+i_1\beta+\cdots+i_{n-1}\beta^{n-1}$$
is called the {\it index} of $u$. For each integer $i\in S_{\beta,n}$, we let $u(i)$ be 
the leaf of $t(n)$ with index $i$. The list $u(0),u(1),\cdots,u(\beta^n-1)$ provides an enumeration
of all leaves of $t(n)$. Let ${\cal V}(t(n))$ be the set of all vertices of $t(n)$ and
let ${\cal V}^+(t(n))$ denote the set of all non-leaf vertices of $t(n)$. 
It is easy to verify the following property:
\begin{itemize}
\item {\bf Property 1:} Let $\phi:{\cal V}(t(n))\to {\mathbb R}$ be any function
such that $\phi(u)=0$ for every leaf $u$ of $t(n)$. Let $\Psi_{\phi}:{\cal V}^+(t(n))\to{\mathbb R}$
be the function
$$\Psi_{\phi}(u) \define \phi(u)-\sum_{i=0}^{\beta-1}\phi(c[u,i]),\;\;u\in {\cal V}^+(t(n)).$$
Then 
$$\phi(u^*) = \sum_{u\in{\cal V}^+(t(n))}\Psi_{\phi}(u).$$
\end{itemize}
Let
$${\cal V}_1(t(n)) = \{u(i):0\leq i < \beta^{n}-1\}.$$
Note that ${\cal V}_1(t(n))$ is the set of all leaves of $t(n)$ other than the leaf $u(\beta^n-1)$,
 which is the only leaf of $t(n)$ whose address has every entry equal to $\beta-1$.
We define mapping $F:{\cal V}_1(t(n))\to {\cal V}^+(t(n))$ as follows. Let $u\in {\cal V}_1(t(n))$,
and let $(i_0,\cdots,i_{n-1})$ be the address of $u$. Let $J$ be the smallest integer $j\in\{0,\cdots,n-1\}$
such that $i_j<\beta-1$. Letting $u_0,\cdots,u_n$ be the vertices in order along to leaf-to-root
path starting at $u$, we define $F(u)=u_{i_{J+1}}$. The function $F$ obeys the following property:
\begin{itemize}
\item {\bf Property 2:} $F$ maps onto ${\cal V}^+(t(n))$. Moreover, 
$$|\{u\in {\cal V}_1(t(n)): F(u)=w\}| = \beta-1,\;\;w\in{\cal V}^+(t(n)).$$
\end{itemize}
(For example, the $\beta-1$ leaves in ${\cal V}_1(t(n))$ mapped into $u^*$ by $F$
have the addresses $(\beta-1,\beta-1,\cdots,\beta-1,i)$ ($i=0,1,\cdots,\beta-2$),
from which it can be worked out that this set of leaves is $\{u(\beta^{n-1}i-1):i=1,\cdots,\beta-1\}$.) Letting
$\phi$ be a function under Property 1, then, combining Properties 1-2, we conclude
\begin{equation}
\phi(u^*) = (\beta-1)^{-1}\sum_{w\in{\cal V}_1(t(n))}\Psi_{\phi}(F(w)).
\label{1dec2015eq2}\end{equation}
By choosing an appropriate function $\phi$ in equation (\ref{1dec2015eq2}), 
we will obtain (\ref{2dec2015eq1}), completing our proof. 
By Lemma 5.1, 
$$\mu(x^u)>0,\;\;u\in {\cal V}(t(n)).$$
This property ensures the existence of 
the particular function $\phi:{\cal V}(t(n))\to{\mathbb R}$ 
which is $0$ over the leaves of $t(n)$ and satisfies
$$\phi(u) = -\log_2\mu(x^u),\;\;u\in{\cal V}^+(t(n)).$$
Let $\omega\in C_n(k,x)$, where $k<\beta^n-1$. The path $\omega_0^{n-1} \in \Pi_D(0,n,x)$
and the leaf $u(k)$ of $t(n)$ have the same address $(i_0,\cdots,i_{n-1})=[k]_{\beta,n}$.
Letting $u_0,\cdots,u_n$ be the vertices of $t(n)$ along to leaf-to-root path starting at
$u(k)$, the label assigned to vertex $u_j$ is $X_j(\omega)$.  
Let $N=N(\omega)$. Letting $J$ be the smallest integer $j\in\{0,\cdots,n-1\}$ such that
$i_j<\beta-1$, we have $J+1=N$, and therefore $F(u(k))=u_N$ and $x^{u_N}=X_N(\omega)$. 
If $N=1$, the children of $u_N$ are leaves of $t(n)$, and we have
$$\Psi_{\phi}(F(u(k)) = \phi(u_N) - \sum_{i=0}^{\beta-1}\phi(c[u_N,i]) =$$
$$ \phi(u_N) = -\log\mu(X_1(\omega)) = (\beta-1)h_{\mu}(\omega)$$
by formula (\ref{21apr2016eq1}). If $N>1$, the children of $u_N$ are not leaves of $t(n)$,
and we have
$$\Psi_{\phi}(F(u(k)) = \phi(u_N) - \sum_{i=0}^{\beta-1}\phi(c[u_N,i]) =$$
$$-\log\mu(X_N(\omega)) + \sum_{i=0}^{\beta-1}\log_2\mu(X_N(\omega)[i]) = (\beta-1)h_{\mu}(\omega)$$
by formula (\ref{21apr2016eq2}).
We have shown that
\begin{equation}
\Psi_{\phi}(F(u(k))) = (\beta-1)h_{\mu}({\omega}),\;\;\omega\in C_n(k,x),\;0\leq k\leq \beta^n-2.
\label{1dec2015eq1}\end{equation}
Now let $i\in S_{\beta,n}$ be arbitrary and let $\omega\in C_n(i,x)$. 
Then a property of the Vershik transformation tells us that
$$T^j\omega \in C_n(i+j,x),\;\;-i \leq j \leq -i+\beta^n-1.$$
Combining this statement with (\ref{1dec2015eq1}), we have
$$\Psi_{\phi}(F(u(i+j))) = (\beta-1)h_{\mu}(T^j{\omega}),\;\;-i \leq j \leq -i+\beta^n-2,$$
and also
$$\{u(i+j):-i \leq j \leq -i+\beta^n-2\} = {\cal V}_1(t(n)).$$
Employing (\ref{1dec2015eq2}), we have
$$(\beta-1)\sum_{j=-i}^{-i+\beta^n-2}h_{\mu}(T^j{\omega}) = \sum_{w\in{\cal V}_1(t(n))}\Psi_{\phi}(F(w)) = $$
$$(\beta-1)\phi(u^*) = -(\beta-1)\log_2\mu_n(x),$$
and thus formula (\ref{2dec2015eq1}) holds, completing our proof. 
\par

{\it Proof of almost sure convergence in Theorem 5.3.} By Lemma 5.5,
for each $n\geq 1$, we have
$$-\log_2\mu_n(X_n) = (\sum_{j=-Z^{(n)}}^0U_j) + (\sum_{j=-1}^{-Z^{(n)}+\beta^n-2}U_j) -U_{-1}-U_0 $$
almost surely $[P]$. 
Both of the sums on the right side of the preceding equation are non-empty, consisting of $Z^{(n)}+1$
terms and  $\beta^n-Z^{(n)}$ terms, respectively. (Since
$0 \leq Z^{(n)}\leq \beta^n-1$, the random variables $Z^{(n)}+1$ and $\beta^n-Z^{(n)}$
are positive integer valued.)  We then have

$$-\beta^{-n}\log_2\mu_n(X_n) = \left(\frac{Z^{(n)}+1}{\beta^n}\right)\left(\frac{\sum_{j=-Z^{(n)}}^0U_j}{Z^{(n)}+1}\right)$$
\begin{equation}
 + \left(\frac{\beta^n-Z^{(n)}}{\beta^n}\right)\left(\frac{\sum_{j=-1}^{-Z^{(n)}+\beta^n-2}U_j}{\beta^n-Z^{(n)}}\right) - \frac{U_{-1}+U_0}{\beta^n}
\label{19noveq1}\end{equation}
Note that
$$Z^{(n+1)}-Z^{(n)} = \beta^nZ_n \geq 0.$$
Since the sequence $(Z_0,Z_1,\cdots)$ is aperiodic, $Z_n$ must be $>0$ for infinitely many $n$.
It follows that $Z^{(n)}\to\infty$ everywhere on ${\Omega}_D$. Thus, 
\begin{equation}
\lim_{n\to\infty} \left(\frac{\sum_{j=-Z^{(n)}}^0U_j}{Z^{(n)}+1}\right) = h_{\mu}^*\;\;{\rm a.s.}\;[P],
\label{19noveq2}\end{equation}
because the pointwise ergodic theorem \cite[Thm.\ 3.3.6]{ashgardner} tells us that
$$\lim_{m\to\infty} (m+1)^{-1}\sum_{j=-m}^{0}U_j = h_{\mu}^*\;\;{\rm a.s.}\;[P].$$
Also, note that
$$(\beta^{n+1}-Z^{(n+1)})-(\beta^n-Z^{(n)}) = \beta^n[(\beta-1)-Z_n] \geq 0.$$
We must have $Z_n<\beta-1$ for infinitely many $n$, so
$\beta^n-Z^{(n)}\to\infty$ everywhere on ${\Omega}_D$. Thus, 
\begin{equation}
\lim_{n\to\infty} \left(\frac{\sum_{j=-1}^{-Z^{(n)}+\beta^n-2}U_j}{\beta^n-Z^{(n)}}\right) = h_{\mu}^*\;\;{\rm a.s.}\;[P],
\label{19noveq3}\end{equation}
because the pointwise ergodic theorem tells us that 
$$\lim_{m\to\infty} m^{-1}\sum_{j=-1}^{m-2}U_j = h_{\mu}^*\;\;{\rm a.s.}\;[P].$$
Statements (\ref{19noveq1})-(\ref{19noveq3}) easily imply that
$$\lim_{n\to\infty}-\beta^{-n}\log_2\mu_n(X_n) = h_{\mu}^*\;\;{\rm a.s.}\;[P].$$

\par

{\it Proof of $L^1[\Lambda_P]$ convergence in Theorem 5.3.} Via equation (\ref{23nov2015eq1}) and the fact that
$0 \leq Z^{(n)} \leq \beta^n-1$, 
\begin{equation}
\left|-\beta^{-n}\log_2\mu_n(X_n)\right| \leq 2\left[\frac{\sum_{j=-\beta^n}^{\beta^n-1}|U_j|}{2\beta^n}\right]
\label{23nov2015eq2}\end{equation}
holds almost surely for each $n\geq 1$. By the $L^1$  ergodic theorem \cite[Thm.\ 3.3.7]{ashgardner}, the sequence
$\{(2m)^{-1}\sum_{j=-m}^{m-1}|U_j|:m\geq 1\}$ is convergent in  $L^1[\Lambda_P]$ norm;  therefore, the sequence is uniformly integrable.
Consequently, appealing to the bound (\ref{23nov2015eq2}), the sequence $\{-\beta^{-n}\log_2\mu_n(X_n):n\geq 0\}$ is also uniformly integrable.
A uniformly integrable sequence of random variables which is almost surely convergent is also
convergent in $L^1$ norm (to the same limit function) \cite[Cor. 6.5.5]{ashdade}. Thus, the sequence $\{-\beta^{-n}\log_2\mu_n(X_n):n\geq 0\}$
converges in $L^1[\Lambda_P]$ norm to $h_{\mu}^*$.

\section{Fixed-Length Lossy Source Encoding}

Let $D=(V,E)$ be a $\beta$-regular Bratteli diagram. Let $n\geq 0$. We define an $n$-th order
fixed-length encoder for $D$ to be any mapping $\phi_n:V_n\to\{0,1\}^*$
in which all of the binary codewords in $\phi_n(V_n)$ have the same length. We do not
require that $\phi_n$ be one-to-one, that is, we are now allowing lossy encoding rather than lossless encoding.
The encoding rate $R(\phi_n)$ of $n$-th order fixed-length encoder $\phi_n$ is defined to be
$L_n/\beta^n$, where $L_n$ is the fixed length of $\phi_n$'s codewords.
\par

Let $\delta\in(0,1)$. Let $\mu$ be any source in ${\cal S}(D)$. We define ${\mathbb E}_{\mu}(\delta)$
to be the set of all sequences $\{\phi_n:n\geq 0\}$ such that
\begin{itemize}
\item {\bf (a):} For each $n\geq 0$, $\phi_n$ is an $n$-th order fixed-length encoder
for $D$.
\item {\bf (b):} For each $n\geq 0$, there is a mapping $\psi_n:\phi_n(V_n)\to V_n$ such that
\begin{equation}
\mu_n(\{x\in V_n:\psi_n(\phi_n(x))=x\}) \geq 1-\delta.
\label{16dec2015eq2}\end{equation}
\end{itemize}
The sequences in ${\mathbb E}_{\mu}(\delta)$ shall be referred to as $\delta$-lossy encoder sequences for $\mu$.
The parameter $\delta$ is called the {\it error level} of the encoders in ${\mathbb E}_{\mu}(\delta)$. 
The error level $\delta$ controls the degree to which an encoder sequence $\{\phi_n\}$ can be lossy; with the
$\delta$-lossy requirement, we are stipulating that a vertex in $V_n$ can be decoded from its binary codeword
except for a set of vertices of $\mu_n$-probability at most $\delta$ ($n\geq 0$). 
\par
{\it Definition.} For a given error level $\delta\in(0,1)$ and source $\mu\in{\cal S}(D)$, we wish to examine how small the rate sequence
$\{R(\phi_n):n\geq 0\}$ can become asymptotically as $n\to\infty$ for the
encoder sequences $\{\phi_n\}$ in ${\mathbb E}_{\mu}(\delta)$. We make this idea precise as follows.
We define source $\mu$ to be {\it stably encodable} at error level $\delta$ if there exists a
(necessarily unique) non-negative real number $R^*(\delta,\mu)$ such that
both of the following two statements hold:
\begin{itemize}
\item {\bf (a):} There exists an encoder sequence  $\{\phi_n\}$ in ${\mathbb E}_{\mu}(\delta)$
for which
\begin{equation}
\varlimsup_{n\to\infty}R(\phi_n) \leq R^*(\delta,\mu).
\label{17dec2015eq3}\end{equation}
\item {\bf (b):} For any encoder sequence in $\{\phi_n\}$ in ${\mathbb E}_{\mu}(\delta)$,
\begin{equation}
\varliminf_{n\to\infty}R(\phi_n) \geq R^*(\delta,\mu).
\label{16dec2015eq3}\end{equation}
\end{itemize}
A question of interest is to determine whether a given Bratteli-Vershik source $\mu\in {\cal S}(D)$ is stably encodable at a given
error level $\delta$, and, if so, to specify how $R^*(\delta,\mu)$ is to be computed. Theorem 6.1
below elucidates this question. It implies that $\mu$ is stably encodable at certain error levels 
related to the SMB theorem limit function, and computes $R^*(\delta,\mu)$ at these levels $\delta$.  The exceptional
error levels not covered by Theorem 6.1 are at most countable in number.
 
\par

{\it Definition.} Let $\mu\in{\cal S}(D)$.  Let $F_{\mu}:{\mathbb R}\to[0,1]$ be the cumulative distribution function 
defined in (\ref{3may2016eq2}). For each $\delta\in(0,1)$, we define
$$R^+(\delta,\mu) \define \inf\{x\in{\mathbb R}: F_{\mu}(x)>1-\delta\}$$
$$= \inf\{x\in {\mathbb R}:\lambda_{\mu}(\{\sigma\in{\cal S}_e(D):H_{\infty}(\sigma)\leq x\})> 1-\delta\}.$$
and we define
$$R^-(\delta,\mu) \define \sup\{x\in{\mathbb R}: F_{\mu}(x)<1-\delta\}$$
$$= \sup\{x\in{\mathbb R}:\lambda_{\mu}(\{\sigma\in{\cal S}_e(D):H_{\infty}(\sigma) < x\})< 1-\delta\}.$$
\par
{\it Discussion.} Fix $\mu\in{\cal S}(D)$ throughout this discussion. 
Both $R^-(\delta,\mu)$ and $R^+(\delta,\mu)$ are nonincreasing functions of $\delta \in (0,1)$.
We have
$$0\leq R^-(\delta,\mu) \leq R^+(\delta,\mu),\;\;0<\delta<1.$$
Suppose we have a particular $\delta\in(0,1)$ for which $R^-(\delta,\mu) < R^+(\delta,\mu)$.
As discussed in \cite{partha} \cite{epsiloncap}, we then have
\begin{eqnarray*}
\lambda_{\mu}(\{\sigma\in{\cal S}_e: H_{\infty}(\sigma) \leq R^-(\delta,\mu)\}) &=& 1-\delta\\
\lambda_{\mu}(\{\sigma\in{\cal S}_e: H_{\infty}(\sigma) \geq R^+(\delta,\mu)\}) &=& \delta,
\end{eqnarray*}
which implies that the open interval $(R^-(\delta,\mu),R^+(\delta,\mu))$
yields an ``entropy rate gap'', meaning that
$$\lambda_{\mu}(\{\sigma\in{\cal S}_e(D): R^-(\delta,\mu) < H_{\infty}(\sigma) < R^+(\delta,\mu)\}) = 0.$$
There can be only countably many such entropy rate gaps since they are pairwise disjoint and
each one contains a rational number. We conclude that
$$R^-(\delta,\mu) = R^+(\delta,\mu)$$
for all but countably many $\delta\in(0,1)$.
\par

The following result tells us about lossy fixed-length encoding of non-ergodic Bratteli sources.
Parthasarathy \cite{partha} proved the analogous result for 
stationary non-ergodic finite-alphabet sequential sources.
\par

{\bf Theorem 6.1.} Let $\mu\in{\cal S}(D)$. Let $\delta\in(0,1)$. Then the following
statements hold.
\begin{itemize}
\item {\bf (a):} There exists an encoder sequence  $\{\phi_n\}$ in ${\mathbb E}_{\mu}(\delta)$
for which
\begin{equation}
\varlimsup_{n\to\infty}R(\phi_n) \leq R^+(\delta,\mu).
\label{17dec2015eq4}\end{equation}
\item {\bf (b):} For any encoder sequence in $\{\phi_n\}$ in ${\mathbb E}_{\mu}(\delta)$,
\begin{equation}
\varliminf_{n\to\infty}R(\phi_n) \geq R^-(\delta,\mu).
\label{17dec2015eq6}\end{equation}
\item {\bf (c):} If $R^-(\delta,\mu) = R^+(\delta,\mu)$, then $\mu$ is stably encodable at
error level $\delta$ and
$$R^*(\delta,\mu) = R^-(\delta,\mu).$$
\item {\bf (d):} If $\mu$ is stably encodable at
error level $\delta$, then
$$R^-(\delta,\mu) \leq R^*(\delta,\mu) \leq R^+(\delta,\mu).$$
\end{itemize}
\par

{\bf Remarks.} 
\begin{itemize}
\item Let $M_n(\delta,\mu)$ be the integer defined by (\ref{3may2016eq1}). It is straightforward
to show that any encoding scheme  $\{\phi_n\}$ in  ${\mathbb E}_{\mu}(\delta)$ for which
$$R(\phi_n) = \beta^{-n}\lceil\log_2M_n(\delta,\mu)\rceil,\;\;n\geq 0$$
yields the
minimum $R(\phi_n)$ for every $n\geq 0$. 
It follows that $\mu$ is stably encodable at
error level $\delta$ if and only if $\lim_n\beta^{-n}\log_2 M_n(\delta,\mu)$ exists, in which case
$R^*(\delta,\mu)$ equals this limit. From these observations, one sees that Theorem 1.10 is equivalent
to Theorem 6.1.
\item Let $\mu\in{\cal S}_e(D)$. Then $\mu$ is
stably encodable at every error level, and
$$R^*(\delta,\mu) = H_{\infty}(\mu),\;\;0 < \delta < 1.$$
Equivalently,
$$\lim_{n\to\infty}\beta^{-n}\log_2M_n(\delta,\mu) = H_{\infty}(\mu),\;\;0<\delta<1.$$
This follows from Theorem 6.1 because $\lambda_{\mu}$ is the
point mass at $\{\mu\}$, which implies
$$H_{\infty}(\mu) = R^-(\mu,\delta) = R^+(\mu,\delta),\;\;0<\delta<1.$$
\item We know how to find non-ergodic B-V sources $\mu$ for which 
the exceptional set 
$$S(\mu) = \{\delta: R^-(\delta,\mu) < R^+(\delta,\mu)\}$$
is non-empty and $\mu$ is stably encodable at
every error level in $S(\mu)$. We also know how to find non-ergodic B-V
sources $\mu$ for which $S(\mu)$ is non-empty and $\mu$ fails
to be stably encodable at every error level in $S(\mu)$. Thus, Theorem 6.1
is not sufficient to completely analyze non-ergodic B-V sources; more work is needed.
\end{itemize}
\par

Parts (c) and (d) of Theorem 6.1 follow from part (a) and part (b). Parts (a) and (b) are proved
in the following two subsections, which complete Section 6.

\subsection{Proof of Theorem 6.1(a)} 
Fix $\mu$ and $\delta$. Let $\epsilon>0$ be arbitrary. It suffices to find
a sequence $\{\phi_n:n\geq 0\}$ in ${\mathbb E}_{\mu}(\delta)$ for which
\begin{equation}
\varlimsup_{n\to\infty} R(\phi_n) \leq R^+(\delta,\mu) + \epsilon.
\label{17dec2015eq5}\end{equation}
Choose $h\in (R^+(\delta,\mu),R^+(\delta,\mu)+\epsilon)$ such that $F_{\mu}$ is continuous at $h$.
By definition of $R^+(\delta,\mu)$, 
$$F_{\mu}(h) > 1-\delta.$$
By the weak form of the SMB theorem (Theorem 1.9), 
$$\lim_{n\to\infty}\mu_n(\{x\in V_n:-\beta^{-n}\log_2\mu_n(x) \leq h\}) = F_{\mu}(h).$$
Letting
$$S_n \define \{x\in V_n: \mu_n(x) \geq 2^{-\beta^nh}\},\;\;n\geq 0,$$
we have
$$\lim_{n\to\infty}\mu_n(S_n) > 1-\delta.$$
Pick non-negative integer $N$ such that
$$\mu_n(S_n) > 1-\delta,\;\;n\geq N.$$
Note that
$$|S_n| \leq 2^{\beta^nh},\;\;n\geq 0.$$
For $0\leq n < N$, let $\phi_n$ be any fixed-length
$n$-th order lossless encoder. For $n\geq N$, let $\phi_n$
be a fixed-length $n$-th order encoder which
is one-to-one when restricted to $S_n$ and employs
the codeword length $\lceil\log_2|S_n|\rceil$. 
Then $\{\phi_n\}$ belongs to ${\mathbb E}_{\mu}(\delta)$ and
(\ref{17dec2015eq5}) holds because
$$R(\phi_n) \leq \beta^{-n}\lceil\log_2|S_n|\rceil\leq \beta^{-n} + h \leq \beta^{-n} + R^+(\delta,\mu) + \epsilon,\;\;n\geq N.$$

\subsection{Proof of Theorem 6.1(b)}
Let $\{\phi_n:n\geq 0\}$ be an encoder sequence in 
${\mathbb E}_{\mu}(\delta)$. For each $n\geq 0$,
we have a decoder mapping $\psi_n:\phi_n(V_n)\to V_n$ such that (\ref{16dec2015eq2}) holds. Define
$$F_n \define \{x\in V_n: \psi_n(\phi_n(x))=x\},\;\;n\geq 0.$$
We have
$$\mu_n(F_n)\geq 1-\delta,\;\;n\geq 0.$$
Let $L_n$ be the codeword length employed by encoder $\phi_n$. Then $\phi_n(F_n) \subset \{0,1\}^{L_n}$, so we have
$$|\phi_n(F_n)| \leq 2^{L_n} = 2^{\beta^nR(\phi_n)}.$$
There is a one-to-one correspondence between the sets $F_n$ and $\phi_n(F_n)$, and so these two sets have the same
cardinality. Thus, we have
$$\log_2|F_n| \leq \beta^nR(\phi_n),\;\;n\geq 0.$$
Inequality (\ref{17dec2015eq6}) will follow once we show that
\begin{equation}
\varliminf_{n\to\infty}\beta^{-n}\log_2|F_n| \geq R^-(\delta,\mu).
\label{18dec2015eq1}\end{equation}
If $R^-(\delta,\mu)=0$, there is nothing to prove, so assume that $R^-(\delta,\mu)>0$.
Let $\epsilon$ be an arbitrary number in the interval $(0,R^-(\delta,\mu))$. Choose $h\in (R^-(\epsilon,\mu)-\epsilon,R^-(\delta,\mu))$ such that $F_{\mu}$ is continuous at $h$.
By definition of $R^-(\delta,\mu)$, 
$$F_{\mu}(h) < 1-\delta.$$
By the weak form of the SMB theorem (Theorem 1.9), 
$$\lim_{n\to\infty}\mu_n(\{x\in V_n:-\beta^{-n}\log_2\mu_n(x) < h\}) = F_{\mu}(h).$$
Let $\delta'=1-F_{\mu}(h)$, and define
$$S_n \define \{x\in V_n: \mu_n(x) \leq 2^{-\beta^nh}\},\;\;n\geq 0.$$
Then
$$\lim_{n\to\infty}\mu_n(S_n) = \delta' > \delta.$$
We have
$$\mu_n(S_n\cap F_n) \geq \mu_n(F_n) + \mu_n(S_n) - 1 \geq \mu_n(S_n)-\delta,$$
and hence
$$\varliminf_{n\to\infty}\mu_n(S_n\cap F_n) \geq \delta'-\delta > 0.$$
We may pick non-negative integer $N$ such that 
$$\mu_n(S_n\cap F_n) > 0,\;\;n\geq N.$$
The sequence $\{\mu_n(S_n\cap F_n):n\geq N\}$
is bounded away from $0$, and hence
$$\lim_{n\to\infty}\beta^{-n}\log_2\mu_n(S_n\cap F_n) = 0.$$
By definition of the set $S_n$, we have
$$|S_n\cap F_n| \geq 2^{\beta^nh}\mu_n(S_n\cap F_n),\;\;n\geq 0.$$
Thus,
$$\beta^{-n}\log_2|S_n\cap F_n| \geq h + \beta^{-n}\log_2\mu_n(S_n\cap F_n),\;\;n\geq N.$$
The last term on the right side drops out as $n\to\infty$, and we thus have
$$\varliminf_{n\to\infty}\beta^{-n}\log_2|F_n| \geq \varliminf_{n\to\infty}\beta^{-n}\log_2|S_n\cap F_n| \geq h > R^-(\delta,\mu)-\epsilon.$$
As the preceding holds for all $\epsilon>0$ sufficiently small,
 (\ref{18dec2015eq1}) holds and our proof is complete.

\bibliographystyle{unsrt}
\bibliography{firstbrattelipaper6}

\end{document}